\documentclass[a4paper,mathleft,final]{an}
\usepackage{epsfig}
\usepackage{array}
\usepackage{graphicx}
\usepackage{times}

\newcommand{\Msun}{M$_{\odot}$}
\newcommand{\Mdot}{M$_{\odot}$\,yr$^{-1}$}
\newcommand{\kms}{km\,s$^{-1}$}
\newcommand{\etal}{et al.}
\newcommand{\nii}{[N\,{\sc ii}]}
\newcommand{\oiii}{[O\,{\sc iii}]}
\newcommand{\Teff}{T_{\rm eff}}
\newcommand{\Lsun}{L$_{\odot}$}

\def\changed{}

\begin{document}

\Pagespan{378}{408}
\Yearpublication{2014}%
\Yearsubmission{2014}%
\Month{2}%
\Volume{335}%
\Issue{4}%
 \DOI{10.1002/asna.201412051}%

\title{A hydrodynamical study of multiple-shell planetary nebulae}  

\subtitle{\large III. Expansion properties and internal kinematics:
               Theory versus observation\thanks{Based partly on observations obtained at the European Southern
Observatory, Paranal, Chile (ESO programme No. 077.D-0652).}}

\titlerunning{A hydrodynamical study of multiple-shell planetary nebulae III.}
\author{D. Sch\"onberner\inst{1}\fnmsep\thanks{Corresponding author: deschoenberner@aip.de} 
       \and R. Jacob\inst{1}
       \and H. Lehmann\inst{2} 
       \and G. Hildebrandt\inst{1}\fnmsep\thanks{Deceased on 2004 December 23.}
       \and M. Steffen\inst{1}
       \and A. Zwanzig\inst{1}
       \and\\ C. Sandin\inst{1}
       \and   R.L.M. Corradi\inst{3,4}
        }

\institute{Leibniz-Institut f\"ur Astrophysik Potsdam,
           An der Sternwarte 16, D-14482 Potsdam, Germany
           \and 
           Th\"uringer Landessternwarte Tautenburg,
           Karl-Schwarzschild-Observatorium,
           D-07778 Tautenburg, Germany
          \and 
           Instituto de Astrof\'isica de Canarias, E-38200 La Laguna, Tenerife, Spain
          \and
           Departamento de Astrof\'isica, Universidad de La Laguna, E-38206 La Laguna, Tenerife, Spain}

\received{2014 Feb 3} 
\accepted{2014 Mar 7} 
\publonline{2014 May 2}


\abstract{%
   We present the result of a study on the expansion properties and internal kinematics 
   of round/elliptical planetary nebulae of 
   the Milky Way disk, the halo, and of the globular cluster M\,15.  
   The purpose of this study is to considerably enlarge the small sample of nebulae 
   with precisely determined expansion properties (Sch\"onberner et al. \cite{SJSPCA.05}).
   To this aim, we selected a representative sample of objects with different evolutionary 
   stages and metallicities and conducted high-resolution \'echelle spectroscopy.    
   In most cases we succeeded in detecting the weak signals from the outer nebular shell 
   which are attached to the main line emission from the bright nebular rim.
   Next to the measurement of the motion of the rim gas by decomposition the main line
   components into Gaussians,
   we were able to measure separately, for most objects for the first time, the gas velocity 
   immediately behind the leading shock of the shell, i.e. the post-shock velocity. 
   We more than doubled the number of objects for which the velocities of 
   both rim and shell are known and 
   confirm that the overall expansion of planetary nebulae is accelerating with time. 
   There are, however, differences between the expansion behaviour of the shell and the rim: 
   The post-shock velocity is starting at values as low as around 20 \kms\
   for the youngest nebulae, just above the AGB wind velocity of $\sim$\,10--15 \kms,
   and is reaching values of about 40 \kms\ for the nebulae around hotter central stars.
   Contrarily, the rim matter is at first decelerated below the typical AGB-wind velocity and
   remains at about 5--10 \kms\ for a while until finally a typical flow velocity of up to 
   30~\kms\ is reached.   
   This observed distinct velocity evolution of both rim and shell is explained 
   by radiation-hydrodynamics simulations, at least qualitatively: 
   It is due to the ever changing stellar radiation field and wind-wind interaction
   together with the varying density profile ahead of the leading shock during the progress of 
   evolution. The wind-wind interaction works on the rim dynamics while the radiation field and 
   upstream density gradient is responsible for the shell dynamics.    Because of these
   time-dependent boundary conditions,
   a planetary nebula will never evolve into a simple self-similar expansion.
   Also the metal-poor objects behave as theory predicts: The post-shock velocities are 
   higher and the rim flow velocities are equal or even lower  compared to disk objects at 
   similar evolutionary  stage.    
   The old nebula around low-luminosity central stars contained in our sample 
   expand still fast and are dominated by reionisation.  
   We detected, for the first time, in some objects an asymmetric expansion behaviour: 
   The relative expansions between rim and shell appear to be different for the receding
   and approaching parts of the nebular envelope.  
   \vspace{1mm}      }
\maketitle

\section{Introduction}                      \label{intro}
\subsection{Historical background}          \label{hist.back}
  The process of formation and evolution of planetary nebulae (PNe) is intimately
  connected with mass-loss processes along the asymptotic giant branch (AGB) 
  and the final contraction and heating of the stellar remnant until the white-dwarf
  cooling path at the hot side of the Hertzsprung-Russell diagram (HRD) is
  reached. Thereby, the slow but very dense AGB wind is heated by photo-ionisation due
  to the intense radiation field of the hot central star and compressed 
  from within by the action of the tenuous but very fast central-star wind.

  There is, however, no direct interaction of the fast wind with the former AGB matter:
  The wind is thermalised by a strong shock, and the system's steady attempt to achieve
  pressure balance between the ionised shell and this shocked wind material on one side,
  and between the shell and the still undisturbed AGB wind on the other side is 
  responsible for
  the formation of what we call a planetary nebula and its evolution with time.
  In this view a PN is not simply the evidence of matter ejected from the stellar
  surface but can instead be described by a thermally driven shock
  wave through the ambient AGB wind envelope, starting at the inner edge of this envelope
  and powered by heating due to photo-ionisation.    Morphology and kinematics
  of PNe are thus the result of shock waves initiated by ionisation and modified by winds
  interaction, and their physics may not be described adequately by static models.

  The expansion of a PN is usually measured by the
  Doppler split components (if the line is resolved) or the line width (if
  unresolved) of strong emission lines.
  Since the pioneering work of Wilson (\cite{Wi.50}) it is known that the
  velocity field within a planetary nebula is not uniform.  Instead, one
  has to assume that the flow velocity generally increases with distance
  from the central star.  Because of this (positive) velocity gradient,
  lines of ions with different ionisation potential exhibit different line
  splits if the ionisation within the nebular shell is stratified.

\begin{figure}
\includegraphics[width=0.92\columnwidth]{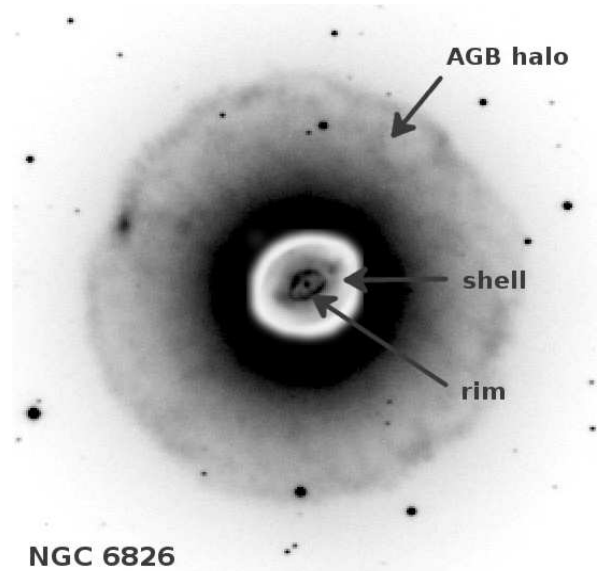}
\caption{\label{ngc2022}
         Combination of two \oiii\ images of the multiple-shell 
         planetary nebula \object{NGC 6826} with different (logarithmic) intensity scales 
         for inset and main image (see Corradi \etal\ \cite{CSSP.03} 
         for  details). The inset picture shows the PN proper, consisting of rim and shell. 
         The nebula is expanding into the rather large, only slowly expanding spherical
         halo consisting of ionised AGB-wind matter.  \vspace{-2mm}
        }
\end{figure}

  Although already known from photographic images, observations with modern,
  sensitive CCD detectors revealed many more PNe which consist of up to three
  distinct shells.  The inner two shells make up the PN proper, and
  Frank, Balick \& Riley (\cite{FBR.90}) coined the terms ``rim'' for the inner,
  bright shell and ``shell'' for the outer one which is usually fainter.
  The rim encloses an inner cavity  containing the hot shocked
  wind gas from the central star.
  The third, extremely faint but mostly round region which embraces the shell 
  is called ``halo'' and consists of the ionised AGB wind. 
  A recent compilation of PNe with detected halos is those of Corradi \etal\ 
  (\cite{CSSP.03}), but see also Frew, Boji\v{c}i\'c \& Parker (\cite{frewetal.12}).
  An example for the
  typical morphology of a multiple-shell PN is rendered in Fig.\,\ref{ngc2022}.
  Since we are dealing with the two main nebular shells only, we use also the term
  ``double-shell'' planetary in the following.

   While the density structure can be deduced from
   monochromatic images, the velocity field requires high-resolution
   spectroscopy of the emission lines, and the interpretation in terms of
   the internal velocity field is anything but straightforward. One
   reason is intrinsic to the objects: Density structure
   and velocity field are intertwined such that a measured line split along the
   central line-of-sight probes the matter velocity at a radial position
   corresponding to the largest emission measure, or more precisely, from the
   line split one derives only an average velocity, where the average along the
   line-of-sight is weighted by density squared.
   The other reason is that normally the emission from the shell is
   rather weak and escapes detection in spectrograms of poor quality.
   This is very unfortunate since the shell contains usually most and the fastest moving
   matter of the whole object!

  According to Chu, Jacoby \& Arendt (\cite{CJA.87}),
  at least 50\,\% of all round/elliptical PNe appear to be of the mul\-tiple-shell 
  (or double-shell) type.   By means of high-resolution \'echelle spectrograms, 
  Chu et al.\ (\cite{CKKJ.84}) and Chu (\cite{Chu.89}) detected also the typical
  spectroscopic signatures of the shell as weak (outer) shoulders attached to the main
  emission-line profile originating from the bright rim and noted that there are
  obviously two different expansion modes one has to deal with.
  Normally, the shell is faster than the rim, but exceptions are also known 
  (cf.\ Sabbadin  et al.\ \cite{SBH.84}).    A good example for normal expansion is 
  the well-known PN NGC 6826 where Chu et al. (\cite{CKKJ.84})
   deduced an expansion velocity of 27~\kms\ for the shell from the outer, 
   weak shoulders of the line profile, in contrast to 8~\kms\ for the rim from
   the strong, doubly peaked central profile components (cf. also our
   Fig.~\ref{line.analyse} in Sect.~\ref{disk.PN}). 

\begin{figure*}
\hskip-1mm
\includegraphics[width=0.99\textwidth]{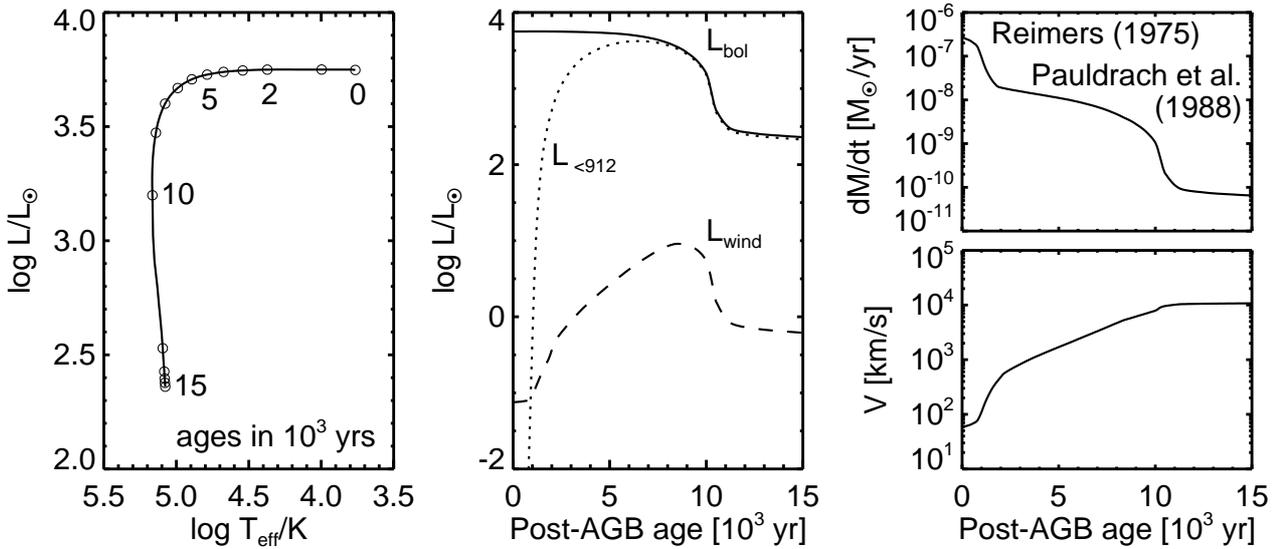}
\vskip-2mm
\caption{\label{mod.prop}
           Example of stellar and wind properties used in our simulations as inner boundary
           conditions.  {\em Left panel}:  evolutionary path for a
           0.595~M$_{\odot}$ post-AGB model with the post-AGB ages ages indicated; 
           \emph{middle panel}:
           the corresponding time evolution of the bolometric, $L_{\rm bol}$ (solid), 
           Lyman continuum, $L_{<912}$ (dotted), and mechanical (wind) luminosity, 
           ${L_{\rm wind}=\dot{M}\, V^2/2}$ (dashed), all in units of \Lsun. 
           {\em Right panels}:  mass-loss rate $\dot{M}$ (\emph{top}) and  wind velocity $V$
	   (\emph{bottom}).  The end of the strong AGB wind sets the zero
	   point of the post-AGB evolution and the beginning of the much weaker
	   post-AGB wind, modelled first by the prescription of Reimers 
          (\cite{R.75}). Later, for ${T_{\rm eff}\ge 25\,000}$~K, or 
           post-AGB ages  $\ge\!2{\times}10^3$ years, the theory of radiation-driven winds 
           as formulated by Pauldrach et al. (\cite{pauldrachetal.88}) is used
           for the rest of the evolution.  \vspace{-0mm}
}
\end{figure*}

   Important more recent observational studies concerning the structure and
   kinematics of PNe, partly employing (static) nebula models together with 
   high-resolution line profiles, 
   are those of Stanghellini \& Pasquali (\cite{StaPa.95}) (only structure), 
   Gesicki, Acker \& Szczerba (\cite{GAZ.96}), G{e}sicki et al.\
   (\cite{GZAS.98}), Guerrero, Villaver \& Manchado (\cite{GVM.98}),
   Gesicki \& Zijlstra (\cite{GeZi.00}) (only kinematics), Neiner et al.\ 
   (\cite{NAGS.00}), and Gesicki, Acker \& Zijlstra (\cite{GAZ.03}). Especially
   interesting is the study by Sabbadin et al. (\cite{Sabbetal.04}) on 
   \object{NGC 7009} in which two distinct velocity laws for rim and shell
   could be derived.

   All these studies agree upon the facts that 
\begin{enumerate}
\item the expansion velocity does not necessarily increase linearly with distance 
      from the central star, and that
\item often the gas velocity reaches a local minimum roughly at the rim/shell interface, 
      i.e. rim and shell have very distinct expansion behaviours, and usually the 
      shell matter reaches the highest expansion velocities.
\end{enumerate}
   One can conclude from these studies that planetary nebulae \emph{do not} 
   expand according to a ${v(r)\propto r}$ law, although this assumption is still often used
   to construct spatiokinematical models (but see also Steffen \& L\'opez \cite{wsteffen.06};
   Steffen, Garc\'ia Segura \& Koning \cite{wsteffen.09}). 

   It has early been realised that deciphering the expansion behaviour of PNe is
   important for understanding their formation and evolution.  For instance, the
   total lifetime, or visibility time, of PNe is an important quantity to determine
   the total PN number of a stellar population, either by observations or theoretically by
   stellar population synthesis calculation (cf. Moe \& de Marco \cite{moedemarco.06}).  
   This, however, is an uncertain endeavour as 
   long as the internal kinematics and especially the ``true'' expansion velocity of PNe
   is not known.  A recent discussion of this subject can be found in Jacob, Sch\"onberner
   \& Steffen (\cite{jacobetal.13}).
   
   Also, the problem of individual distances is an obstacle if observed peak line
   separations or line half widths are plotted against nebular radii, i.e. if one wants
   to deduce any evolution of the expansion with time.    From the
   early works of Bohuski \& Smith (\cite{BoS.74}) and Robinson et al. (\cite{RRA.82})
   a certain trend of ``expansion'' velocities with radii is detectable
   (see, e.g., Fig.~2 in Bianchi \cite{Bi.92} for a more recent study of this kind),
   but the only safe statement one can make is that PNe start their evolution with
   comparably low expansion rates. How and why this expansion increases with
   time remained quite obscure. These analyses were additionally hampered by the 
   fact that no distinction regarding nebular morphology and/or central-star type 
   (Wolf-Rayet vs. \hbox{O-type)} was made.
   
   The first convincing observational evidence about the increase of nebular expansion 
   rates (as measured from half widths of strong emission lines) with
   time or evolution was presented by Dopita \& Meatheringham (\cite{dopetal.91})
   for Magellanic Clouds PNe and by M\'endez, Kudritzki \& Herrero (\cite{MKH.92})
   for Milky Way objects.  In both studies  a systematic increase
   of the emission line widths with stellar temperature was found, based on the
  (distance-independent) central-star temperatures as a proxy of evolution. 

   Medina et al. (\cite{medetal.06}) discriminated, for the first time, between WR and
   normal spectral types of the central stars and found indications that nebula around
   WR central stars expand faster than those around O-type central stars.
   Dopita \& Meatheringham (\cite{dopetal.91}) and also Medina et al. (\cite{medetal.06}) 
   used the 10\,\% level of the line profile in order to get hold of the fastest 
   expanding matter.     Richer \etal\ (\cite{richeretal.08}, \cite{richeretal.10}) 
   studied the kinematics of PNe of the Milky Way bulge and found also that the 
   HWHM velocities increase with the pace of evolution.  
   These authors used various distance-independent indicators to 
   discriminate the evolutionary states, as, e.g., the strength of \ion{He}{ii} 
   $\lambda$4686 \AA. 

\begin{figure*}[t]
\vskip-2mm
\hskip-5mm
\includegraphics[width=0.99\textwidth]{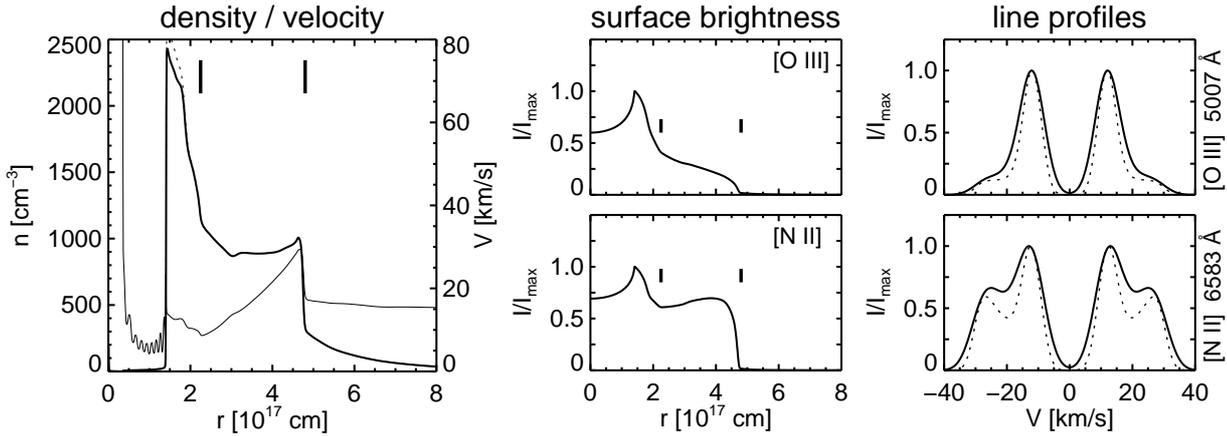}
\vskip-2mm
\caption{\label{model.595}
         Snapshot of a typical middle-aged nebular model from a 1D-radiation-hydrodynamics 
         simulation around a 0.595 \Msun\ central star whose properties are shown in 
         Fig.\,\ref{mod.prop} (cf. Sch\"onberner et al. \cite{SchJSt.05}).
         The stellar parameters are: post-AGB age ${t= 6106}$ yr, $\Teff = 80177$ K,
         and $L = 5.057{\times}10^3$ \Lsun. \emph{Left panel}: heavy particle density (thick),
         electron density (dotted), and gas velocity (thin); \emph{middle panels}:
         (normalised) surface brightnesses in \oiii\ $\lambda$5007 \AA\ and \nii\ 
         $\lambda$6583 \AA; \emph{right panels}: the corresponding normalised line 
         profiles computed for the central line-of-sight with infinite spectral resolution
         (dotted) and broadened with a Gaussian of 6 \kms\ FWHM (solid),
         both with a circular aperture of ${1\!\times\!10^{16}}$ cm. The thick vertical marks
         (\emph{left and middle}) indicate the positions of the leading shocks of the rim and
         shell, respectively. 
         The nebular mass enclosed by the rim's shock is $M_{\rm rim} = 0.07$ \Msun, that
         enclosed by the shell's leading shock is $M_{\rm shell} = 0.47$ \Msun\ 
         (= total nebular mass).
         \vspace{-0mm}
         }
\end{figure*}

\subsection{Theoretical considerations}    \label{theo.cons}

  Radiation-hydrodynamics simulations provide a rather detailed picture how a PN is 
  being formed (see, e.g., Schmidt-Voigt \& K\"oppen \cite{SK.87a, SK.87b}; 
  Marten \& Sch\"onberner \cite{MS.91}; Mellema \cite{Mellema.94}, 
  \cite{Mellema.95}; Villaver et al. \cite{villetal.02}; Perinotto et al. \cite{PSSC.04}):
   Ionisation creates a rarefaction wave (the shell) which expands into the ambient medium (the 
   former AGB wind), led by a shock.  
   The innermost, only very slowly expanding part of this 
   wave is being compressed and accelerated into a dense rim     
   by the (thermal) pressure of the  so-called ``hot bubble'' which 
   consists of wind matter from the central star heated to very high temperatures
   (${\simeq\!10^6}$--${>\!10^7}$ K) by the reverse wind shock
   and is separated from the nebula proper by a contact discontinuity.\footnote
{Mellema (\cite{Mellema.95}) used a different notation: The rim is called ``W-shell'' 
 (W\,=\,wind) because it
 is the signature of interactions between the fast stellar wind and the older, slowly
 expanding former AGB material, and our shell is named ``I-shell'' (I\,=\,ionisation)
 because it is originally generated by ionisation.
 Consequently, the rim is bounded by a ``W-shock'', and the shell by an ``I-shock''.} 
   Kinematics and shape of the wind-compressed rim is controlled by     
   the wind power of the central star and the density and velocity of the ambient medium, 
   which is here given by the low-velocity inner tail of the shell  
   (Koo \& McKee \cite{komc.92}).   In contrast,
   the propagation speed of the shell's shock, $\dot{R}_{\rm out}$,
   is exclusively determined by the electron temperature 
   ${(\dot{R}_{\rm out} \propto \sqrt{T_{\rm e}})}$ \emph{and}\/ the upstream 
   density gradient (Franco, Tenorio-Tagle \& Bodenheimer \cite{FTB.90};
   Chevalier \cite{ch.97}; Shu et al. \cite{SLGCL02}).

   Thus, the typical PN consists, next to the hot bubble which is only seen in X-rays, 
   of two important dynamical subsystems, the shell and the rim. 
   In the simplest case of a spherical configuration,  morphology and kinematics  
   are ruled by  (i) the radial run of the density gradient of the 
   ambient matter together with a changing nebular electron temperature, and 
   (ii) the evolution of stellar wind power and radiation field with time.
   This means that the kinematics of shell and rim are expected to be quite 
   independent of each other: The shell's expansion depends mainly on the previous 
   mass-loss rate variation along
   the tip of the AGB because the electron temperature increases only slowly 
   with evolution, while the expansion of the rim depends mainly on the stellar wind 
   evolution and, to a lesser extend, on the shell's expansion properties. 

   As the shell's shock expands faster 
   than the rim, the shell becomes diluted while at the same time the slowly moving rim 
   becomes much denser and brighter: the typical double-shell structure emerges
   very soon. The real expansion speed of a PN is, of course,
   defined by the propagation of the shell's leading shock, but this velocity
   cannot be measured spectroscopically! For more details, see 
   Sch\"onberner \etal\ (\cite{SchJSt.05}, \cite{SJSPCA.05}, \cite{SchJSaSt.10}).
  
\begin{figure*}
\vskip-1mm
\hskip-2mm
\includegraphics*[bb= 0cm 0.7cm 20cm 10cm, width=0.97\textwidth]{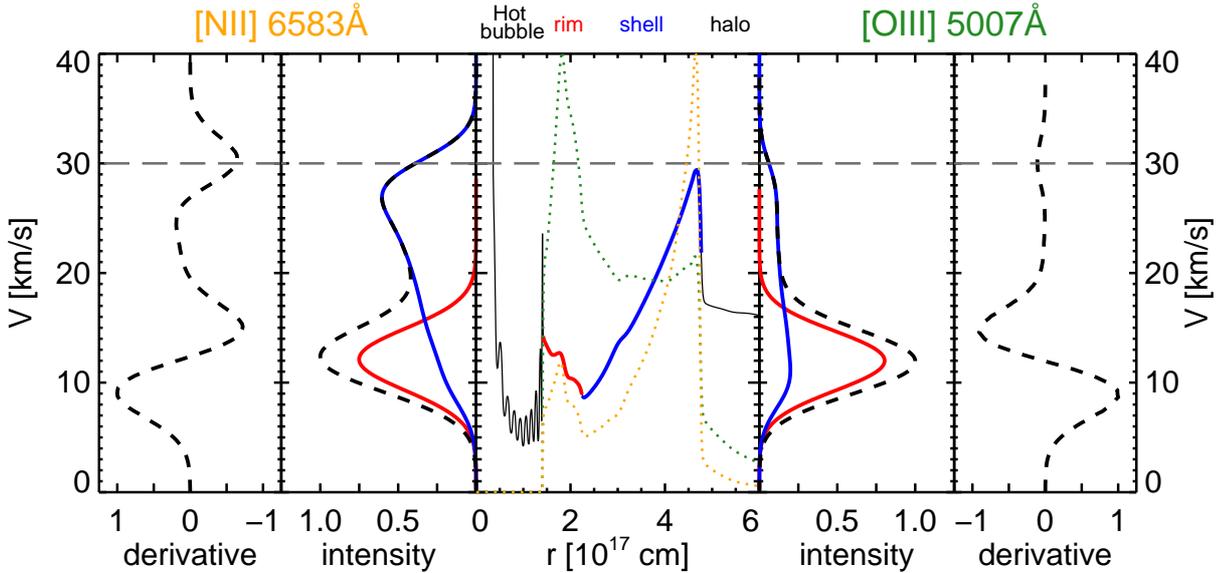}
\caption{\label{decomposition} 
         Receding-shell components of the normalised lines \nii\ $\lambda$6583 \AA\
         (\emph{next to left}) and \oiii\ line $\lambda$5007 \AA\ (\emph{next to right}), 
         computed (black dashed) for the model nebula shown in the \emph{middle panel} 
         and decomposed into the different contributions 
         of rim (red) and shell (blue).  The model is the same as displayed in 
         Fig.~\ref{model.595}, but here we render the velocity
         profile only (thick solid red and blue), supplemented by the normalised 
         densities of N$^+$ (orange dotted) and O$^{++}$ (green dotted).  
         Thin lines indicate the velocity field belonging the hot bubble and the halo.
         Both the outermost panels display the derivatives of both line profiles.  
         The (relative) minima which belong to the profiles' inflection points at 
         30 \kms\ are marked by the horizontal dashed line.  
         The value of the model's post-shock velocity is 29.5 \kms.
         \vspace{-1.5mm}
          }
\end{figure*} 
  
  During its expansion, a PN ``sees'' orders of magnitude changes of   
  the stellar UV-radiation field and the wind power.
   An illustration is given in Fig.~\ref{mod.prop} for a 0.595~\Msun\ post-AGB model 
   where the changes of important stellar quantities are displayed: luminosities 
   of the UV radiation and the  wind (middle panel), mass-loss rate 
   and wind velocity (right panel).\footnote
{\changed{All model sequences shown in this work are based on the 
          \textsc{nebel} code whose physical and numerical details are described in
          Perinotto et al. (\cite{PKSM.98}). The atomic data used for the (time-dependent)
          ionisation/recombination calculations are listed in Marten \& Szczerba (\cite{MS.97}). 
          An update of these data has only a negligible influence on the hydrodynamical
          properties of the models and is not necessary as long as one is not interested in 
          the determination of chemical abundances.}
 }

  Noticeable is the following:  While the power of the UV-radiation field increases 
  rapidly with time (and effective temperature), the wind power increases more gradually
  with time from a value as low as $\simeq$10$^{-4} L_{\rm bol}$ to a
  maximum value of about $\simeq$10$^{-2.5} L_{\rm bol}$ at the end of the horizontal 
  part of evolution through the HRD, at about maximum stellar temperature. 
  This increase of the wind power 
  is entirely due to the growing wind speed as the star shrinks, which more than 
  compensates for the decrease of the mass-loss rate: 
  from typical AGB-wind velocities up to 10\,000 \kms\ at the white-dwarf stage.
  Only beyond maximum stellar temperature we see a rapid decline of the wind power
  in parallel with the drop of stellar luminosity while the wind speed remains constant.
 
  The low wind power at the beginning suggests that
  formation and early shaping of PNe is mainly ruled by photoionisation and not by 
  interacting winds, as is the general assumption in the literature. 
  The effect of photoionisation becomes even more important in systems with low metallicity, 
  hence also with weaker winds, as demonstrated by Sch\"onberner \etal\ (\cite{SchJSaSt.10}).

  An illustration of the typical nebular structure and 
  the corresponding observable quantities like radial intensity and line profiles that
  are predicted by 1D-hydrodynamical simulations is rendered in Fig.\,\ref{model.595}.
  This particular model belongs to
  a radiation-hydrodynamics PN simulation around a 0.595~\Msun\ 
  central star whose properties are illustrated in Fig. \ref{mod.prop}. 
  The initial shell configuration is based on hydrodynamical 
  simulations of dusty AGB envelopes during the final AGB evolution (see Steffen
  \etal\ \cite{steffenetal.98} for details). 
  Theoretical line profiles are computed for central lines of sight with a numerical
  aperture of 1$\times$10$^{16}$ cm radius, broadened by a Gaussian of FWHM of 6 \kms\
  to mimick a typical finite spectral resolution.  
   More details about this particular sequence can be seen in Sch\"onberner et al. 
  (\cite{SchJSt.05}, Figs. 1, 2, and 3 therein). 

   Figure \ref{model.595} renders a typical middle-aged and fully ionised 
   nebular structure where we
   see, next to the ionisation-generated shell, already a well-developed rim.
   The shell is somewhat diluted because it has already expanded into the former AGB wind,
   while the rim is of much higher density because of the bubble's comparatively high
   pressure.  Consequently,
   both the rim and the shell show a very distinct behaviour in their radial 
   intensity distributions and their contributions to the line profiles.
   Note especially the rather faint signatures of the shell in \oiii\ in both
   the brightnesses and line profiles, although the shell contains most of the nebular 
   mass, i.e. 85\,\% of the 0.47 \Msun\ of ionised matter contained in the PN model 
   shown in the left panel of Fig.\,\ref{model.595}.  Note especially the rather low flow
   velocities within the rim, ${\approx\!12}$ \kms, as compared to the post-shock velocity 
   of the shell, 29.5 \kms.
   
   It is interesting to look in more detail at the different contributions from rim 
   and shell to the total emission profiles seen in the right panels of Fig.~\ref{model.595}.  
   Therefore, we have decomposed numerically the total nebular line emission  of
   this model snapshot into the
   contributions from rim and shell and show them, together with the relevant model
   structures (velocity field and ion densities) in Fig.~\ref{decomposition}.
   First of all, the rim contribution is similarly strong and narrow in both ions and 
   reflects the low velocities but high densities of the rim gas.  This line 
   contribution can well be approximated by a Gaussian profile.  The shell contribution is 
   weaker but much wider because of the velocity spread within the shell: the larger this
   velocity spread, the wider the shell contribution in velocity space (cf. middle panel of
   Fig.~\ref{decomposition}). 
   
   The difference in shape between \nii\ and \oiii\ lines is caused by the 
   different radial density profiles of the respective ions (cf. middle panel of 
   Fig.~\ref{decomposition}).  It is obvious that the shell profile resembles only poorly
   a Gaussian profile, but it is the only part of the total nebular emission line that
   allows an estimate of the true nebular expansion speed:  The inflexion point at the
   outer flanks of the profiles (either from N$^+$ or O$^{++}$) mark closely the post-shock
   velocity which is, under the physical conditions found here, usually 20--25\,\% below the
   shock propagation speed (see discussion in Sect.~\ref{subsect.post.shock} and 
   Corradi \etal\ \cite{CSSJ.07}, hereafter Paper~II; 
   Jacob, Sch\"onberner \& Steffen \cite{jacobetal.13}).

   We note in passing that the double-shell structure seen in Fig. \ref{model.595}
   develops from an initial configuration with a rather smooth radial density profile 
   which falls off outwards with a gradient varying between $\alpha$\,=\,2 and 4 
   if one approximates the density by power-law distributions with variable $\alpha$,
   ${\rho \propto r^{-\alpha(r)}}$.  This smooth initial     
   configuration reflects the assumed evolution of the mass loss along the tip of
   the AGB (see, for details, Steffen et al. \cite{steffenetal.98}, Fig. 19 therein).
   This rather simple initial structure is completely reshaped firstly by 
   ionisation and then by the stellar wind into the much more complicated configuration
   as seen in Fig. \ref{model.595}.

   Given the theoretical considerations above, we conclude that the term ``expansion velocity''
   is by no means unambiguous, even if it is based on high-resolution spectroscopy. 
   It characterises observationally nothing else than the \emph{spectroscopically} 
   determined flow of matter at a certain position within an expanding shell with a radial
   density and velocity profile (see Fig.\,\ref{model.595}, left panel), and
   it is quite important for any interpretation to know whether this position is within
   the rim or the shell.     
 
   Likewise, the radiation-hydrodynamics simulations confirm the observations discussed in  
   Sect.~\ref{hist.back} that the velocity field even for spherical systems is by no means
   as simple as a ${v(r) \propto r}$ law.  Instead, one has to consider different velocity laws
   for the rim and shell matter.

\subsection{Kinematic ages}                \label{subsect.kin.age}

   A quantity very often used in the literature in various contexts is the kinematic age
   of a PN.  Its usual definition is nebular radius, taken from images, divided by a 
   spectroscopically measured ``expansion velocity''.   Irrespective of the fact that one
   combines a (distance-dependent) quantity measured in the plane of sky with one measured 
   along the line-of-sight, most age determinations suffer from an internal inconsistency:
   The nebular outer edge, i.e. that of the shell, is combined with a velocity measured 
   for the bright rim!
   Additional complications are the velocity gradient within the expanding shell, the 
   acceleration of expansion with time, and the fact that the real expansion velocity, 
   i.e. that of the outer, leading shock, cannot be measured spectroscopically at all.
   
   Even if one believes in a good kinematic age of the PN,    
   the conversion of a PN's kinematic age into a post-AGB age is, by no means, straightforward. 
   One has to consider 
\begin{itemize}
\item that the measured kinematic age is just the current time scale of evolution which is 
      subject to change because of a possible accelerated expansion,
\item that the nebula forms at some distance from the star at the inner edge of the receding 
      AGB wind, and that
\item there is an offset between the (assumed) zero point of post-AGB evolution and the 
      begin of PN formation, the so-called ``transition time''.  The latter is quite 
      uncertain and depends severely on the post-AGB mass-loss history and remnant mass.     
      Different definitions exist in the literature, but
      a typical value may be 2000 years (cf. Fig.\,\ref{mod.prop}). 
\end{itemize}   

\begin{figure}
\vskip-1mm
\hskip-1mm
\includegraphics[width= 8.3cm, height= 10.8cm]{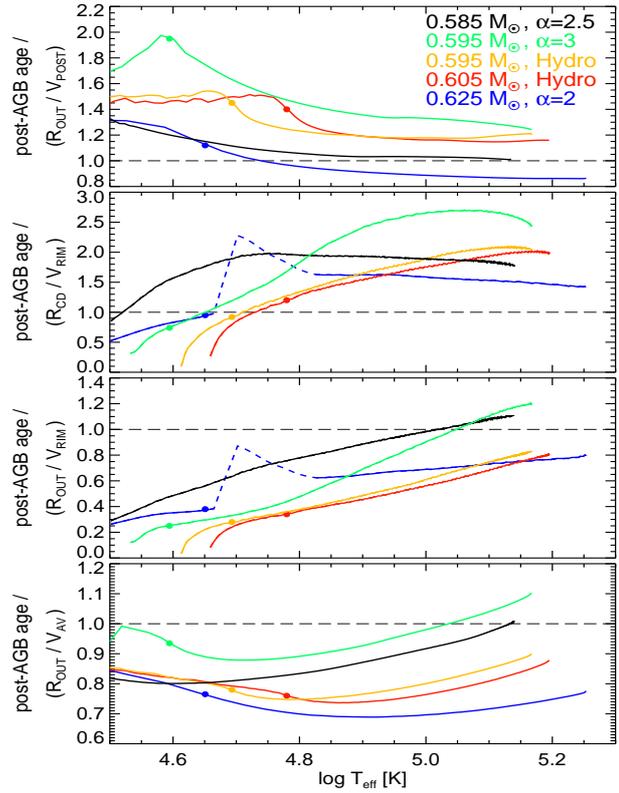}
\vskip-1mm
\caption{Ratios of post-AGB ages to various definitions of the kinematic age as predicted from
         a number of hydrodynamical simulations for different post-AGB remnants and initial
         configurations, parameterised by stellar mass and envelope structure (slope $\alpha$ or
         ``hydro''), vs. stellar effected temperatures (see, for more details, Perinotto
         et al. \cite{PSSC.04}; Sch\"onberner et al. \cite{SchJSt.05}).  Plotted are the 
         age ratios only for the horizontal part of the HRD evolution until maximum stellar
         effective temperatures.  
         Velocities were determined from computed line profiles broadened by a Gaussian of 
         6 \kms\ FWHM to account for the limited spectral resolution:  $V_{\rm rim}$ from 
         the peak separation of the strongest line components of \oiii, and $V_{\rm post}$ 
         from the (outer) inflexion points.
         Dots along the sequences indicate the position of the optically thick/thin transition
         if not below ${\log\Teff = 4.5}$.  The color coding of the 
         tracks will remain the same throughout the paper.  The dashed part of the 0.625 \Msun\
         track seen in the two middle panels will be explained later (Sect.~\ref{comp.rim}).
         Note also the different ranges of the ordinates.
         \emph{Top panel}: kinematic age defined as (outer) shell radius, $R_{\rm out}$, over
                           post-shock velocity $V_{\rm post}$.
         \emph{Second panel}: kinematic age defined as radius of contact discontinuity, 
                              $R_{\rm cd}$ (as proxy of $R_{\rm rim}$), over velocity of the
                               rim, $V_{\rm rim}$.
         \emph{Third panel}: kinematic age defined as shell radius, $R_{\rm out}$ over
                              $V_{\rm rim}$. 
         \emph{Bottom panel}: kinematic age defined as shell radius, $R_{\rm out}$, over
                              mass-averaged velocity, $V_{\rm av}$.                           
       \vspace{-2mm}      
        }
\label{fig.kinages}
\end{figure}

  An insight into the correction one has to deal with for different definitions of the kinematic
  age gives Fig.~\ref{fig.kinages}.  There the ratios of true model ages (= post-AGB ages) to
  various definitions of kinematic ages are plotted over effective temperatures of the
  respective central stars as predicted by our various 1D-model simulations whose principal
  properties are already explained above.  We assume that the mass-loss history, as
  devised by Bl\"ocker (\cite{B2.95}), comes close to reality.  The tracks have quite 
  some spread caused by the different initial conditions and central-star properties, 
  but they all show a similar run with effective temperature (or time).

  We begin the discussion with the kinematic age definition 
  $R_{\rm out}/V_{\rm post}$ which appears to us the best choice
  one can get because $V_{\rm post}$ characterises the flow immediately behind
  the shock at $R_{\rm out}$ which is usually also the fastest within the entire nebula 
  (cf. Fig.~\ref{decomposition}), and because $V_{\rm post}$ changes only slowly with time
  during a significant part of the nebular expansion.  
  However, this choice for the kinematic age leads, at least in principle for expansion with
  constant rate, to a systematic \emph{overestimation} of the true age since 
  ${V_{\rm post}\simeq\dot{R}_{\rm out}/1.3}$ (cf. Fig.~\ref{shock.postshock}).
  In reality, we have an age offset due to the finite transition time, a rapid shock
  acceleration during the optically-thick stage, and a low shock acceleration during the
  following evolution across the HRD (cf. Fig.~\ref{shell.comp} below), 
  all of which conspire for \emph{underestimating} the post-AGB age.  
  
  The situation is illustrated in the top panel of Fig.~\ref{fig.kinages}.  At  early stages
  of evolution, transition time and shock acceleration dominate and result in age 
  underestimates by factors between 1.3 and 2.0, depending on model sequence and evolutionary
  stage.  Later, the nearly constant expansion dominates, and the correction factors
  decrease slowly, but still remain above unity.  Only for the 0.625 \Msun\ sequence with its
  short transition time and the virtually constant expansion rate after the thick/thin
  transition, the correction factor falls below unity quite soon.

  The second panel of Fig.~\ref{fig.kinages} relates the model ages to rim radius divided by
  rim velocity, which is a consistent combination, but not used in practice.  Because the
  rim radius is sometimes difficult to detect numerically, we used here the radius of the 
  contact discontinuity as a proxy instead.\footnote
{The rim is geometrically quite thin, so that $R_{\rm rim} \simeq R_{\rm cd}$.}
  Despite its internal consistency, the use of this age can not be recommended: 
  The correction factors range from about 0.5 up to about 2, on the average. 
  The reason lies in the kinematics of the rim: first the rim gas is nearly stalling, 
  pushing the kinematic age above the models' age; 
  later when the rim begins to accelerate (for ${\ga\!50\,000}$ K), the kinematic ages 
  falls below the true ages.
  
  This panel is qualitatively similar to Fig.~20 in Villaver et al. (\cite{villetal.02}), 
  although a quantitative comparison is impossible because of the different assumptions
  about the zero point of the post-AGB evolution and the different definitions of 
  ``expansion'' velocity and radius made in both studies.
 
  The third panel of Fig.~\ref{fig.kinages} relates the model ages to shell radius divided by
  rim velocity, an inconsistent method but used frequently in the literature.  As expected, the
  result is disastrous: The mean correction factor runs from 0.2 up to close to unity for 
  the older models.  
  Because the rim expansion is so slow at the beginning, kinematic ages based on this method
  overestimate grossly the true (post-AGB) ages.  Only at later stages of evolution, when 
  the rim accelerates, the discrepancy to the true post-AGB ages decreases.
  
  The bottom panel of Fig.~\ref{fig.kinages} relates the model ages to shell radius 
  divided by $V_{\rm av}$, the mass-weighted velocity. This method, introduced by Gesicki
  et al. (\cite{GZAS.98}), is very elaborate because it
  demands knowledge of (radial) density and velocity profiles of the PN.  The former can 
  be deduced from the surface brightness distribution, e.g. from H$\alpha$, if spherical
  symmetry is imposed, but the latter has to be derived from emission lines and is not
  always unambiguously possible.  
  It is mandatory that density profile and velocity field are consistent with the 
  physics of expanding shock waves and wind-wind interaction, a constraint which is
  difficult to judge from observations alone. 
  Nevertheless, neglecting the more extreme case of the ``${\alpha = 3.0}$'' sequence, 
  a value of about 0.8 seems reasonable for typical PNe, i.e. post-AGB age = 
  $(0.8\pm0.1){\times} R_{\rm out}/V_{\rm av}$, but holding only for ${\log\Teff \la 4.9}$.
   
  It is obvious that our hydrodynamics models provide only a rather simplistic view of 
  true nebulae.  Still, the results presented in Fig.~\ref{fig.kinages} suggest that
  measurements of flow velocities right behind the leading shock of the PN's shell 
  is a robust and the most reliable method for estimating true post-AGB ages. 
  But even so, errors due to asphericity, and especially the uncertainty of the
  transition time, make any PN age determination to a risky endeavour.
  
  We add that the ad hoc assumption to measure a relevant expansion velocity at the 10\,\%
  level of the line profile (resolved or not resolved) is, of course, also a reasonable
  choice if the post-shock velocity is not measurable,
  but one must recognise that this criterion is unphysical and does not take care
  of ionisation differences.  This can be seen in Fig.~\ref{decomposition}  
  where the 10\,\% criterion delivers a flow velocity of 33 \kms\ from the \nii\ line,
  while \oiii\ gives 28 \kms.  Only the profile's inflexion point provides the same 
  value for both ions: 30 \kms.   

\subsection{The relation between shock and post-shock velocity}
\label{subsect.post.shock}

   The post-shock velocity, i.e. the flow velocity immediately behind the shell's leading 
   shock (or the shell's outer edge), is a well-defined physical quantity and is related 
   to the shock's propagation speed via the well-known jump conditions.  
   For instance, the relation between both velocities depends on
   the shock properties (adiabatic, isothermal, or intermediate) which are
   ruled by the physical state of the flow and the environment into which the shock
   expands.  A thorough discussion for the specific case of planetary nebulae has been 
   presented in Mellema (\cite{Mellema.04}) and Sch\"onberner et al. (\cite{SJSPCA.05}).  

\begin{figure}
\vskip-4mm
\hskip-1mm
\includegraphics[width=1.01\linewidth]{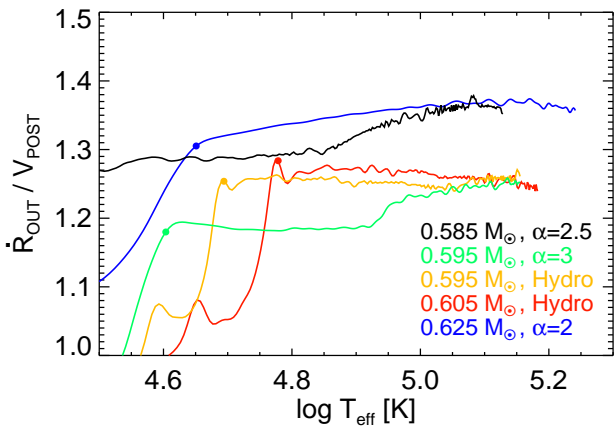}
\vskip-1mm
\caption{\label{shock.postshock}
         Ratio of outer shock velocity, $\dot{R}_{\rm out}$, and post-shock velocity, 
         $V_{\rm post}$,
         vs. stellar effective temperature as predicted by hydrodynamical sequences with 
         different initial conditions and central stars with various masses (see legend).
         Only the results for the high-luminosity part of evolution through the HRD
         until maximum stellar temperatures are reached are shown.   
         The rapid increase of $\dot{R}_{\rm out}/V_{\rm post}$ from about unity to 1.2--1.3  
         belongs to the optically thick part of evolution whose end is marked by a dot. 
         }
\end{figure}

   Figure~\ref{shock.postshock} shows how the relation between shock and post-shock 
   velocities, as predicted by our hydrodynamical PN simulations, depend on evolutionary 
   state (again measured by the stellar effective temperature) and model parameters.
   One sees that the ratio of both velocities does not vary much between the different model 
   sequences:  typical values for optically-thin models are between 1.2 and 1.35, with very 
   little variation during the evolution across the HRD.  The only
   exceptions are the transition phases from optically-thick to an optically-thin model
   where the velocity ratio ``jumps up''.  
   
   The ratios of shock to post-shock velocity shown in Fig.~\ref{shock.postshock} agree 
   very well with the value estimated by Mellema (\cite{Mellema.04}).  Based on the
   analytical solutions of Chevalier (\cite{ch.97}) and Shu et al. (\cite{SLGCL02}), he
   concluded that this ratio is, for the case of typical PN conditions, around 1.20.
   
   The close correspondence between the propagation speed of the shell's shock and the
   post-shock velocity seen in Fig.~\ref{shock.postshock} allows us in the following 
   to use the post-shock velocity as a proxy for the nebular expansion:  
   All expansion properties found from spectroscopic determinations hold also for the 
   outer shock, apart from a scaling factor of about 1.3.

\subsection{A modern observational approach}       \label{new}

   From the previous sections it becomes evident that a more thorough discussion of 
   the general problem of the expansion of PNe is urgently needed, as well as a  
   physically sound interpretation of the observed line profiles.   This implies the necessity 
   to observe line profiles with high signal-to-noise in order to get hold of weak outer
   signatures of fast moving gas immediately behind the leading shock front 
   combined with radiation-hydrodynamics simulations for a proper
   interpretation.   Knowing the true expansion behaviour of PNe is mandatory for
   estimating total lifetimes of PNe or mean visibility times belonging to PN ensembles
   in stellar populations (cf. Jacob, Sch\"onberner \& Steffen \cite{jacobetal.13}).

   A first study to gain closer insight into the expansion behaviour of planetary nebulae, 
   especially in terms of the different properties of shell and rim, 
   has already been presented by Sch\"onberner \etal\
   (\cite{SJSPCA.05}).  Basically, these authors determined typical bulk
   velocities of the rim and the shell matter by means of Doppler
   decomposition of strong emission lines gained by high-resolution spectroscopic
   observations into four Gaussians and compared them with theoretical line profiles 
   computed from appropriate radiation-hydrodynamics simulations. The theoretical 
   profiles were decomposed into four Gaussians in exactly the same way as the observed
   ones.

   Sch\"onberner et al. (\cite{SJSPCA.05}) found a significant increase of typical bulk 
   velocities of rim and shell with progress of evolution, 
   the latter measured by the effective temperature of the central star.  The expansion of
   the shell turned out to be very fast, up to about 30--40~\kms.   
   Sch\"onberner \etal\ were then able to interpret this 
   expansion by means of radiation-hydrodynamics simulations:
   A PN expands into a circumstellar environment whose radial slope is much steeper than
   usually assumed, ${\rho \propto r^{-\alpha}}$ with ${\alpha > 2}$.   
   This result is fully consistent with 
   the very few existing direct measurements of the radial density
   profiles of halos from their surface brightness distribution
  (Plait \& Soker \cite{PS.90}; Corradi \etal\ \cite{CStSchP.03}; Sandin et al. 
  \cite{sandetal.08}).

   The PNe sample investigated by Sch\"onberner \etal\ (\cite{SJSPCA.05}), however, was
   rather small, and additional data from the literature was not available at the time. 
   Since a detailed velocity information of a much
   larger sample of round/elliptical PNe appears to be necessary to strengthen the 
   conclusions mentioned above, the 
   Sch\"onberner \etal\ sample was increased by new high-resolution spectroscopic 
   observations with very good signal-to-noise ratios.
   Only if the sample is sufficiently large, a meaningful comparison with (spherical) 
   hydrodynamical models can be made which then will give important constraints on 
   the physical conditions responsible for the formation and evolution of planetary nebulae.

   In the present work we took up the method recommended in Paper II.
   There, we showed that the derivative of the line profile allows an rather
   accurate determination of the post-shock gas velocity of the shell.  Knowing the
   post-shock velocity, the conversion to the true expansion velocity, i.e. the propagation
   velocity of the outer shock, is simple (see discussion in Sect.~\ref{subsect.post.shock}).  
   This is the optimum one can get since, 
   as already said, the propagation speed of the shock itself cannot be measured 
   spectroscopically. 

   In Sect.\,\ref{obs.red} we
   report on our observations, the profile extraction and the determination of the
   respective expansion velocities which are then presented in Sect.~3.  
   Section 4 is devoted to detailed comparisons with predictions of radiation-hydrodynamics  
   simulations.  In Sect.~5 we present, for the first time, apparent expansion asymmetries 
   along the line-of-sight seen in some objects. Next, in Sect.~6, the expansion behaviour
   of very evolved PNe harbouring hot white dwarfs as central stars is discussed.
   Section~7 deals then with the expansion of metal-poor objects.   
   A discussion of the results follows in Sect.~8, and the conclusions (Sect.~9)  
   close the paper. -- We note that preliminary results of this study have been published in 
   Jacob et al. (\cite{jacobetal.12}).

\section{Observations and data reduction}                       \label{obs.red}

\subsection{Planetary nebulae in the Galactic disk}            \label{disk.PN}

  We selected our targets from a list of Galactic disk PNe with well developed double-shell
  structures which do not deviate too much from a spherical shape.  We verified
  that the central stars are not belonging the Wolf-Rayet spectral class since
  the evolution of these hydrogen-poor objects is still poorly understood and
  certainly not adequately described by the evolutionary calculations available.

  We observed our sample in 1999 July and 2000 January with the Coud\'e-\'echelle
  spectrograph attached to the 2-m telescope in Tautenburg, Germany.
  The (projected) slit length was 29\arcsec\ with a width of 0\farcs9,
  providing a resolving power of ${\lambda/\Delta\lambda=48\,700}$, corresponding to 6\,\kms. 
\changed{The slit was positioned such that it contained the central star in order to get
         the full velocity separation for rim and shell.
         The slit width was always very small compared to the object sizes.}  
  The useful spectral ranges were 4700--5800\,{\AA} and 5490--6760\,{\AA}, respectively.
  The detector used
  was a 15\,$\mu$m SITe-CCD, giving a spatial resolution of 0\farcs5\,pix$^{-1}$.
  The typical seeing during the observations was between 2\arcsec\ and 3\arcsec.  
  We tried to chose the exposure times such that also the fainter
  shoulders of the emission lines have sufficient signal.  Most objects were
  observed at least twice.

The spectra were reduced and calibrated using standard ESO-MIDAS packages. Spectrum reduction 
included bias and straylight subtraction, filtering of cosmic ray events, flat fielding, 
order extraction, and wavelength calibration using a ThAr lamp. In the result, we obtained 
one 2D-image for each \'echelle order. From these, we extracted the profiles of 
the strong \nii\ and \oiii\ lines, and those of H$\alpha$ as well, using the central pixel row
along the cross-dispersion direction. 
Where necessary, the spectra of the adjacent rows on either side were also 
extracted and co-added in order to improve on the S/N.\footnote
{\changed{We did not make use of the velocity ellipses because the extent of many targets
          were not fully covered by the slit.  The stellar continuum is much too faint as 
          to disturb the strong emission lines, especially for the high spectral 
          resolution used here. }  }
  
  Most of our targets are still in the stellar high-luminosity phase
  of their evolution, and examples of the line profiles are displayed in Fig.\,\ref{multi.1}  
  of the Appendix, altogether for 15 PNe.   Few objects  are in a later phase of evolution 
  where recombination of the outer nebula parts had already occurred and which appear now
  to be in the reionisation stage. 
  Analysis and discussion of these objects are postponed to Sect.~\ref{evolved.PNe}.
  
  The appearances of the line profiles are very similar and reflect the similarity
  of the apparent morphologies. The two main line components of the receding and 
  approaching parts of the nebular shell are usually split into two subcomponents with 
  separation and relative strength that reflect the kinematical state of the respective 
  object:   The inner of these subcomponents, which is the signature of the bright rim, 
  are usually flanked by much fainter shoulders
  (more pronounced in the light of \nii) whose positions   
  indicate velocities much higher (up to about seven times, see Fig.~\ref{comb.diff}) 
  than those typical of the bright rim.
  The line emission of these shoulders originating from the shell 
  is sometimes very weak and can easily escape detection in spectrograms of 
  insufficient S/N, although they are, as we have seen, of utmost importance
  for interpreting expansion properties of PNe. 
  It is obvious that any detailed information on nebular kinematics is \emph{not\/} 
  retrievable from hydrogen (or helium) lines because of their high thermal broadening
  (cf. H$\alpha$ profiles shown in Fig.~\ref{multi.1}).  

\begin{figure}
\includegraphics[width=0.96\columnwidth]{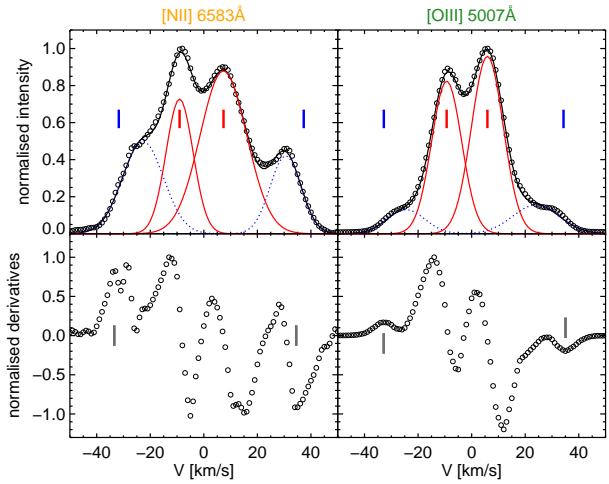}
\caption{\label{line.analyse}
         Examples of high-resolution, high signal-to-noise line profiles (open symbols) and 
         their decompositions into 4 Gaussians (blue dotted and red solid) 
         (\emph{top}) and line profile
         derivatives (\emph{bottom}) from \nii\ (\emph{left}) and \oiii\ (\emph{right})
         for the PN \object{NGC 6826}, from spectrograms taken at Tautenburg observatory. 
         The sum of the four Gaussians is shown in black solid (\emph{top}).
         Vertical dashes in the \emph{top panels} mark either the positions of the two rim
         components (as follows from the decomposition), or the outer inflection points of the 
         shell Gaussians (center of Gaussian\,+(--)\,sigma).  The dashes in the \emph{bottom
         panels} indicate the outer inflexion points of the observed line profile 
         (corresponding to the outer relative extremes of the derivatives).
         \vspace{1mm}      }
\end{figure}

  We retrieved the kinematical informations from the line profiles in the following way:
  Firstly, we decomposed the whole line profile into four Gaussians and identified the two
  innermost, strongest components to be representative for the flow of the gas belonging to 
  the rim.  Then we used the line profile derivatives to
  determine the (outer) inflexion points of the shoulder profiles, i.e. we determined the
  outermost extremes of the derivatives.  It was demonstrated in Paper~II 
  that, based on radiation-hydrodynamics models, these reflexion points correspond very closely
  ($\pm$1~\kms) to the post-shock velocity of the shells (see also Fig.~\ref{decomposition}). 
  The method is illustrated in detail in Fig. \ref{line.analyse} for \object{NGC 6826}.
 
  We note that this method demands an extremely high quality of the observations, which
  was not always met. In these cases where the derivative could not be used 
  we took the two additional Gaussians as substitutes of the
  shell emissions and used their (outer) inflexion points as representative
  of the post-shock velocity. We verified by means of high-quality
  line profiles that the positions of the shell inflexion points from the line profile agree 
  reasonably well with those derived from the Gaussians.    
  In the example shown in  Fig. \ref{line.analyse} there exists for \nii\ only a small
  difference between both methods, $\Delta V\la$\,2 \kms.
 
  In passing we note the obvious asymmetry seen in the line profiles of NGC 6826 as shown in
  Fig.~\ref{line.analyse}: (i) The velocity difference between rim and shell is higher for the
  receding part of the nebula compared to the approaching part, and (ii) this velocity
  asymmetry seems to be correlated with a difference in the strength of the two rim components,
  most notably in the \nii\ profile.  We will discuss this issue in more detail in 
  Sect.~\ref{asymm}. 

  All individual velocities measured by profile decompositions and derivatives
  are collected in Table\,\ref{data.tautenburg}, with references
  to previous works.  The typical error for an individual velocity measurement
  is $\pm 1$\,\kms\ for velocities of the rim components (peak separation of the
  Gaussian fits) and $\pm2$\,\kms\ for the post-shock velocities of shells 
  (separation of the derivative extremes).  
  Systematic errors due to slit position and poor seeing can be larger, i.e.\ up to
  $\pm3$\,\kms.

\begin{table}[!t]                 
\caption{Mean values of measured rim and shell (post-shock) velocities, deduced from the 
         profiles of \oiii\ 4959/5007 \AA\ and \nii\ 6548/6583 \AA, and effective temperatures 
         of the central stars for Galactic disk objects.  
         The velocity data are from Table~\ref{data.tautenburg} (Tautenburg),
         supplemented by reanalysed data from observations taken at ESO or La Palma 
         (Sch\"onberner et al. \cite{SJSPCA.05}).
         }
\label{tab.gesamt}
\tabcolsep=5.0pt
\begin{tabular}{lccccc}
\hline\noalign{\smallskip}
Object  & \multicolumn{2}{c}{$V_{\rm rim}$ (km\,s$^{-1}$)}
        & \multicolumn{2}{c}{$V_{\rm post}$ (km\,s$^{-1}$)}
        & $\log\left(T_{\rm eff}/{\rm K}\right)$\\[1.5pt]
\cline{2-3}
\cline{4-5}
\noalign{\smallskip}
       &[O\,{\sc iii}]&[N\,{\sc ii}]&[O\,{\sc iii}]&[N\,{\sc ii}]&           \\[1.5pt]
\hline
\noalign{\smallskip}
Tautenburg:        &            &           &        &       &                   \\[3pt]
\object{IC 289}    &        25  &        -- &  47    &  --   &    5.00$^{\,2}$   \\
\object{IC 3568}   &\enspace 9  &\enspace 9 &  33    &  31   &    4.70$^{\,1}$   \\
\object{IC 4593}   &\enspace 3  &\enspace 4 &  22    &  24   &    4.60$^{\,1}$   \\
\object{M2-2}      &        11  &        11 &  36    &  --   &    4.90$^{\,5}$   \\
\object{NGC 2022}  &        29  &        -- &  --    &  --   &    5.00$^{\,2}$   \\
\object{NGC 2610}  &        33  &        -- &  --    &  --   &    5.00$^{\,2}$   \\
\object{NGC 3242}  &        16  &        18 &  39    &  40   &    4.87$^{\,1}$   \\
\object{NGC 6543}  &        16  &        21 &  --    &  45   &    4.82$^{\,6}$   \\
\object{NGC 6826}  &\enspace 8  &\enspace 9 &  32    &  34   &    4.70$^{\,1}$   \\
\object{NGC 6884}  &        --  &        19 &  --    &  --   &    4.94$^{\,3}$   \\
\object{NGC 6891}  &\enspace 7  &\enspace 7 &  32    &  34   &    4.70$^{\,1}$   \\
\object{NGC 7009}  &        18  &        18 &  36    &  --   &    4.91$^{\,1}$   \\
\object{NGC 7354}  &        24  &        27 &  44    &  44   &    5.00$^{\,3}$   \\
\object{NGC 7662}  &        26  &        27 &  --    &  --   &    5.00$^{\,3}$   \\
\object{Vy\,2-3}   &\enspace 9  &        -- &  35    &  --   &    4.78$^{\,3}$   \\
\hline
\noalign{\medskip}
\multicolumn{2}{l}{ESO or La Palma:}   &            &       &       &                \\[3pt]
\object{IC 418}$^{\,*}$    &  $<$\,5    &        --  &  --   &  13   &  4.56$^{\,1}$  \\
\object{IC 2448}$^{\,*}$   &  18        &        18  &  35   &  --   &  4.81$^{\,1}$  \\
\object{M1-46}             &  --        &        12  &  --   &  32   &  4.65$^{\,4}$  \\
\object{My 60}$^{\,*}$     &  23        &        24  &  38   &  --   &  5.02$^{\,3}$  \\
\object{NGC 1535}$^{\,*}$  &  22        &        --  &  32   &  --   &  4.85$^{\,1}$  \\
\object{NGC 2610}$^{\,*}$  &  34        &        --  &  42   &  --   &  5.00$^{\,2}$  \\
\object{NGC 5882}          &  --        &        22  &  --   &  47   &  4.83$^{\,3}$  \\
\object{NGC 6629}          &  --        &\enspace 6  &  --   &  34   &  4.67$^{\,1}$  \\
\object{NGC 7662}$^{\,*}$  &  26        &        --  &  35   &  --   &  5.00$^{\,3}$  \\
\object{Tc 1}              &  --        &        12  &  --   &  30   &  4.52$^{\,1}$  \\
\hline
\end{tabular}
\\[3pt]
\emph{Notes}: $^*$\,$V_{\rm post}$ from long-slit spectrograms (Paper~II).                                                        
Stellar temperatures are from:
$^1$\,M\'endez, Kudritzki \& Herrero (\cite{MKH.92}),
$^2$\,McCarthy, M\'endez \& Kudritzki (\cite{MMK.97}),
$^3$\,He\,{\sc ii} Zanstra temperature (Gorny, priv.\ comm.),
$^4$\,Guerrero et al. (\cite{Guerrero.96}),
$^5$\,Aller \& Keyes (\cite{AK.87}),
$^6$\,Georgiev et al. (\cite{georgetal.08}).
\vspace{-2mm}
\end{table}

  All velocities for a line component and ion listed in Table\,\ref{data.tautenburg} 
  were (if possible) averaged over the individual measurements and  compiled 
  in Table \ref{tab.gesamt}, together with the respective central-star temperatures taken 
  from the literature.   We supplemented the Tautenburg sample         
  by objects from the work of Sch\"onberner et al. (\cite{SJSPCA.05}, Table 4 therein).
  These additional objects were observed either at La Silla (ESO/CAT or NTT) or on
  La Palma (NOT), and the line profiles were reanalysed  
  in the same way as described here.  For some objects (annotated in the table), $V_{\rm post}$
  was taken from the analyses of long-slit spectrograms reported in Paper~II.

\subsection{Planetary nebulae in the Galactic halo }            
\label{halo.PN}   

   The mean nebular oxygen abundance of the objects of our sample listed in
   Table~\ref{tab.gesamt} is, according to Henry et al. (\cite{HKB.04}), 
   $\langle$12 + log\,(O/H)$\rangle$ = 8.65, with a 1$\sigma$ dispersion 
   of 0.10 dex only.\footnote
{More precisely, this mean abundance is based on 14 objects out of the 27 listed in 
 Table~\ref{tab.gesamt} which are in common with the compilation of 
 Henry et al. (\cite{HKB.04}).
   The more recent determinations by Pottasch \& Bernard-Salas (\cite{PB.10})  yield 
   $\langle$12 + log\,(O/H)$\rangle$ = 8.54, also with a 1$\sigma$ dispersion 
   of 0.1 dex, but from 7 common objects only.}   
   This average oxygen abundance can be considered as typical for Galactic disk PNe.  
   Since oxygen is the main cooling agent in a PN, we use it here as a convenient proxy 
   for the overall metallicity $Z$.
   
   The rather small variation of the oxygen abundance within our PN sample in
   Table~\ref{tab.gesamt} renders any study of the dependence of the expansion velocity 
   (i.e. of $V_{\rm post}$) on metallicity impossible.   
   A better control sample is certainly provided by metal-poor 
   nebulae, such as found in the Galactic halo. 
   We selected four targets from the list of metal-poor (halo) PNe provided by Howard et al.
   (\cite{HHMcC.97}), \object{BoBn~1}, \hbox{\object{M2-29}}, \object{PRMG~1}, and \object{Ps\,1},
   the nebula around K648 in M\,15.   
  
  These four metal-poor PNe were observed in August 2006 by C.\,Sandin at the VLT
  with the FLAMES/ ARGUS integral-field unit  (Program 077.D-0652, PI D. Sch\"onberner).
  Two spectral regions were observed, 4911--5158 \AA\ (HR8) and 6383--6626 \AA\ (HR14B),
  with resolutions of 32\,000 and 46\,000, resp.  The seeing was below 1\arcsec, sufficiently
  good to ensure full spatial resolution of the targets.  The reduction of the data was
  performed using the standard Giraffe reduction pipeline 2.5.3, and from the resulting
  2D spectra the (final) 1D spectrum for the central line-of-sight was retrieved for
  further analysis.   The presentation of relevant line profiles and their detailed 
  interpretation in terms of expansion velocities  is postponed to Sect.~\ref{met.vel}.

%
\section{General expansion properties}                 
\label{exp.vel}
\subsection{Rim and shell velocities of high-luminosity PNe}                             
\label{int.kin}

  A comparison with results from the literature, notably from the catalogue of
  Weinberger (\cite{We.89}), is not straightforward since no reference to faint 
  outer wings is given, and also the details of the
  measurements and the errors are difficult to judge. Assuming that the listed
  velocities refer to the peak separation of the line profiles, there exists fair agreement
  with our rim velocities (cf. Table~\ref{data.tautenburg}). 
  Only in rare cases more recent measurements of much higher velocities,
  comparable to our post-shock results, are available.  
        
  For three objects (NGC 2022, NGC 3242, and NGC 7662) a comparison of the results from 
  the Tautenburg spectrograms with the previous 
  measurements by Sch\"onberner \etal\ (\cite{SJSPCA.05}), which are also based on
  high-resolution spectra and treated in the same way, was possible.
  The agreement is excellent, with deviations of less than 2~\kms\ (rim only).     
    
\begin{figure} 
\vskip-1mm
\hskip-1mm
\includegraphics[width=1.01\columnwidth]{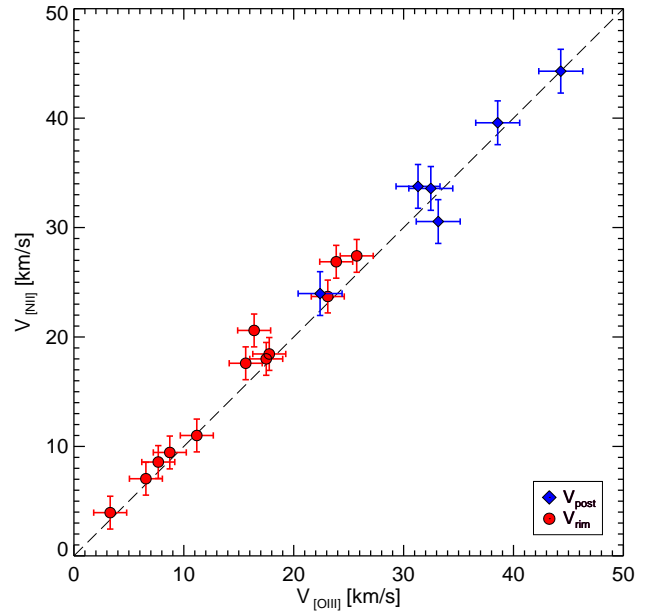}
\caption{Comparison of measured rim velocities (red dots) and shell post-shock velocities 
         (blue diamonds) based on \nii\ and 
         \oiii\ lines as listed in Table\,\ref{tab.gesamt} for those objects for which values
         from both ions are available, with appropriate error bars.  
          The dashed diagonal line is the 1:1 relation.  Note also that not for all rim entries
          a corresponding shell post-shock value does exist.
        }
\label{vergleich}
\end{figure}

   More interesting is the internal consistency between results from \nii\ and \oiii.
   This is illustrated in Fig.\,\ref{vergleich} for those objects where both velocities  
   are available (cf. Table\,\ref{tab.gesamt}). 
  A close 1:1 relation between the velocities from \nii\ and \oiii\ lines
  is to be expected for optically thin nebulae since both ions extend over the whole  
  PN, although with different abundances.  This is indeed the  case for the post-shock
  velocity: Within the errors, the values from \nii\ and \oiii\ do agree.  
  The same holds for $V_{\rm rim}$ of young objects with low rim velocities.

  There appears, however, a small offset between velocities from \nii\ and \oiii\  
  at higher rim velocities in the sense that the \nii\ values are higher by a few \kms. 
  This indicates the development of a positive velocity gradient within the rim
  during the PN evolution (cf. Wilson \cite{Wi.50}).
  However, in the following discussion we neglect these small differences and use 
  averages if both velocities are available.


\paragraph{Notes on individual objects}
\label{indiv.objects}

  For five objects (IC 289, IC 3568, IC 4593, NGC 6543, and NGC 7009)   
  we detected the signatures of the attached shells in the central line profiles 
  for the first time and measured the corresponding post-shock velocities.
  In four cases (\object{NGC\ 2022}, \object{NGC 2610}, \object{NGC\ 6884}, and
  \object{NGC 7662}) we were unable to detect/measure the faint shell
  emission in the central line profile.  
 {Attached shells are well-known from the images, but their signatures in the 
 line profiles are masked
 because of too small velocity differences between rim and shell.	
 In these cases post-shock velocities can only be determined by means of long-slit 
 spectrograms (see Paper~II            
 and \object{NGC 2610} and \object{NGC 7662} in Table~\ref{tab.gesamt}). }  
  
  \object{NGC 2022} belongs morphologically also to the class of nearly round 
  double-shell PNe, yet it is an exception because its shell edge
  obviously expands slower than the rim, as is evident from the long-slit 
  spectrograms taken by Sabbadin et al. (\cite{SBH.84}).   We estimate a post-shock 
  velocity of about 20 \kms\ from their position-velocity diagrams.

  \object{IC 418} from the La Palma sample is a very young,
  compact PN with a cool central star ($\simeq$\,36\,000 K) which is just at the verge 
  of becoming optically thin (Morisset \& Georgiev \cite{MG.12}; Ramos-Larios et al. 
  \cite{ramosetal.12}).
\changed{For this very reason, our method for the determination of the post-shock
  velocity is not going to work since the regular double-shell structure is not yet
  existent. Instead, we proceeded as follows: We know from our simulations that this
  particular phase is characterised by a maximum of the shock velocity (cf. 
  Sch\"onberner et al. \cite{SchJSt.05}, Fig. 4, left panels). 
  From the models we estimated a correction of 5--10\,\% necessary for converting the 
  velocity from the \nii\  peak separation (12.5 \kms, Sch\"onberner et al. 
  \cite{SJSPCA.05}) into the post-shock velocity, i.e. $V_{\rm post} = 13.5$ \kms. 
  The shock velocity itself is then $13.5{\times}1.25 = 17$ \kms.}

\subsection{Kinematics and evolution}                             
\label{evol.kin}

  The total combined sample of double-shell PNe with accurate velocity information 
  for both the rim (bulk gas velocity) and shell (post-shock velocity) spans, 
  as one can judge from the stellar temperatures, 
  a wide range of evolutionary stages and is about twice as large as the sample 
  used by Sch\"onberner et al. (\cite{SJSPCA.05}).  
  For the following interpretations we prefer here again, as in Sch\"onberner
  et al. (\cite{SJSPCA.05}), the use of the effective temperature of the central stars 
  as a distance-independent proxy for the evolution in time.\footnote 
  {The use of the stellar temperature is of advantage because both the radiation field and 
  the wind power of a central star are directly related to its effective temperature
  during the horizontal part of evolution across the HR diagram, and 
  both are also the main drivers of PN evolution.
  A second advantage is that the total effective temperature range varies only rather modest
  with central star mass, which means that the effective temperature is not closely related
  to age, unless all central stars have the same mass.}

\begin{figure}
\vskip-2mm
\hskip-2mm
\includegraphics[width=1.02\columnwidth]{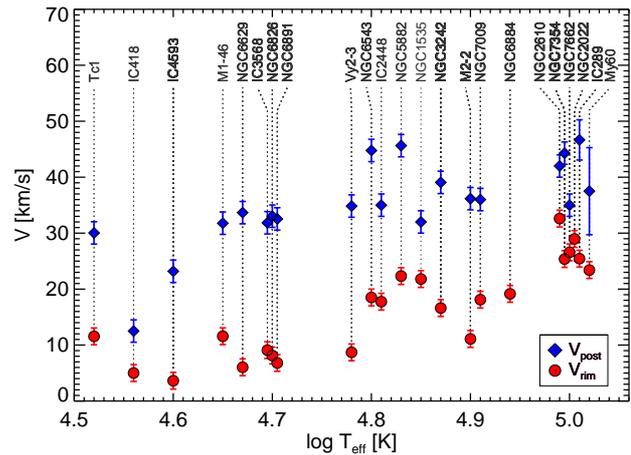}
\vskip-1mm
\caption{\label{evol.teff}
         Shell post-shock velocities, $V_{\rm post}$ (blue diamonds) and rim velocities,
         $V_{\rm rim}$ (half of line peak separation, red dots) of the objects  
         listed in Table\,\ref{tab.gesamt} vs. the effective temperatures of the 
         respective central stars. There are generally two velocity 
         entries for each object, except for those for which the post-shock velocity 
         could not be measured (NGC 2022 and NGC 6894). Velocities are either from 
         \nii\ or \oiii, or averaged 
         if information from both ions is available (cf. Table~\ref{tab.gesamt}).  
         Vertical dotted lines connect the two velocities with the object's name.
         Some object entries are slightly shifted in temperature as to avoid overlapping. 
         }
\end{figure} 

  Figure \ref{evol.teff} is an improvement of Fig. 12 (top) in Sch\"on\-berner
  et al. (\cite{SJSPCA.05}) and presents a general velocity evolution of
  rim and shell while the star crosses the HRD.
  A qualitatively similar diagram (many objects in common with our 
  sample) can be found in M\'endez et al. (\cite{MKHHG.88}, Fig.~5b therein), 
  but it gives the velocity from the half-width of H$\gamma$ only, i.e. without any
  discrimination between rim and shell velocities. 
  Nevertheless, the basic trend is the same as here: The velocities increase
  with stellar temperature. The main results from Fig.~\ref{evol.teff} are the following:
\begin{itemize}
\item  The shell's edge expands considerably faster than the rim, 
       but both accelerate during the evolution,   
       albeit in a different way.
       The post-shock velocity of the shell increases  with $\Teff$
       from about 25~\kms\ up to about 40 \kms\ at ${\log \Teff \ga 4.8}$.   
       In contrast, the rim is
       stalling at rather low velocities ($\simeq$\,5--10 \kms) for quite a while
       until it accelerates up to about 30 \kms.

       Because of this different evolutionary behaviour,
       the velocity difference between rim and shell changes from initially 
       $\sim$\,20--25 \kms\ to about 15~\kms\ only, making it more difficult to detect
       the shell signature in the line profiles of evolved objects (cf. notes on 
       individual objects).
\item The initial shell expansion
      starts with post-shock flow speeds \emph{higher} than the    
      typical AGB outflow velocities of $\approx$\,10--15 \kms).\footnote 
{The real expansion velocity of the shell is given by the shock propagation which
 is even higher than the post-shock velocity (see Sect. \ref{comp.hydro}).}  
      The rim gas, however, has  
      initial velocities which are roughly equal or even well 
      well \emph{below} the typical 
      AGB outflow velocities. Typically for these young PNe is
      that the post-shock velocity of the shell exceeds the rim expansion by factors up
      to more than six ($\simeq$\,3.5 \kms\ vs. $\simeq$\,23~\kms\ for IC 4593)!
\end{itemize}

   Our results are consistent with those in Fig.\,12 of Sch\"onberner et al.
   (\cite{SJSPCA.05}), and the interpretation of Fig. \ref{evol.teff} is then the following: 
   At first, as the AGB remnant heats up, photoionisation is at work and forms an 
   ionised shell out of the neutral AGB matter which is still slowly expanding
   with about ${V_{\rm agb}\simeq}$\,10--15 \kms. The leading edge of this shell --
   the D-type shock ahead of the ionisation front -- accelerates into the AGB wind 
   and attains velocities well above $V_{\rm agb}$.
   At the same time, the inner part of the shell becomes decelerated      
   because the thermal pressure of the wind-blown bubble falls short of the increasing shell
   pressure.   The reason is the initially rather low
   wind speed (or wind luminosity) (see, e.g., Fig. \ref{mod.prop}, right panels). 
   Nevertheless, a (proto-) rim forms slowly as the  stellar wind gains power with
   time (cf. left panel in Fig.~\ref{model.595}), but its outward velocity remains rather 
   low for quite a while.
  
   As evolution across the HRD progresses, the nebula becomes
   optically thin, which means that the ionisation front overtakes the shock and
   enters the so-called ``champagne phase'' of expansion (Tenorio-Tagle \cite{TT.79};
   Chevalier \cite{ch.97}).   Since the wind power of the central star, 
   ${L_{\rm wind} = \dot{M} V_{\rm wind}^2 /2}$, increases steadily 
   (cf. Fig.\,\ref{mod.prop}, middle panel), and because a certain part of the
   kinetic wind energy is converted into thermal energy of the hot bubble,  
   the rim is being accelerated into the shell, thereby sweeping up shell matter. 
   This is the double-shell phase during the evolution of a PN, depicted also
   in Fig.~\ref{model.595}.

\section{Comparisons with radiation-hydrodynam- ics simulations}
\label{comp.hydro}

    Before continuing with this section, we emphasise that none of our 
    1D-radiation-hydrodynamics simulations presented in this work are aimed at fitting 
    any particular object.  They are too simple with respect to geometry, and they are thus
    only used in helping to interpret our observations. But we believe that, due to the very
    detailed physics implemented, our models give, on the average, a realistic description
    of the most important aspects of real PN evolution.

\subsection{The shells}
\label{comp.shell}

\begin{figure}
\vskip 1mm
\hskip-1mm
\includegraphics[width=0.99\columnwidth]{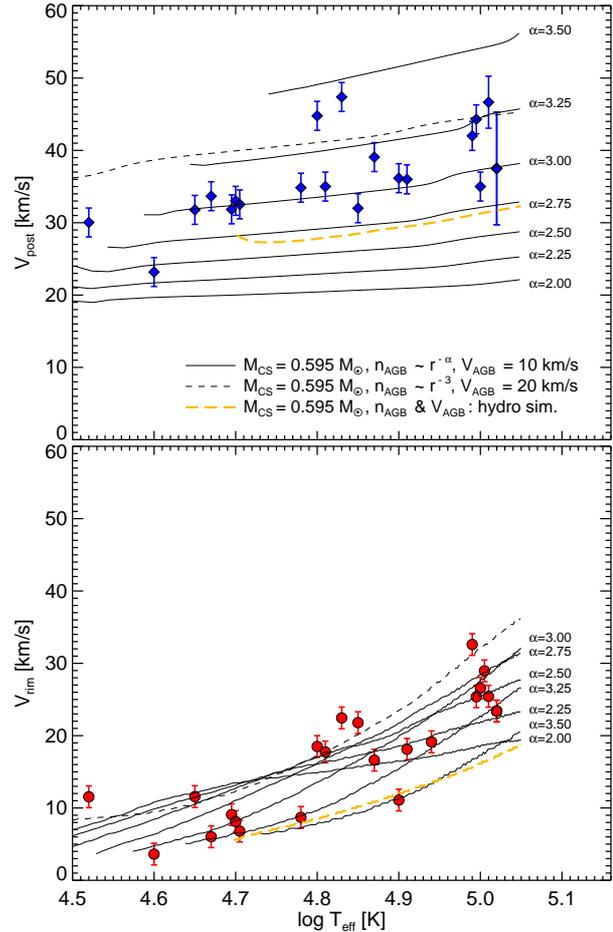}
\caption{\label{shell.comp} 
         Observed post-shock velocities (\emph{top}) and rim velocities (\emph{bottom}), 
         both from Fig.~\ref{evol.teff}, compared with predictions from a series of 
         model simulations with the same central-star model, as indicated in the figure
         legends.  Only the optically thin models are plotted (consequently, IC 418 is omitted).
         In both panels,  the thin lines belong to models that expand into ambient matter
         with power-law density profiles, characterised by the initial exponent $\alpha$. 
         The original AGB-wind velocity is 10 \kms\ (solid) or 20 \kms\ (short dashed),  
         for $\alpha=3$ only in the latter case.  The long-dashed lines belong 
         to a ``hydro'' model.  For more details, see text.   \vspace{-1mm}
         } 
\end{figure}

   Figure \ref{shell.comp} is again an extended and modified version of Fig.\,12 (bottom) of 
   Sch\"on\-ber\-ner et al. (\cite{SJSPCA.05}), displaying now the post-shock velocities 
   for our sample of PNe from Table~\ref{tab.gesamt} together 
   with the predictions of (optically thin) hydrodynamical models with simplified initial
   conditions as explained in Sch\"on\-ber\-ner et al.. In short, all  model 
   nebulae have the same central star of 0.595~\Msun\ and
   expand into an environment whose density falls off according to a power law,
   ${\rho \propto r^{-\alpha}}$, with constant $\alpha$ (the only exception is 
   the ``hydro'' sequence where $\alpha$ increases with distance from the star). 

   Figure \ref{shell.comp} (top panel) illustrates convincingly how strong the velocity behind 
   the leading shock depends on the radial density slope of the upstream matter.
   The modest acceleration is due to
   the electron temperature increase while the star heats up (cf. Figs.~3 and 5
   in Sch\"on\-ber\-ner et al. \cite{SJSPCA.05}) and to a slight steepening of the
   density gradient with time (see below).
   The initial AGB wind velocity must be considered, too, because theory predicts 
   shock (and post-shock) velocities \emph{relative} to the upstream flow.
   An ${\alpha=3}$ model sequence with higher AGB-wind speed (20 \kms) is plotted
   in Fig.~\ref{shell.comp} (top panel) for comparison.
   The range of AGB wind velocities observed
   is, however, only a few \kms\ ($\la\,$5 \kms) to either side of a typical
   value (10 \kms\ is assumed here). Correspondingly, part of the 
  the observed velocity spread may result from different AGB velocities.

   The interpretation of Fig. \ref{shell.comp} (top panel) is then the following: 
   Most objects expand like the ``${\alpha = 3}$'' model, but some have steeper
   density gradients, one (IC 4593) obviously a flatter upstream density profile.   
   We note that our ``hydro'' model expand a bit too slowly 
  (cf. long-dashed line in Fig.~\ref{shell.comp}):  The reason is the still rather
   modest (initial) density gradient which is comparable to an $\alpha \simeq 2.7$ model for
   that part of the evolution that is shown in the figure.

   Direct density profile determinations of halo matter have been performed by Plait \& Soker 
   (\cite{PS.90}) and, more recently, by Sandin et al. (\cite{sandetal.08}).
   Both groups converted the halo surface-brightness profiles of a few objects 
   (\object{IC 3568}, \object{M2-2}, \object{NGC 6826}, \object{NGC 7662})  into density
   profiles by means of an Abel transformation. Plait \& Soker (\cite{PS.90}) found 
   $\alpha^\prime$ = 4 for the halo of \object{NGC 6826}, whereas Sandin et al. 
   (\cite{sandetal.08}) determined the following average values: $\alpha^\prime$ = 5.6 
   (IC 3568), 4.5 (M2-2), 4.2 (NGC 6826), and 4.6 (NGC 7662).\footnote
   {We used $\alpha^\prime$ (instead of $\alpha$) to characterise the actual halo 
   density gradient because the upstream density slope steepens continuously with time
   once the PN becomes optically thin and the halo heated by photoionisation.  
   For instance, starting with a power law exponent of $\alpha =3.0$, one ends up
   with an (upstream) exponent ${\alpha^\prime = 3.7}$ once the stellar temperature 
   has passed 10$^5$~K! This effect is the higher the steeper the initial density slope.} 
               
   Using the images presented in Corradi et al. (\cite{CSSP.03}),
   we were able to measure the surface brightness gradient of the halo   
   in front of the leading shell edge for six objects from Table\,\ref{tab.gesamt}: 
   IC 2448, NGC 1535, NGC 2022, NGC 3242, NGC 6826, NGC 7009, and NGC 7662.   
   We adopted an inner and outer radius for the gradient determination at
   roughly  1.1 and 1.3 shell radii.  
   The slope was then measured at 36 position angles in the haloes, and averaging 
   the results for H$\alpha$\,+\,\nii\ and \oiii\ where
   both are available (there are no significant differences between the H$\alpha$ and 
   \oiii\ measurements).   Excluding NGC 2022 (see below),
   the average surface brightness slope is characterised by ${\gamma = 5.2 \pm 0.2}$,
   with ${{\rm SB} \propto p^{-\gamma}}$, where $p$ is the impact parameter.
   Individual values range between 5.0 and 5.5, with corresponding errors between 0.1 
   and 0.7.  The main sources of these errors are brightness irregularities and small
   deviations from sphericity. 

   For converting the average halo intensity slope $\gamma$ into a corresponding density
   slope $\alpha^\prime$ we used the approximation $\alpha^\prime = (\gamma + 1)/2$
   and got $\alpha^\prime = 3.1\pm0.1$ for the average upstream halo density slope of the five
   PNe listed above (NGC 2022 excluded).\footnote 
{We assume an infinitely extended 
 halo with density slope ${\rho \propto r^{-\alpha^\prime}}$ 
 and approximate its surface brightness/intensity in an emission line at impact parameter $p$,
 ${\rm SB} \propto p^{-\gamma}$, to be proportional to $(1/p^{\alpha^\prime})^2\,  p$.} 
   We tested this approximation by using our
   radiation-hydrodynamic models and proceeded in the same way as for the real objects.
   We found a very good consistency between $\gamma$ and 
   $\alpha^\prime$ for ${\alpha = 2}$ models, 
   but for  ${\alpha>2}$ the above approximation tends to underestimate
   the density slope. For instance, our  ${\alpha = 3}$ models suggest that the present 
   density slope is underestimated by about 0.3.  Keeping this in mind, we estimate an
   average value of ${\alpha^\prime \simeq 3.4}$, i.e. a bit higher than the mean 
   initial value ${\alpha \approx 3.1}$ read off from Fig. \ref{shell.comp} (top panel). 
   The five considered PNe have, on the average, a post-shock velocity
   of 35 \kms, with a 1$\sigma$ dispersion of 3 \kms.
      
   The direct comparison of our measurements with those of 
   Sandin et al. (\cite{sandetal.08})
   is possible only for two objects, NGC 6826 and NGC 7662.  Our individual values for
   $\alpha^\prime$, 3.3$\pm$0.2 and 3.2$\pm$0.3, fall short of Sandin et al.'s values
   4.2 and 4.6, resp., even if one corrects our values by $+$0.3.  One must keep in mind,
   however, that both methods trace different halo regimes: 
   The Abel transformation used by Sandin et al. (\cite{sandetal.08}) considers the entire 
   but finite halo brightness distribution
   and corrects implicitly also for possible projection effects, under the assumption
   of strictly spherical symmetry.  Our method traces only the halo region
   immediately ahead of the shell and averages over possible asymmetries.  Also, the
   role of straylight from the bright nebula must be investigated and possibly considered.
   Given all these uncertainties and the individual errors, we believe that both methods 
   give fairly consistent results.
               
   A separate discussion is needed for NGC 2022 whose post-shock velocity is, quite unusually,
   lower than  $V_{\rm rim}$ ($\simeq$\,20 \kms\ vs. 29\ \kms; see Table\,\ref{tab.gesamt} and
   notes at end of Sect.\,\ref{int.kin}).  We measured a halo 
   surface-brightness slope of ${\gamma = 2.7 \pm 0.6}$, resulting in a density slope 
   ${\alpha^\prime = 1.9 \pm 0.3}$ only!  Our ${\alpha = 2.0}$ model displayed in  
   Fig.~\ref{shell.comp} (top panel) expands, without virtually any acceleration, 
   with a post-shock velocity of $\simeq$\,20 \kms, in excellent agreement.  Note that the
   rim of NGC 2022 expands with 29 \kms\ as is typical for objects with hot central stars
   (cf. Table~\ref{tab.gesamt} and bottom panel of Fig.~\ref{shell.comp}). 
                
   The determination of halo density profiles and their comparison with the respective
   expansion (= post-shock) velocities provide a direct observational proof that the
   nebular expansion is ruled by the upstream density profile, as radiation-hydrodynamics
   predicts.   The implication of the observed high expansion rates 
   ($V_{\rm post}$ up to ${\approx\! 40}$ \kms) 
   is that the shells expand into halos whose initial
   density profiles can be characterised by a power-law ${\rho \propto r^{-\alpha}}$ 
   with ${\alpha\approx 3}$. 
   Assuming constant outflow velocities, AGB mass-loss rates  
   must steadily increase until they stop suddenly at the very end of AGB evolution
   when the remnant is already rapidly evolving off the AGB.
   More details can be found in Sandin \etal\ (\cite{sandetal.08}).
   Exceptions like NGC 2022 seem to be quite rare. 

   The hydrodynamical simulations along the final AGB evolution performed by Steffen et al.
   (\cite{steffenetal.98}) suggest that the mass-loss increase is large when the
   star recovers from the luminosity dip after a thermal pulse (cf. their Fig. 2). 
   In contrast, immediately before the following pulse, luminosity, and hence mass-loss rate,
   remains nearly constant.  If a star evolves off the AGB shortly before the next pulse,
   the density slope of the AGB wind envelope should be close to an ${\alpha = 2}$ case.
   One is then tempted to assume that NGC 2022 is in such a thermal-pulse cycle phase,
   implying that it is a candidate for a late or very late thermal pulse, i.e. for a
   thermal pulse occurring during its way to become a white dwarf, like, e.g., the case
   of FG Sge.   The normal case would then be that stars leave the AGB well before 
   the next thermal pulse and develop AGB envelopes with steep density gradients.

\subsection{The rims}
\label{comp.rim}

   The rim is thought to be an example of a wind-compressed 
   shell of matter whose expansion is controlled by the (thermal) pressure of the
   shock-heated stellar wind matter (i.e. the hot bubble) from within and the density and 
   velocity of the shell matter immediately ahead of the rim's shock (cf. Koo \& McKee
   \cite{komc.92}). Hence, with given properties of the ambient medium, only the evolution 
   of the stellar mechanical luminosity (or wind power) is relevant for the kinematics of
   the rim gas.

   The bottom panel of Fig.~\ref{shell.comp} compares our measured rim velocities
   with the model predictions.  Generally, the rim evolution is determined by the
   accumulated rim mass and the shell's expansion property:  Low-$\alpha$ sequences
   start with low rim masses, hence with higher speeds, but later the rim does not
   accelerate much because of the rather slow shell expansion.  
   For high-$\alpha$ sequences,
   the situation is opposite: The rim starts slowly, but accelerates later.  
   In any case, the models cover the observed range of rim velocities and their
   evolution with stellar temperature. The rim velocities of our ``hydro'' sequence
   constitute the lower bound of the observations. 

\begin{figure}
\vskip-3mm
\hskip-2mm
\includegraphics[width=0.98\columnwidth]{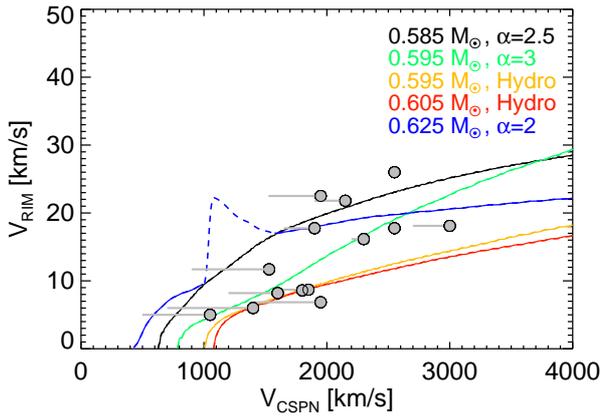}
\vskip-1mm
\caption{\label{rim.comp}
         Rim matter expansion, $V_{\rm rim}$, of objects from Fig. \ref{evol.teff} 
         for which also wind velocities, $V_{\rm wind}$, of their central stars are available 
         from the literature (cf. text). The length of the horizontal bars indicates 
         the range of stellar wind velocities reported in the literature, and
         the object symbols are placed at the high-velocity end of the respective bars. 
         The curves give the corresponding model predictions for various sequences as
         listed in the legend. 
         The theoretical profiles are broadened by a Gaussian with 6 \kms\ FWHM, and 
         $V_{\rm rim}$ is determined from the peak separation of the strongest line components. 
         The dashed part of the 0.625 \Msun\ simulations indicates a transient phase 
         in which the shell component gets stronger than the rim component, i.e. 
         the rim velocity could not be properly traced.      \vspace{-2mm}   
         } 
\end{figure}

   Whereas central-star mass-loss rates are only rather poorly known quantities, the 
   wind velocities are much better observationally constrained.  Also, from the theory of 
   radiation-driven winds follows that mass-loss rates do not change very much
   during the evolution across the HRD with constant luminosity, in contrast 
   to the wind velocity which scales with the stellar escape velocity, i.e. with 
   $(M/R)^{0.5}$, where $M$ and $R$ are stellar mass and radius, respectively.
   Thus the wind velocity increases by more than a factor of 100 during the 
   entire post-AGB evolution (cf. Pauldrach \etal\ \cite{pauldrachetal.88}; Pauldrach \etal\ 
   \cite{pauldrachetal.04}, and Fig.\,\ref{mod.prop}, right panels).

   Because of the dominant influence of the wind velocity on the wind power
   we considered it worthwhile to correlate the rim velocities, $V_{\rm rim}$, 
   also with the actual stellar wind speeds.  
   The result is displayed in Fig.\,\ref{rim.comp}. The stellar wind velocities are taken
   from the recent analysis and compilation of FUSE and IUE  \hbox{P-Cygni} line profiles
   by Guerrero \etal\ (\cite{GRM.10}). Usually, different ions give different answers,
   and the observed ranges of wind velocities are indicated in the figure by 
   horizontal bars.  Since wind speeds increase with distance from the stellar surface,
   we placed the object symbol at the high-velocity end of the respective velocity bar. 

   Figure \ref{rim.comp} reveals indeed a strong correlation between rim expansion and 
   wind velocity of the respective central star, proving thereby that it is the 
   stellar wind power that supports the nebula matter against collapsing and 
   also sweeps up shell matter into a rim which is then accelerated against the more slowly 
   expanding ambient shell gas.        
   The comparison with the predictions of the hydrodynamical model sequences already
   introduced in Sect. \ref{int.kin} is interesting:  
   The models fully encompass the observations and show the same trend with increasing
   central-star wind velocity.  

   Our Fig. \ref{rim.comp} is similar to Fig. 1 in Patriarchi \& Perinotto 
   (\cite{PP.91}) but displays a much tighter correlation because we
   used only objects with well known structure and carefully determined expansion property.
   The sample of Patriarchi \& Perinotto is quite inhomogeneous and contains
   also objects with Wolf-Rayet central stars. These are known to have much 
   higher mass-loss rates at similar wind velocities
   than stars with normal, hydrogen-rich surface composition,
   i.e. about 10$^{-7}$--10$^{-6}$ \Mdot\ (cf., e.g., Marcolino et al. \cite{marco.07},
   and references therein) as compared to about 10$^{-9}$--10$^{-8}$ \Mdot.
   The rim velocities of PNe with Wolf-Rayet nuclei are thus higher  
   for the same stellar temperature.

\subsection{The combined evolution of rim and shell}
\label{comb.evol}

   We have proven observationally in the previous subsections 
   that both the formation and evolution of the rims and shells occur quite 
   independently from each other, but nevertheless there 
   exists some sort of interaction: while the shell expands as given by the density 
   profile of the ambient medium, the rim
   must expand into the ambient shell gas.  Thus not only the velocities proper are
   of relevance, but also their differences or ratios, and a good theory should predict 
   their correct variation with time/evolution.  

\begin{figure}
\vskip-4mm
\hskip-3mm
\includegraphics[width=0.99\linewidth]{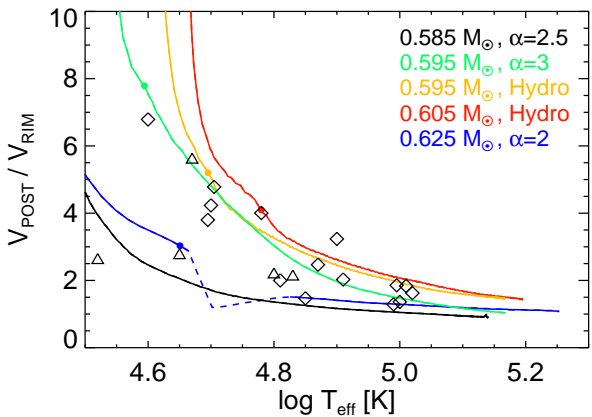}
\vskip-1mm
\caption{\label{comb.diff}
         The ratio of $V_{\rm post}$ and $V_{\rm rim}$ vs. stellar effective temperature,
         $T_{\rm eff}$, for the objects of Table~\ref{tab.gesamt} and as predicted by
         the same hydrodynamical sequences as shown in Fig.~\ref{shock.postshock} 
         (see legend).   Again, only the evolution towards maximum stellar temperature is
         of relevance.  The optically thick/thin transition (if not outside the 
         displayed range) is marked by a dot.   Diamonds are for \oiii\ measurements, and 
         triangles for \nii.  The dashed part of the 0.625\ \Msun\ sequence has the same 
         meaning as in the previous figure.  \vspace{-2mm}
        }
\end{figure}

   For this reason we show the ratio $V_{\rm post}/V_{\rm rim}$ in Fig.\,\ref{comb.diff}
   where we used again the stellar effective temperature as proxy of the evolutionary
   progress.   This figure elucidates impressively how large the ratio between the fastest
   flow velocities within a nebula and the usually measured flow velocities of the rim can be:
   For very young objects, a ratio of even about seven is possible!\footnote
{Note that $\dot{R}_{\rm out}/V_{\rm rim}$ is even larger, viz. by about 25\,\% (cf.
 Fig.~\ref{shock.postshock}).}
   There is, however, a significant evolution with effective temperature: The ratio 
   $V_{\rm post}/V_{\rm rim}$ approaches values around 1.5  
   for PNe with the hottest central stars.
   
    Figure~\,\ref{comb.diff} demonstrates again the close correspondence between theory and 
    observations.  One can state that our radiation-hydrodynamics simulation based on debatable
    assumptions concerning, e.g., the stellar winds and initial configurations, give already
    a rather satisfying description of the real world.

\section{Asymmetric expansion}   
 \label{asymm}
    
   Our precise measurements of rim (peak separation) and shell (post-shock) velocities revealed 
   a fact that, to our knowledge, has not found attention so far:  
   In several objects an asymmetry between
   the approaching and receding parts of the nebula is evident, in the sense that the
   differences between $V_{\rm rim}$ and $V_{\rm post}$ are not equal for both parts.   
   A typical  example with a rather obvious asymmetry is \object{NGC 6826} whose 
   \nii\ and \oiii\ line profiles are displayed in Fig.~\ref{line.analyse}. 
   The differences between both velocities are 29 \kms\ for the receding nebular
   shell and 23 \kms\ for the approaching nebular shell, measured in the \oiii\ lines,  
   a discrepancy of 6 \kms.  Nearly the same value follows from the \nii\ lines.
   It is also apparent from the line profiles of \object{NGC 6826} that it is the 
   stronger (receding) rim component that lags behind the expanding shell's shock, 
   as compared to the weaker approaching rim component.   

\begin{table}                              
\caption{\label{asymm.table}
         Parameters describing the expansion asymmetry, \hbox{RimAs} and ExpAS, given by the average
         of $N$ measured emission lines in \oiii. Values in parenthesis
         give the 1$\sigma$ errors.
         }
\tabcolsep=10pt
\begin{tabular}{l@{\qquad} cc c}
\hline\noalign{\smallskip}
Object   &    RimAs              &  ExpAs                  & $N$ \\[1.5pt]
          &                      &  (\kms)                &     \\
\hline\noalign{\smallskip}
IC 3568  &    0.61 (0.11)       &          --5.9 (0.5)   &  4  \\
M2-2     &    0.96 (0.09)       &          --5.1 (1.8)   &  8  \\
NGC 3242 &    0.79 (0.05)       &          --2.4 (1.2)   &  8  \\
NGC 6826 &    1.26 (0.05)       &\phantom{--}4.9 (0.5)   &  3  \\
NGC 6891 &    2.06 (0.01)       &\phantom{--}2.7 (0.5)   &  2  \\
Vy 2-3   &    0.51 (0.05)       &          --7.4 (0.2)   &  4  \\
\hline
\end{tabular}
\vspace{-2mm}
\end{table}

   We defined an ``expansion asymmetry'', ExpAs, as follows:  
\begin{equation}
    {\rm ExpAs} = |(V_{\rm post} - V_{\rm rim})|_{\rm rec} - |(V_{\rm post} - V_{\rm rim})|_{\rm app} .
\end{equation}   
    We defined also a ``rim asymmetry'', RimAs, as
\begin{equation}
    {\rm RimAs} = A_{\rm rec} / A_{\rm app},
\end{equation}    
    where $A$ is the emission-line area enclosed by the rim component resulting from our
    multi-Gaussian fit. The definitions (1) and (2) ensure that
    ${{\rm RimAs} < 1}$ and ${{\rm ExpAS} < 0}$ or ${{\rm RimAs} > 1}$ and  ${{\rm ExpAS} > 0}$ 
    if the PN side where the expansion of the stronger rim, $V_{\rm rim}$,
    shows also a higher difference to $V_{\rm post}$. 

   All objects for which we were able to see a measurable expansion asymmetry in their 
   line profiles are compiled in Table~\ref{asymm.table}, altogether 6 objects.
   We used data exclusively from \oiii\ because only this ensures an unambiguous 
   determination of the strengths of the two rim components. 
   The \nii\ line profiles show generally stronger shell components which affects rather the
   strength than the position of the rim Gaussian. 
   
   Our measurements reveal that each of the six objects in Table~\ref{asymm.table}  
   behave in the same way:   The nebular side with the stronger rim component shows always 
   the larger velocity difference to $V_{\rm post}$ than the
   weaker component.  There is, however, not always a clear correlation between the size of both
   asymmetries.  \object{M2-2} has a strong expansion asymmetry, yet its rim asymmetry appears to
   be marginal.  On the other hand, \object{NGC 6891} shows one of the largest rim asymmetries,
   but only a rather modest expansion asymmetry.
   
   Our interpretation of this phenomenon in terms of hydrodynamics and winds interaction is 
   then the following:  As noted in the introduction, 
   the shell shock expands with a velocity $\dot{R}_{\rm out}$ according to the upstream
   density gradient and the electron temperature of the shell, but independently of the 
   density itself.  Since the upstream matter (i.e. the halo) is rather spherically distributed, 
   as can be seen in Corradi et al. (\cite{CSSP.03}) in general and in Sch\"onberner et al.
   (\cite{SchStJ.04}) in particular for \object{NGC 6826} (Fig. 4 therein), 
   it is thus rather save to assume that this PN expansion 
   velocity $\dot{R}_{\rm out}$ is, to a first approximation, the same in all directions 
   (at least for the objects considered here).  
   The same holds then, of course, for the post-shock velocity.  On the other hand, the rim is
   formed and accelerated by winds interaction, or more precisely, by compression of the inner
   shell gas by the expanding, wind-blown and shock-heated bubble.  Any non-axisymmetric local 
   inhomogeneities/differences of the gas densities in both shells along the line-of-sight 
   must lead automatically to expansion differences between rim and shell of 
   the receding and approaching parts of the PN, according to the 
   work on interacting winds by  Koo \& McKee (\cite{komc.92}).
   
   The sometimes asymmetric expansion seen in some objects appears 
   thus as a naturally consequence of the rather inhomogeneous and non-spherical appearance of
   wind-blown/bubble-compressed rims seen in the images of most PN.  We note that usually
   the PN's radial velocity is determined by centering the bright rim emission.  If an expansion
   asymmetry exists for a particular object, the absolute mean radial velocity (but not the 
   ``expansion" velocity) can be wrong by up to a few \kms. 
   Instead, centering the line emission using $V_{\rm post}$, or the 10\,\% level, would yield
   better radial velocities in these cases.

\section{Evolved objects}
\label{evolved.PNe}

   Five objects from our target list, \object{Hen\,1-5}, \object{IC 1454}, \object{K\,1-20}, 
   \object{NGC 2438}, and \object{NGC 6894} do not belong to the group of younger middle-aged 
   PNe with a clear double-shell structure and high-luminosity central stars and are not
   listed in the Tables~\ref{tab.gesamt} and \ref{data.tautenburg}.  The
   shells of these objects are recombining/reionising because of low stellar temperature
   but still high luminosity (\object{Hen 1-5}) or low stellar luminosity and high
   stellar temperature (the other four objects).   In such cases a clear interpretation of
   the line profiles in terms of rim and shell velocities is difficult. 
   
\begin{figure*}[!ht]
\vskip-2mm
\includegraphics[width=1.0\columnwidth]{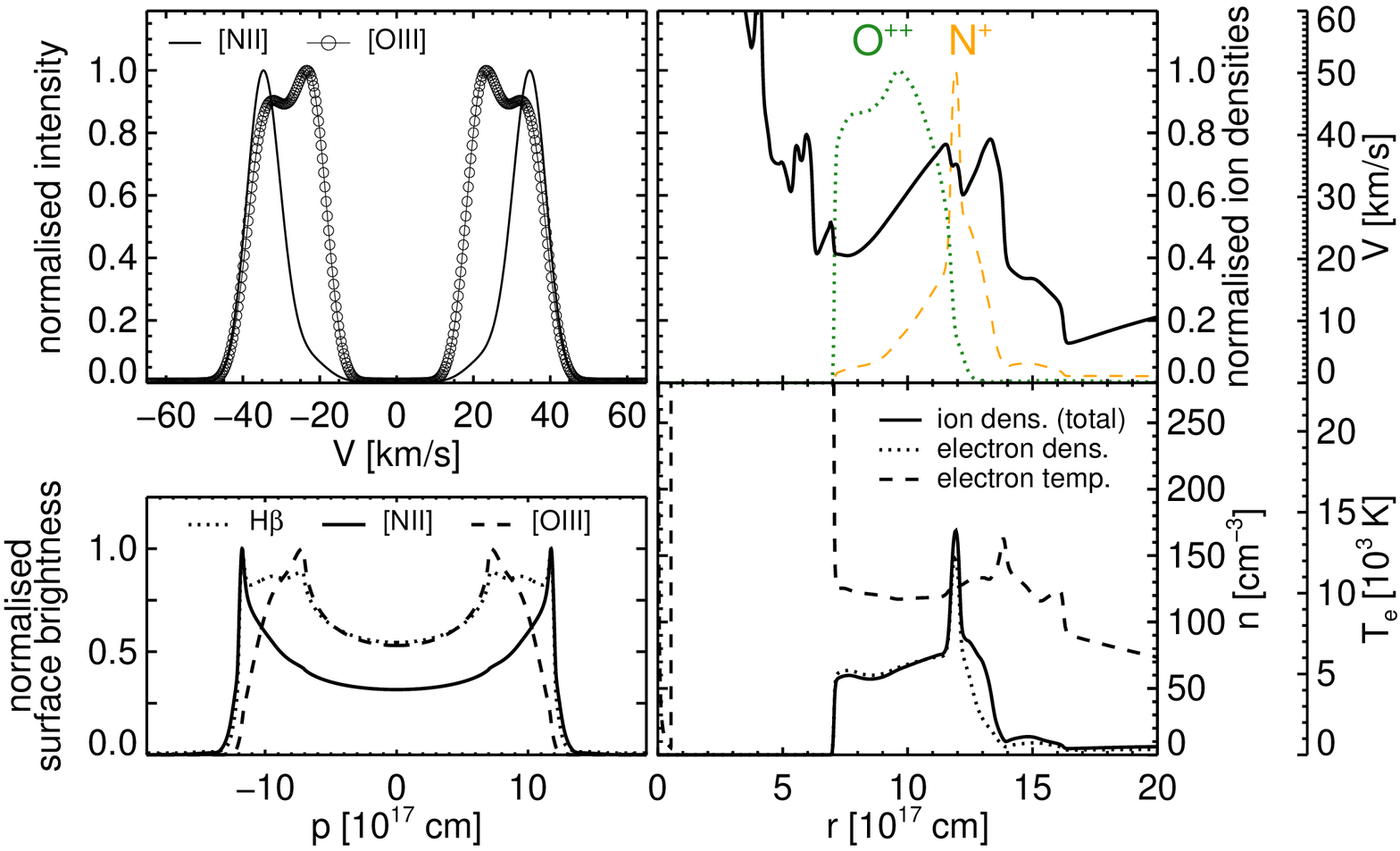}
\includegraphics[width=1.0\columnwidth]{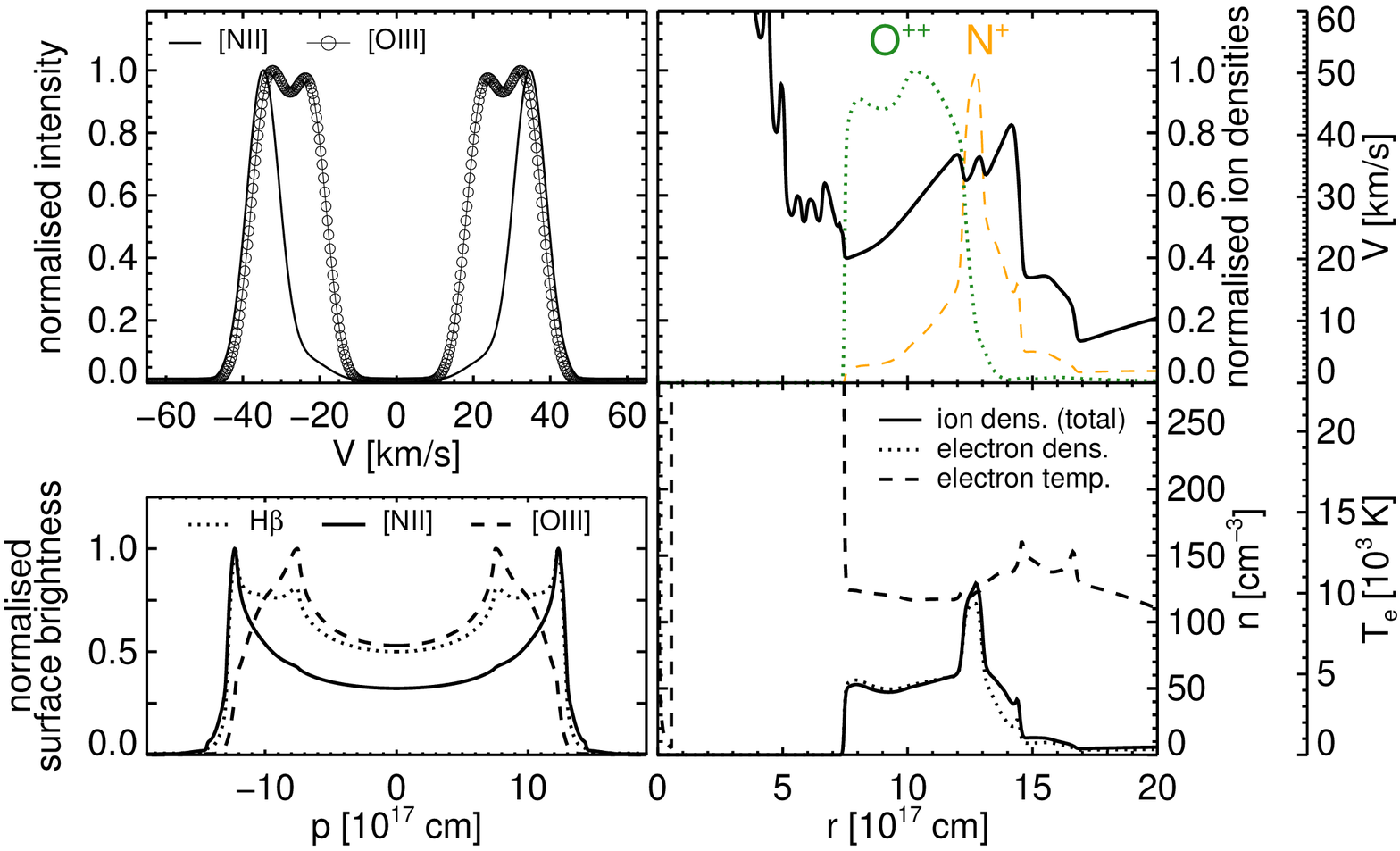} \\[-2mm]
\includegraphics[width=1.0\columnwidth]{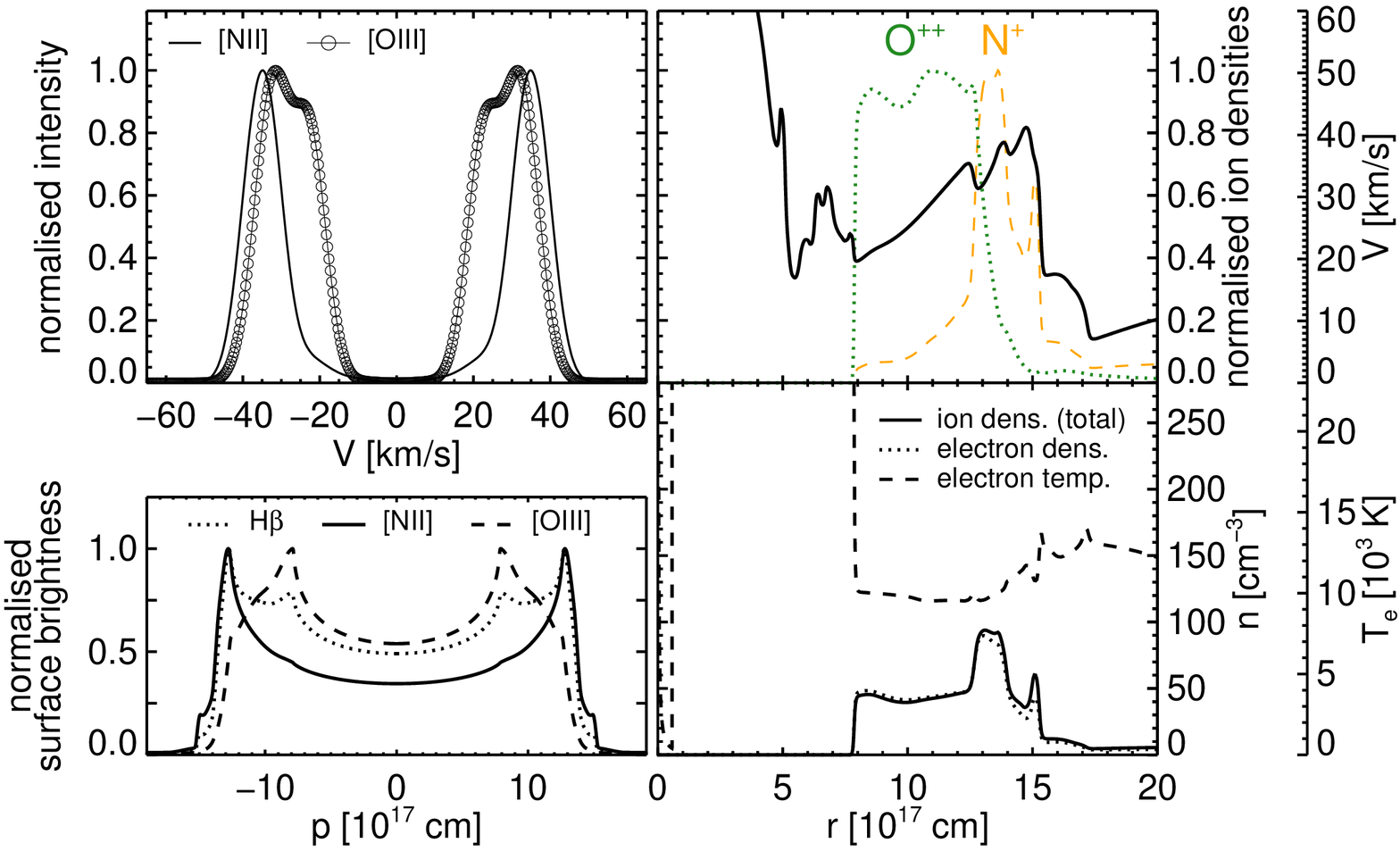}
\includegraphics[width=1.0\columnwidth]{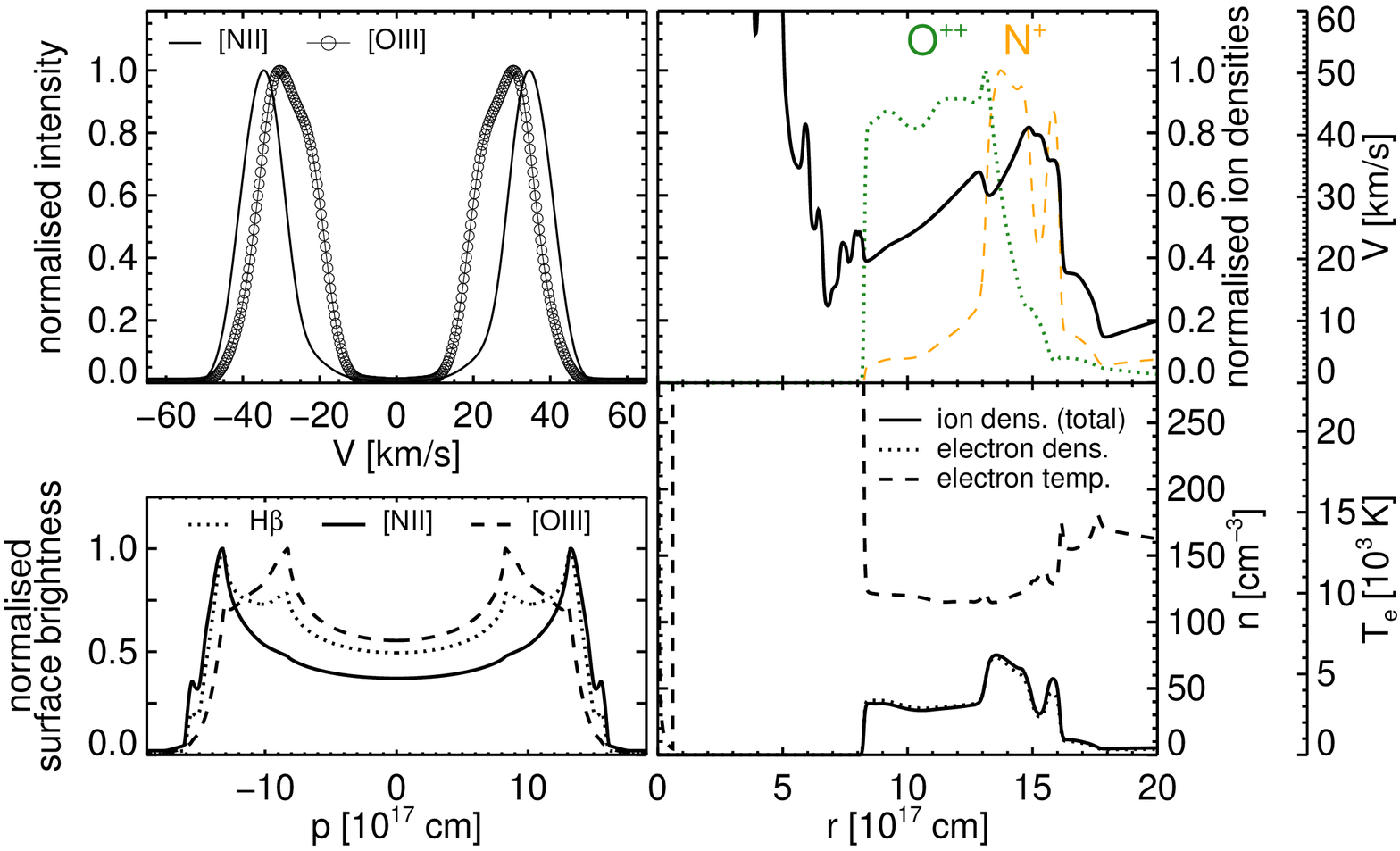}
\vskip-1mm
\caption{\label{fig.evolved.theory}   
          Snapshots of four model structures and respective line and intensity profiles 
          during the post-recombination stage of evolution, taken from our hydrodynamical model 
          sequence with a 0.605 \Msun\ central star.  The models cover a total time span of
          1921 years of  evolution.  The model parameters are as follows: 
          \emph{Top left}:
          $L = 215$ \Lsun, ${\Teff = 120\,370}$ K, at post-AGB age ${t = 13\,136}$ yr,
          corresponding to Fig.~A9 of Paper~I);
          \emph{top right}:  
          $L = 211$ \Lsun, ${\Teff = 120\,117}$ K, at post-AGB age ${t = 13\,778}$ yr;
          \emph{bottom left}:
           $L = 206$ \Lsun, ${\Teff = 119\,814}$~K, at post-AGB age ${t = 14\,419}$ yr; 
          \emph{bottom right}:
          $L = 202$ \Lsun, ${\Teff = 119\,453}$ K, at post-AGB age ${t = 15\,057}$ yr.
          The content of each snapshot is the following:
          \emph{Top left panels}: normalised line profiles of \nii\ (thin solid) and of \oiii\ 
          (connected circles).  The \nii\ line is stronger than the \oiii\ line by   
          factors of about 2--4, depending on the model.  The spectral resolution corresponds 
          to 6~\kms\ in all cases.
          \emph{Bottom left panels}: normalised surface brightness profiles for H$\beta$
          (dotted), \nii\ (solid), and \oiii\ (dashed).         
          \emph{Top right panels}: the velocity field together with the (normalised) ion
          densities of N$^+$ and O$^{++}$.
          \emph{Bottom right panels}: heavy particle density (solid), electron density (dotted),
          and electron temperature (dashed) vs. radius. 
          The PN model proper of each snapshot is confined by the contact discontinuity (density
          jump at the inner PN edge) which propagates from 7 to 8$\times 10^{17}$ cm during the
          1921 years shown here, and the position of the ``old'' leading shock (sharp
          temperature spike) propagates from 14 to 16$\times 10^{17}$ cm in the same time.
          The  new shock/reionisation front is at  12${\times}10^{17}$ cm (\emph{top left}), 
          progresses to 13$\times 10^{17}$ cm (\emph{top right}), but then the reionisation
          ``front`` is passing the old shock between the \emph{bottom left} and 
          \emph{bottom right} snapshots, leaving the new shock behind (at 
          15$\times 10^{17}$ cm, \emph{bottom right}).
          \vspace{-2mm}
          }
\end{figure*}   
     
   The morphology of these objects (except Hen\,1-5) is different from that of the usual
   double-shell PNe with luminous central stars.\footnote
   {Images of these objects, next to others, can be found in 
   http:// www.astro.washington.edu/users/balick/PNIC/ .}  
   The latter show a relatively small but bright rim with a large and faint (attached)
   shell which, because of its size, contains most of the ionised nebular mass (up to 
   about 80\,\%, depending on evolutionary stage).   
   The situation is reversed in the evolved objects: The main body of the PN 
   consists of a bright ring-like structure and an attached small, faint shell containing now 
   obviously only a minor fraction of the whole nebular mass.   This attached shell is usually
   at the 10\,\% level, except for NGC\ 2438 where its brightness is at about 1\% of the
   maximum value only (Fig.~3 in Corradi et al. \cite{CSSP.00}, henceforth Paper~I).
   
   The possibility of nebular recombination which may occur during the very rapid fading of 
   hydrogen-burning central stars was recognised first by Tylenda (\cite{tylenda.86}). 
   He followed the expansion of simple, constant-density model nebulae coupled to evolving 
   central-star models and described the ionisation/recombination fully time-dependently.  
   Tylenda (\cite{tylenda.86}) coined also the term ``recombination halo'' although it was 
   the nebular shell, or part of it, that faded to halo brightness due to recombination.  
   He provided a candidate list of objects which are supposed to be
   in the recombination/reionisation stage, and two of our targets    
   (NGC 2438 and NGC 6894) are on that list.     
   
   \vspace*{-1mm}
\subsection{Model simulations}   
   \vspace*{-1mm}   
   
   Of course, our hydrodynamical simulations cover the phase of recombination/reionisation
   as well in a fully time-dependent fashion, 
   and they were already extensively used in a study of NGC 2438 in Paper~I.
   There we put our emphasis on the development of surface brightnesses during 
   recombination/reionisation, but here we will concentrate on the signatures of the 
   different line profiles (\nii\ vs. \oiii).
 
  We show in Fig.~\ref{fig.evolved.theory} four snapshots of model structures taken from our 
  hydrodynamical model sequence,
  beginning with the last model that has been shown and discussed in Paper~I (Fig.~A9 therein).   
  This particular model depicts the situation during reionisation
  more than 5000 years after the beginning of strong recombination, and the following snapshots, 
  not discussed in Paper~I, illustrate the model evolution for about further 1921 years 
  of evolution.     

   Recombination is, of course, not total, i.e. the recombination ''front'' does not reach
   the bubble/rim interface, although it may penetrate well into the rim.
   At some radial position, however, the local rates of recombination and 
   ionisation become equal, and because of the continued expansion, this point will soon
   move again outwards, i.e. a reionisation ''front'' expands   
   while more distant regions are still recombining.\footnote
{The less dense halo gas ahead of the ``old'' shock usually remains at a much higher
 degree of ionisation because of its lower recombination rates.} 
    Positive density and velocity gradients are soon   
    established again, in conjunction with a strong ionisation stratification,
    as is similar for young, still optically thick PNe.   
    Thus, N$^+$ traces the outer parts of the (re)ionised shell, close to a newly created
    leading shock/ionisation front where the flow velocity is high ($\sim$\,40 \kms), 
    while O$^{++}$ traces preferably the inner parts where the flow velocity is lower.
    In particular, the inner, strong peaks of the \oiii\ profile corresponds to 
    a newly forming shell of weakly compressed gas, i.e. a ``new'' rim (cf. lower right 
    panels in the individual snapshots of Fig.~\ref{fig.evolved.theory}) which expands 
    with about 20 \kms\ only.    
    
    Line profiles and intensity distributions seen in the left panels of the individual frames
    of Fig.~\ref{fig.evolved.theory} reflect the quite extreme ionisation stratification in the 
    models.  In projection, the models appear as broad ring-like structures in the light of
    H$\beta$ and \oiii, but as much narrower ones in \nii.   
    Correspondingly, the \oiii\ line is broad and covers nearly the whole velocity range of the 
    reionised shell.  Typically is a double-peak structure caused by the influence of the
    radial profiles of O$^{++}$ ion density and flow velocity in conjunction with projection
    effects.     
    The relative strengths of these \oiii\ subcomponents change during evolution, reflecting
    obviously the change of the  density profile in that part of the model
    nebula where O$^{++}$ is present, while the velocity gradient remains similar.       
    In contrast to \oiii, the \nii\ line has always no structure, is very narrow and traces 
    the fastest flowing matter immediately behind the new shock. 
 
    The ``old'' leading shock of the former shell, now being mostly neutral, is still 
    moving ahead but with lower speed.  Its position is marked in all snapshots by a
    sharp temperature spike, moving from $14{\times}10^{17}$   
    to $16{\times}10^{17}$ cm at the last snapshot.   As evolution progresses, reionisation 
    passes this old shock, and the former fairly neutral shell becomes visible in, e.g. 
    H$\alpha$ (or H$\beta$) and \nii\ (left bottom panel of left bottom frame in 
    Fig.~\ref{fig.evolved.theory}, between ${r =  14{\times}10^{17}}$ cm and  
    $15.5{\times}10^{17}$ cm).  The intensities of this ``rekindled" attached shell is 
    between 20 and 10\,\% of maximum nebular
    emission, in fair agreement with the images of IC 1454, NGC 6894, and K\,1-20.
    NGC 2438 constitutes obviously an earlier phase of evolution in which the old shell is still 
    mostly neutral and thus ``invisible''.  
    It is interesting to note that the signature of this rekindled shell is not seen in the 
    theoretical line profiles of Fig.~\ref{fig.evolved.theory} (bottom frames).     
    The reason is simple: the  shell is faint, and its flow velocities are too close to
    those behind the new shock.
    
   Our models show that reionisation fronts are not very sharp. The low densities
   ($\approx\!100$ cm$^{-3}$) in combination with the free path lengths of the very abundant 
   high-energy photons emitted by the hot central star ($\simeq$120\,000 K) lead to
   a more gradual decrease of the degree of ionisation with distance.  This can be seen
   by the run of the electron density (dotted) between $12{\times}10^{17}$ and 
   $15{\times}10^{17}$ cm in the lower right panels of the two upper frames of
   Fig.~\ref{fig.evolved.theory}.  
  
    The model simulations predict also that the reionisation stage is long-lasting: For the 
    sequence shown in Fig.~\ref{fig.evolved.theory}, reionisation     
    started about 5000 years ago and is still continuing:  In the last model shown in 
    Fig.~\ref{fig.evolved.theory}, about 7000 years after recombination, $\simeq$\,20\,\% of    
    hydrogen behind the old shock is still neutral.  Even about 4500 years later, at 
    ${t= 19\,480}$ yr, there is about 6\,\%  neutral hydrogen left.   In contrast, 
    recombination occurred on a much shorter time scale:  within only less than 1000 years
    for the simulations shown in Fig.\,\ref{fig.evolved.theory}.  We emphasise here that
    \emph{recombination and reionisation may occur at the same time but in
    different regions of a PN}.  For instance, by comparing Figs. A2, A3, and A4 in Paper\,I         
    one sees that, while the shell is still recombining, the old rim changes into a 
    density structure which is typical for reionisation.
       
    Whether the new shock, created by reionisation, will merge with the old one depends 
    entirely on their further velocity evolution.  
   The last computed model with ${t= 19\,480}$ yr (not shown) has still two outer shock fronts,
   the old one at ${r = 20{\times} 10^{17}}$ cm and the new one at ${18{\times} 10^{17}}$ cm, 
   both propagating with nearly the same speed. 
         
    We noted already that 
    the situation during reionisation as depicted in Fig.~\ref{fig.evolved.theory} is similar 
    to the earliest, optically-thick phase of nebular evolution.  
    There is, however, an important difference: In young PNe the 
    central star is cool, and hence N$^+$ constitutes a major ionisation stage throughout the
    whole nebula, while O$^{++}$ is confined only to the inner region close to the star.   
    Here, the central star is very hot, and the situation is reversed: O$^{++}$ is a
    major species, and N$^+$ is only present close to the reionisation front.\footnote
{The same situation holds, of course, for the relation between O$^+$ and O$^{++}$.}

\subsection{Observations and their interpretation}   
    
\begin{figure}
\vskip-2mm
\hskip-2mm
\includegraphics[width=1.02\columnwidth]{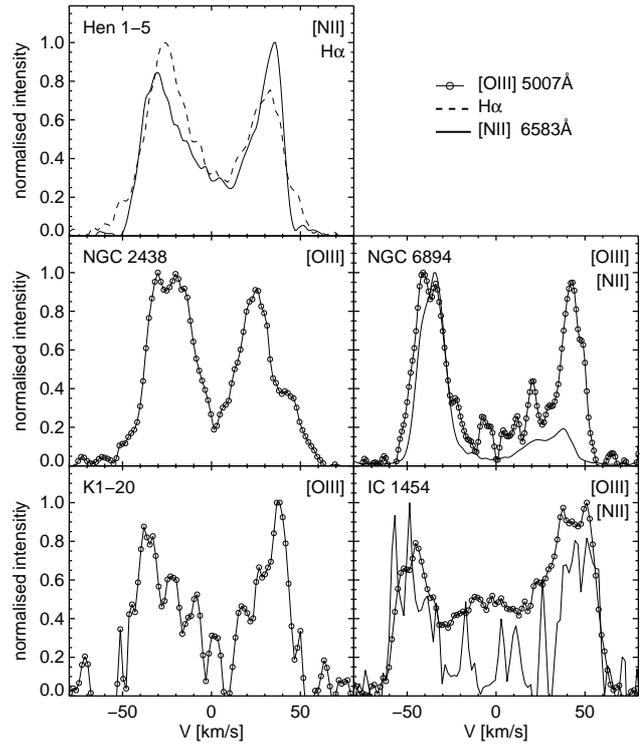}
\caption{\label{fig.evolved.PNe}     
         Central line-of-sight lines profiles of 5 evolved objects, Hen\,1-5 (\emph{top left}),
         NGC 2438 (\emph{middle left}), NGC 6894 (\emph{middle right}), \object{K\,1-20} 
         (\emph{bottom left}), and IC 1454 (\emph{bottom right}) 
         which are not contained in the sample listed in Table~\ref{tab.gesamt}.  
         All profiles are normalised at their maximum strength;  
         solid for \nii, dashed for H$\alpha$, and connected circles for \oiii.  
         The profiles of  \object{K\,1-20} and \object{IC 1454} are averaged over 7 pixel 
         rows for improving S/N.            
        }
\end{figure}

   The central line-of-sight profiles extracted from the spectrograms of our five targets
   are displayed in Fig.\,\ref{fig.evolved.PNe}. 
   The profiles appear quite noisy, even after averaging along dispersion over several
   pixel rows.  Despite this noise there appears a marked difference 
   between these profiles and the ones discussed earlier in this paper.  The individual
   \oiii\ shell components are broad and often split while the corresponding \nii\ line 
   components are narrower and \emph{always} unsplit.  From the total line split follows that 
   maximum flow velocities must be in all cases quite high, ranging from  
   30~\kms\ (Hen\,1-5) up to over 50~\kms\ (IC 1454).   
   
   We will discuss and interpret the line profiles of these five objects in turn in the 
   following paragraphs.   
    
\paragraph{Hen 1-5}

   This object is a singular case in our sample: Its central star, \object{FG Sge}, 
   obviously recovering from a late thermal pulse during the transition from the 
   tip of the AGB to the white dwarf domain, is now a cool object, a so-called 
   ``born-again'' giant with ${\Teff \simeq 5500}$ K  
   (Bl\"ocker \& Sch\"onberner \cite{BS.97}; Jeffery \& Sch\"onberner \cite{JS.06}). 
   Without {any} ionising source for about 130 years, the nebular shell must 
   be fairly well recombined.  
 
   We measured from our \nii\ profiles a peak-to-peak separation of 30$\pm$3 \kms, in good 
   agreement with the determination of Flannery \& Herbig (\cite{FH.73}).   
   The H$\alpha$ line at our disposal is very similar, which 
   indicates that both H$^+$ and  N$^+$ are tracing the same still partly ionised region. 
   A more thorough discussion cannot be done here because such an evolutionary stage is not
   covered by our models.

\paragraph{NGC 2438}

   This object shows a bright, ring-like and somewhat diffuse main shell, attached 
   with two round halo-like structures, placing this object 
   into the small group of PNe with apparently two halos (cf. Corradi et al. \cite{CSSP.03}).  
   It is debated whether the inner halo-like structure is just the faint remnant of 
   the shell which suffered from severe recombination (Paper~I) 
   or a real halo (Rauch et al. \cite{rauchetal.99}; Dalnodar \cite{daln.12}).   
   The central star of 
   \object{NGC 2438} is very hot (${\simeq\!140\,000}$~K) and faint (${\simeq\!160}$ \Lsun, 
   Malkov \cite{malkov.97}), a fact that makes the recombination scenario quite plausible.
   
   Because there exists the possibility that a more detailed velocity information may help in
   clarifying the status of NGC 2438, we took additional spectrograms by placing the slit 
   right in the middle of the bright \oiii\ ``ring''.  We will discuss and interpret the
   resulting line profiles in turn.     
   
   The central profile of the \oiii\ line is seen in Fig.~\ref{fig.evolved.PNe}, middle left.
   The total peak-to-peak separation corresponds to a flow expansion velocity of 
   about 30 \kms.  However, the approaching component shows an unusual feature: 
   It is split into two subcomponents separated by about 14 \kms, and the whole appearance
   of this component is very similar to theoretical profiles shown in 
   Fig.~\ref{fig.evolved.theory} which have, in some cases, components clearly split 
   by about 10 \kms, too (cf. top right frame of Fig.~\ref{fig.evolved.theory}).
   Additionally, there appear weak shoulders (mainly for the receding part) suggesting 
   expansion flow velocities up to about 50 \kms.     
   
\begin{figure}
\vskip-3mm
\includegraphics[width=\linewidth]{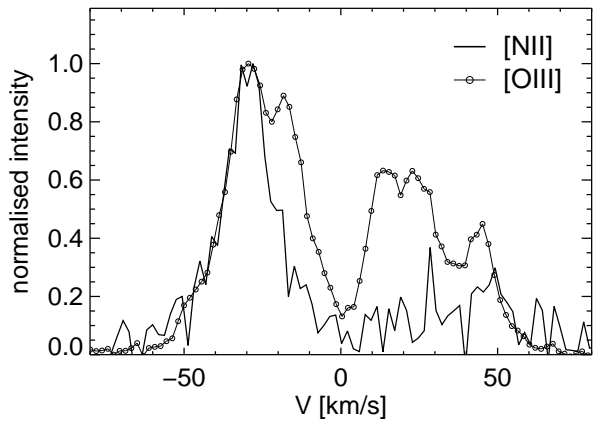}
\vskip-1mm
\caption{\label{fig.2438.romano}
         Central line-of-sight line profiles (normalised) of \nii\ and \oiii\ as extracted 
         from the long-slit 
         spectra of NGC 2438 presented in Fig.~2 of Paper~I. 
         The profiles are averaged over 5 (\nii) and 3 (\oiii) pixel rows, resp. 
         }       
 \end{figure}   

   Since no \nii\ profile from Tautenburg Observatory is at our disposal, we employed long-slit
   spectrograms taken at the 3.5\,m NTT telescope of ESO and already discussed in Paper I.  
   The central profiles of both \nii\ and \oiii\ are shown in Fig.~\ref{fig.2438.romano}.
   The better spectral resolution of the NTT spectrogram 
   reveals that also the \oiii\ component of the receding shell is split.
   The velocity range within the (re)ionised shell as measured by this line split extends from 
   about 17 \kms\  (from the separation of the inner \oiii\ line peaks seen in 
   Fig.~\ref{fig.2438.romano}) to a velocity of at least 26 \kms\ (from the 
   separation of the outer \oiii\ components).\footnote
{More precisely, the velocity range extends to about 30 \kms\ since the post-shock velocity is 
 not traced by the line peak, as judged from our models shown in Fig.~\ref{fig.evolved.theory}.}

   The \nii\ line component (approaching shell only) is single and its velocity position
   coincides, as predicted by our models in Fig.~\ref{fig.evolved.theory}, with the 
   corresponding outer \oiii\  component.  
   In the receding shell N$^+$ is obviously absent, indicating a possibly higher stage of 
   (re)ionisation. In this part of the shell we see in \oiii\ also significant emission at 
   flow velocities between about 40--50 \kms, corresponding
   to the shoulder present in our Tautenburg spectrogram (cf. Fig.~\ref{fig.evolved.PNe},
   middle left). Such velocities are only marginally seen in the approaching part of the nebula.
   It is tempting to identify this flow signature as belonging to the old shell now being
   reionised in the direction of the central line-of-sight.  We have seen earlier in
   Sect.~\ref{exp.vel} that the PN shells attain such high flow velocities (post-shock
   velocities) during the end of the horizontal evolution across the HRD.

   Such a strong line strength difference 
   between approaching and receding parts of the nebula (also seen in NGC 6894, cf. below) 
   is quite unusual during the evolution prior to recombination.   
   Here, however, the reionised shell is shaped out of the already 
   ``patchy''/inhomogeneous rim material, and this fact is most likely the cause of the  
   now very asymmetrical appearance of the line profiles.     

   If the bright ring of NGC 2438, as seen in projection onto the plane of sky, has indeed 
   a positive velocity gradient combined with an ionisation stratification as it is typical
   for the stage of reionisation, line profiles taken at a position centred on this ring
   should reflect this fact:  Both components 
   should be split/structured, at least in \nii\ because N$^+$ is concentrated only at the 
   outer edge of the reionised shell.
   
\begin{figure}
\includegraphics[width=\linewidth]{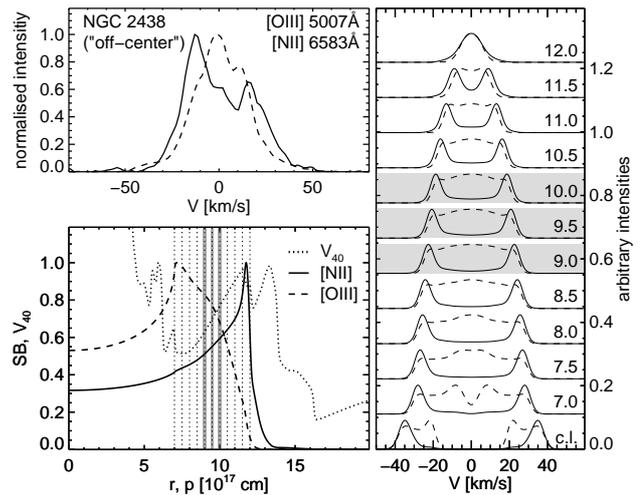}
\caption{\label{fig.2438}
         Line profiles of \object{NGC 2438} taken at the radial position 
         of the bright ring and their interpretation by the hydrodynamical model
         shown in Fig.~\ref{fig.evolved.theory}, top right.
         \emph{Top left}: observed profiles of \nii\ (solid) and \oiii\ (dashed).
         \emph{Bottom left}: modelled normalised surface brightness profiles of \nii\ 
         (solid) and \oiii\ (dashed), and the flow velocity in units of 40 \kms\ (dotted).  
         Vertical lines (thick gray and dotted) indicate the radial 
         positions of the numerical slit with width of $2{\times}10^{16}$ cm.  
         \emph{Right}: normalised line profiles of \nii\ (solid) and \oiii\ (dashed) 
         as computed from the model.  
         The numbers ($\times 10^{17}$ cm) at the individual profiles correspond
         to the radial positions of the numerical slit belonging to the respective profiles.  
         The bottom profiles are for the central line-of-sight (cf. top left
         panel of top right snapshot in Fig.~\ref{fig.evolved.theory}).  
         The profiles underlaid by a grey shade 
         belong to the slit positions marked in thick grey in the \emph{bottom left} panel
          which correspond closely to the slit position used for the observations. 
          }  
\end{figure}   
    
   A comparison between the observation and theoretical prediction is depicted
   in Fig.~\ref{fig.2438}.  Indeed, the observed line profiles of \nii\ and \oiii\ correspond
   to our expectations: The \oiii\ line is unsplit, but with shoulders, whereas the 
   \nii\ line is split, and its peak-to-peak separation corresponds closely to the shoulders of 
   the \oiii\ line (top left panel of Fig.~\ref{fig.2438}).   
   The \nii\ peak separation, 2$\times$15 \kms\   
   corresponds, if the projection effect is taken into account, 
   to a central line-of-sight separation of about 2$\times$27 \kms, in fair agreement with the
   central line-of-sight measurement (2$\times$30 \kms).
   
   The bottom left panel of Fig.~\ref{fig.2438} renders the intensity profiles
   in \nii\ and \oiii\ together with the positions of the numerical slits used
   for the computation of the profiles seen in the right panel.  The radial gas velocity 
   distribution is shown as well.  
   These line profiles, and the intensity profiles as well, are the consequence of the  
   ionisation stratification of the reionising model, as is evident by the very different
   intensity profiles of \nii\ and \oiii\ in combination with the positive velocity
   gradient (cf. top right frame of Fig.~\ref{fig.evolved.theory}).
   The shaded part highlights those modelled profiles with the closest similarity to 
   the observed ones.  
      
   We state that the nebular model presented in this figure provides a
   satisfactory description of the evolutionary stage of NGC\,2438.    
   The observed line profiles (together with the surface brightness profiles shown
   in Fig.~3 of Paper~I) can be interpreted such that the shell of NGC 2438 has experienced a 
   rather strong recombination in the very recent past and that the object is now in the slow
   phase of reionisation due to the continuing expansion.  
   NGC 2438 seems to be in a quite early phase of reionisation
   because its former (attached) shell is still very faint, i.e. must still be quite neutral 
   (cf. top left frame of Fig.~\ref{fig.evolved.theory}).  
   
   The recently measured high electron temperature beyond the boundary of the bright ring (rim), 
   i.e. beyond the reionisation front (Dalnodar \cite{daln.12}), is not in contradiction to this 
   statement as our models show (bottom right panel in the top left frame of 
   Fig.~\ref{fig.evolved.theory}):  Despite its still lower degree of ionisation, the 
   upstream region of the reionisation shock at $12{\times}10^{17}$ cm
   is even hotter than the fully reionised regions behind (about 11\,000--12\,000 K 
   versus about 10\,000 K).  The reason is heating beyond the reionisation front proper 
   by the most energetic photons which also have the highest free path lengths.    
   Our model prediction is fairly consistent with the Dalnodar measurement of
   about 15\,000~K versus 10\,000--12\,000~K in the bright ring.

\paragraph{NGC 6894}  

    The image of this object displays a bright, round inner
    structure with a hardly to be seen (attached) shell.  It is a typical very far evolved 
    object supposed to be in the recombination/reionisation stage as well   
    with a central star of ${T_{\rm eff} \simeq 120\,000}$~K and ${L \simeq 120}$ \Lsun\ 
    (Malkov \cite{malkov.97}).   
     
     The \oiii\ line component of the approaching shell is clearly split,  but by only 
     about 9 \kms.
     Again, the corresponding \nii\ is not split (Fig.~\ref{fig.evolved.PNe}).  
     Here, however, the \nii\ line-peak position corresponds to the inner \oiii\ 
     component (cf. explanation below).     
     We can only estimate the flow velocity close to the edge of the reionised shell from the
     total \oiii\ line separation, 42 \kms, in good agreement with the entry in Weinberger's 
     catalogue (Weinberger \cite{We.89}) of 43 \kms.  
     
   The velocity range within the (re)ionised shell is smaller than that in NGC 2438, which is
   clearly seen by the different spacings of the split \oiii\ lines: 
   $\simeq$\,9 vs. $\simeq$\,14 \kms\ (Fig.~\ref{fig.evolved.PNe}).  
   The corresponding line split for the model rendered in Fig.~\ref{fig.evolved.theory} is 
   25 \kms\ (top frames).  
          
    It appears surprising and not in line with our model predictions that the
    \nii\ peak separation gives a lower velocity than \oiii\
     (cf. Fig.\,\ref{fig.evolved.PNe}, middle right, $\simeq$\,38 vs. 42~\kms). 
    We must state, however, that the \oiii\ and \nii\ spectrograms are taken on different days.
    Given the high expansion velocity and the nebula's moderate apparent size, already a small 
    difference in slit position might lead to a measurable velocity separation.

\paragraph{K\,1-20}

   \object{K\,1-20} is an object with a very low surface brightness and appears on the sky
   as a broad ring with an attached shell (cf. Guerrero et al. \cite{GVM.98}).  According to 
   G\'orny (priv. comm.), the central star is very faint ($<$\,100 \Lsun), too.  Our noisy
   \oiii\ line profile indicates, from the peak separation, flow velocities up to 
   about 40 \kms, in very good agreement with the result of Guerrero et al. (\cite{GVM.98}) 
   from their  \nii\ measurements.\footnote
{The authors list $39\pm 3$ \kms\ in their Table~4, but the corresponding \'echellogram 
 is not presented in their Fig.~1.}

\paragraph{IC 1454}

   \object{IC 1454} (= \object{A 81}) is another far evolved object with a very hot but faint
   central star (${L \approx 150}$ \Lsun,  G\'orny, priv. comm.).  The image consists of 
   a wide, bright rim-like structure with a small attached shell as well. 
   Our \oiii\ line profile shown in Fig.~\ref{fig.evolved.PNe} (bottom right) is rather noisy, 
   but appears to have the (expected) double components, with a split of about 10 \kms.
   The total line separation corresponds to a radial flow velocity up to at least 50 \kms.
   Our extremely noisy \nii\ spectrogram is consistent with such high velocities.
   
   Our measurement is in fair agreement with the corresponding entry in Weinberger's
   catalogue (46 \kms\ from the \oiii\ peak separation; Weinberger \cite{We.89}).  
   An ``expansion'' velocity as low as 20 \kms, as advocated by 
   Guerrero et al. (\cite{GVM.98}), is not confirmed.

\subsection{Remarks}     
   
   We note here that recombination, if it occurs at all, is only a transient phenomenon during 
   the total life of a PN, and that a PN does not  become ``extinct'' by recombination because
   the latter never progresses towards the innermost regions of a PN where the radiation field 
   even of a low-luminosity central star is sufficient to keep up ionisation.   
   Rather, expansion continues, accompanied by reionisation, until the PN's 
   surface brightness becomes too low for detection.  
   The final reionisation phase of nebula evolution is characterised by 
   strong ionisation stratification, resulting in smaller peak-separation velocities for
   \oiii\ as compared to \nii\ (see Fig.~\ref{fig.evolved.theory}), at least temporarily. 
   Velocities based on \oiii\ alone would then suggest a nebular deceleration 
   (Pereyra et al. \cite{pereyra.13}) which, however, is not real .
  
   We note also that all five objects investigated here expand rather fast (${>\! 40}$ \kms\
   post-shock velocity, except Hen\,1-5),  
   which is comparable to the phases of evolution immediately prior to recombination 
   (cf. Fig.~\ref{evol.teff}).   This is somewhat surprising since the common believe is that, 
   once the central star has faded, the nebular expansion will slow down (cf., e.g., discussion 
   in Pereyra et al. \cite{pereyra.13}).  Indeed, our model calculations show that
   the ``old'' leading shock decelerates during recombination, but due to reionisation
   a new shock is created further inwards, starting with a propagation velocity based on that
   of the rim matter, i.e. of about 30 \kms\ (instead of 10--15 \kms\ during the ``first''
   ionisation).  This new shock can therefore easily be accelerated to about 50 \kms\ during
   reionisation, and it will merge eventually with the old shock, maintaining its 
   high velocity.
    
   We emphasise that recombination/reionisation as is discussed here is only an issue if 
   significant parts of the nebula recombine.   Only then a new shock is created which
   leads the expansion of the reionised shell. 
   Thus this phenomenon is restricted to more massive central stars that
   evolve fast enough such that the nebula is still sufficiently dense to allow for severe 
   recombination during stellar fading. The sequence shown in Fig.~\ref{fig.evolved.theory}
   describes the case of rather strong recombination.  
  
   Another interesting aspect of fast expanding nebulae around evolved, very faint 
   central stars is worth noting: These objects are in a phase of very weak or maybe 
   even nearly absent stellar winds, yet they expand very fast.  
   Hence interacting winds cannot be responsible for this expansion!
   We conclude again that it is exclusively the thermal pressure gradient established by 
   excessive gas heating during the process of reionisation.  
   Indeed, the models shown in Fig.~\ref{fig.evolved.theory} reveal a very slow expansion of 
   the hot bubble of less than 20 \kmsß only, which is clearly not enough 
   to compress the inner gas shells into a rim as is seen in the younger nebulae, not to speak
   of accelerating the rim matter to the observed high velocities.
    
   In this context we note that the close resemblance of our theoretical line profiles with
   the observations indicates that the hot bubbles must not necessarily collapse in 
   reality, at least for the objects investigated in this work.   This implies that, as
   assumed in the modelling, the central-star (or hot white dwarf) is still able to
   sustain a wind sufficiently strong not only for preventing the collapse of the bubble
   but also for keeping up a slow expansion of the bubble.  
    
    
\section{Expansion and metallicity}                       
 \label{met.vel}
  
  Thanks to the spatial and spectral resolution capability of the FLAMES/ARGUS IFU at the VLT, 
  we were able to determine separate velocity information for the rims and shells for a 
  sample of four metal-poor PNe.  We analysed the line profiles in the same
  way as for the Galactic disk objects, and the results are presented in the following.

\begin{figure}
\vspace*{-1mm}
\hskip-1mm
\includegraphics[width=1.01\linewidth]{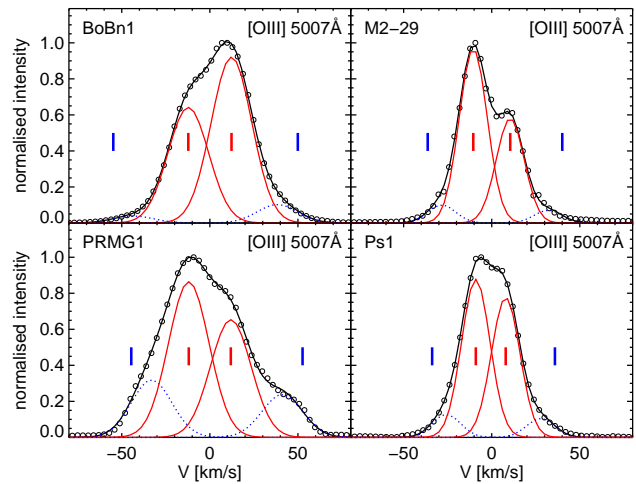}
\caption{\label{fig.4halo}
         Central line-of-sight line profiles of \oiii\,5007\,\AA\ for four 
         metal-deficient PNe observed  by us with VLT FLAMES /ARGUS  and their
         decompositions into four Gaussians.  
         The two rim Gaussians (red solid) are forced to have the same half widths,
         and the post-shock velocities are marked, too.  
         The latter are based on the sigmas of the respective Gaussians, except for
         \object{PRMG 1} where the derivative method could be applied.  \vspace{-2mm}
         }
\end{figure}
  
\begin{table} 
\caption{Values of measured rim and post-shock velocities deduced from the line
         profiles of \oiii, oxygen abundances (Col. 4) and stellar temperatures 
         (Col. 5), both from Howard et al. (\cite{HHMcC.97}), and additional stellar
         temperature information (Col. 6) from Malkov (\cite{malkov.97}). 
         }
\label{tab.metal.poor}
\tabcolsep=1.6pt 
\begin{tabular}{lccccc}
\hline\noalign{\smallskip}
Object  & \multicolumn{1}{c}{$V_{\rm rim}$ }
        & \multicolumn{1}{c}{$V_{\rm post}$ }
        &  12 + 
	    & $\log\left(T_{*}/{\rm K}\right)$
	    & $\log\left(T_{*}/{\rm K}\right)$ \\[1.5pt]  
 	& [km\,s$^{-1}$]  & [km\,s$^{-1}$] &  log\,(O/H)   &  \\[1.5pt]
\hline
\noalign{\smallskip}   
\object{BoBn 1}    &   12      &  52  &        7.83\rlap{$\,^3$}    & 5.10\rlap{$\,^3$} & 4.98 \\
\object{M2-29}     &   10      &  38  &        7.71\rlap{$\,^1$}  & 4.86\rlap{$\,^1$} & 4.70   \\ 
\object{PRMG 1}    &   12      &  49  &        8.06               & 5.01              & 4.91   \\
\object{Ps\,1} &\enspace 8&  35  &        7.61               & 4.66\rlap{$\,^2$} & 4.59   \\  
\hline
\end{tabular}
\\[3pt]
\emph{Notes}:
         All velocities are based on one spectrogram only, hence
         individual velocity errors are estimated to $\pm$2 \kms\ for $V_{\rm rim}$ and $\pm$3~\kms\ 
         for $V_{\rm post}$. \\
         $^1$\,Miszalski et al. (\cite{miszal.11}) determined ${\Teff = 50\pm 10}$ kK for the central 
              star of \object{M2-29} from detailed matching of photospheric lines by means of
              NLTE model atmospheres and a nebular oxygen abundance of 
              12\,+\,log\,(O/H) = 8.3 from the nebular lines. \\               
         $^2$\,Heber, Dreizler \& Werner (\cite{heberetal.93}) 
               determined ${\Teff = 37}$ kK, Rauch, Heber \& Werner (\cite{rauchetal.02})
               ${\Teff = 39}$ kK.  \\  
         $^3$\,According to Otsuka et al. (\cite{OTHI.10}), 
               ${\log \Teff = 5.10}$ and 12\ +\ log\,(O/H) = 7.74.  
 \vspace{-2mm}              
\end{table}

\subsection{Our sample of metal-deficient PNe}
\label{four.result}   
    
   The results of the line analyses of the four metal-poor objects are illustrated in 
   Fig.~\ref{fig.4halo}.  All profiles appear nearly resolved, allowing a
   measurement of the rim velocities which are similar and around 10 \kms\ for all 
   four objects. Also, in all cases weak but somewhat extended shoulders are visible, 
   allowing in each object a firm determination of the shell's post-shock velocity, 
   either by a Gaussian decomposition or the derivative method (PRMG 1 only).
    
\begin{figure}
\vspace*{-1mm}
\hskip-1mm
\includegraphics[width=1.01\linewidth]{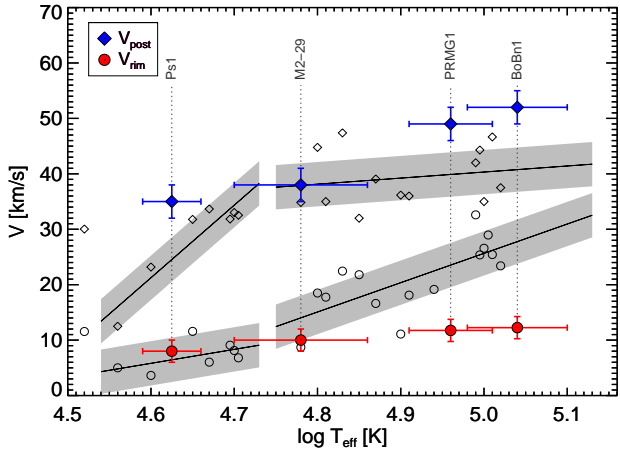}
\caption{\label{fig.metal.poor.teff}
         Positions in the $T_{\rm eff}$-$V$ plane of our metal-poor PNe, with horizontal
         bars, indicating the range of temperature determinations 
         (cf. Table~\ref{tab.metal.poor}).
         The general run of the velocities with effective
         temperatures for the PNe in the Milky Way disk are shown in a schematic way as
         well.  The objects from Fig.~\ref{evol.teff} are underlaid as small 
         open diamonds and circles.    \vspace{-2mm}
         }
\end{figure}

   Our velocity measurements are listed in Table~\ref{tab.metal.poor}, 
   together with the objects' oxygen abundances and stellar effective temperatures, 
   both taken from Howard et al. (\cite{HHMcC.97}), for consistency.  
   For comparison, values from other sources are mentioned as well.       
   The positions of these four metal-poor objects in the $T_{\rm eff}$-$V$ plane is,
   together with our sample of Galactic disk objects (from
   Fig.~\ref{evol.teff}), also illustrated in Fig.~\ref{fig.metal.poor.teff}.        
    Despite the considerable uncertainties of the stellar temperatures 
    (except for Ps 1), the  following facts are noticeable: 
\begin{itemize}
\item  \object{Ps\,1} with about 1/10 of solar oxygen is a young object with a cool
        central star, yet its $V_{\rm post}$ is with 42~\kms\ unusually high.
\item  \object{BoBn 1} and \object{PRMG 1} are less depleted in oxygen and have  
       hot central stars, i.e. they are the most evolved objects in our sample.
       Yet both have a very low rim velocity ($V_{\rm rim} = 12$ \kms), but their 
       post-shock velocity of about 50 \kms\ is above the upper 
       boundary for Galactic disk objects with similarly hot central stars.   
\item  \object{M2-29} is obviously a middle-aged PN with a $V_{\rm rim}$ at the lower 
       boundary compared to the Galactic disk PNe.  Its  $V_{\rm post}$ (38 \kms) 
       is still well within the range of typical values for Galactic disk PNe.      
\end{itemize}

    The four objects with metal-poor composition analysed here comply perfectly with the
    expectations from theory, as outlined in the Introduction and in more detail in
    Sch\"onberner et al. (\cite{SchJSaSt.10}).  
    Although the initial rim expansion can be assumed to have been about the same as in 
    metal-rich objects (close to the original AGB-wind speed), 
    there is hardly any further rim acceleration with increasing stellar 
    temperature which is obviously
    evidence for a rather weak stellar wind, certainly a consequence of the low metallicity.  
    On the other hand, the velocities of the shells' outer edges 
    are just as high or even higher than those of Galactic disk objects with similar 
    evolutionary state, reflecting the higher electron temperatures caused by the 
    reduced line cooling: The nebular electron temperatures range from 13\,000 K for the 
    young PN \hbox{Ps\,1} to about 15\,000--16\,000 K for the evolved PNe PRMG~1 and BoBn~1
    (Howard et al. \cite{HHMcC.97}).  For comparison, the electron temperatures for the 
    Galactic disk PN sample in our  Table~\ref{tab.gesamt} range from about 8000 K up to 
    about 12\,000 K for the most evolved objects.  
    
\subsection{Velocities from the literature}  
\label{prev.mes}  
    
   Other measurements of ``expansion'' velocities of metal-poor objects are rare. 
   We restrict our comparison to those
   measurements only that are based on spatial and/or spectroscopic resolutions 
   similar to ours.

\paragraph{BoBn 1}   
     Otsuka et al. (\cite{OTHI.10}) observed \object{BoBn 1} with Subaru/HDS and used archival
     data from VLT/UVES for a detailed chemical and kinematical analysis.  
     The line profiles appear hardly resolved
     (see their Fig.~5).  A two-component Gaussian decomposition of [\ion{O}{ii}] yields,
     in their notation, ${V_{\rm exp} = 16}$ \kms, while we found ${V_{\rm rim}=12}$ \kms\ 
     (see Table~\ref{tab.metal.poor}).  From the half widths
     of the unresolved lines the authors estimated expansion velocities up to about 30 kms,
     well below our value for the post-shock velocity of 52 \kms.  Using, however, the 
     10\,\%-level of the line profiles of ions with medium degree of ionisation, velocities
     of 40--45 \kms\ can be estimated from their figure, much closer to $V_{\rm post}$ as
     derived here.

\paragraph{M2-29}    
     Gesicki et al. (\cite{gesetal.10}) analysed the same data that we used here, 
     but unfortunately they are not showing any line profiles.  
     Instead, they present only position-velocity diagrams with isophotes which render a
     detailed comparison impossible.  Still, the authors combined 
     line profiles with a radial density profile derived from the nebula's intensity
     distributions (in the 1D approach) for determining the radial velocity field: 
     Their best fit is a constant velocity of 12 \kms\ plus an additional turbulent 
     velocity component of 7~\kms\ which is necessary to fit the line wings.  
    
     Given the physics of expanding shock waves, a radially constant velocity field is
     unphysical.  Also the introduction of a turbulent velocity field with amplitudes of 
     the same order as the general flow speed  and encompassing the entire nebula appears 
     to us as an unphysical approach as well.  The value of 12 \kms, however, 
     is consistent with our measurement of $V_{\rm rim}$ = 10 \kms,
     but the wings of the \oiii\ profile shown in Fig.~\ref{fig.4halo} (top right) 
     are certainly the signature of an attached shell with a positive velocity gradient and 
     a rather high post-shock velocity (38 \kms, cf. Table~\ref{tab.metal.poor}).\footnote
{The nebular model of Gesicki et al. (\cite{gesetal.10}) consists of a dedicated structure, i.e.
 of a dense inner rim and a very faint attached shell, mainly visible in \nii\
 (cf. their Fig.~7).  Given an electron temperature of 10\,000~K,
 an AGB-wind velocity of 10 \kms,
 and an AGB-wind density profile that can be described by an $r^{-\alpha}$ profile with 
 exponent ${\alpha=2}$, the shell's post-shock velocity must already be at 20 \kms.  With a more
 realistic ${\alpha=3}$ value, a post-shock velocity between 30 and 40 \kms\ follows 
  (cf. Fig.~\ref{shell.comp}, top panel).  }
     
     Miszalski et al. (\cite{miszal.11}) redetermined the nebular abundances and 
     found only a mild metal deficiency (e.g. 12\,+\,log(O/H) = 8.3), 
     not enough to justify a halo classification.  
     Together with a distance estimate of 7.7 kpc and the coordinates they concluded that 
     \object{M2-29} belongs to the Milky Way bulge, although its chemical abundances are a bit
     below typical bulge values (see Table 6 in Miszalski et al. \cite{miszal.11}).
     The only modest metal deficit explains our measurement of the shell's 
     post-shock velocity, the lowest of all the four objects listed in
     Table~\ref{tab.metal.poor} and in fair agreement with typical values for Galactic disk
     objects at the same evolutionary state (Fig.~\ref{fig.metal.poor.teff}).
     
     With known distance and a set of ``expansion velocities'' it appears interesting to discuss   
     their relevance for an estimate of the object's post-AGB age.  With distance  
     to the bulge (centre) and angular size of the object (4\farcs9), a nebular radius 
     of $2.8{\times}10^{17}$~cm (0.09~pc) follows.  Hence a kinematical age of 2300 years 
     is derived from simply applying $R_{\rm out}/V_{\rm post}$ and using our measurement of 
     ${V_{\rm post} = 38}$ \kms. 
     The real post-AGB age is higher because (i) the transition time   and (ii) the 
     nebular acceleration must be considered (cf. Sect.~\ref{subsect.kin.age}): 
     From our hydrodynamical models we estimate an age increase of up to 40\,\%, i.e. to 
     a (reasonable) post-AGB age of up to 3300 years, depending on the model used 
     (cf. Fig.~\ref{fig.kinages}).  With an expansion speed of only 12 \kms, as advocated 
     by Gesicki et al. (\cite{gesetal.10}),
     the post-AGB age would be about three times higher, viz. about 10\,000 years. 
     The possible central-star mass range given by Miszalski et al. (\cite{miszal.11}), 
     $0.6^{+0.28}_{-0.10}$ \Msun, is very large and accommodates (theoretical) post-AGB ages 
     from a few hundreds to up to more than 100\,000 years (cf. Bl\"ocker \cite{B2.95}).

\paragraph{Ps 1}  
 
     Bianchi et al. (\cite{bietal.01}) measured for \object{Ps\,1} an ``expansion'' velocity  
     of 12 \kms\ from the HWHM of high\-resolution \nii\ lines, which corresponds 
     to our ${V_{\rm rim} = 8}$ \kms\ result gained from the decomposition of the \oiii\ 
     line (cf. Table~\ref{tab.metal.poor}). 
     More recently, Tajitsu \& Otsuka (\cite{TO.06}) observed Ps\,1 with the High Dispersion
     Spectrograph of the 8.2\,m Subaru telescope.  Their central line-of-sight profile is
     very similar to ours (see Tajitsu \& Otsuka \cite{TO.06}, their Fig.\,1), and the
     authors were able to resolve the \oiii\ profile into four Gaussian components: 
     in two strong, inner components separated by 15 \kms\ and two faint outer ones 
     separated by about 60 \kms. The latter were called ``high-velocity components'' 
     pointing in opposite direction in the line-of-sight.  Our interpretation is that these
     high-velocity components are nothing else but the signature of the faint outer shell
     seen in the image of Ps\,1.  They appear much stronger in \nii\ 
     (see again Tajitsu \& Otsuka \cite{TO.06}, bottom left of their Fig. 1), 
     which is typical for the \nii\ emission
     from nebulae with still cooler central stars.\footnote
{An example is seen in Fig. \ref{line.analyse} for the case of NGC 6826.} 
       
     Extracting the Gaussians given in Fig.\,1 of Tajitsu \& Otsuka (\cite{TO.06}), we found 
     from the \oiii\ profile ${V_{\rm rim} = 8}$ \kms\ (from the strong inner components) and 
     ${V_{\rm post} = 31}$ \kms\ from the outer, faint shell components.  The \nii\ profile
     is centrally unresolved, but from the outer Gaussian components we extracted
      ${V_{\rm post} = 36}$ \kms.
     Altogether we have excellent agreement between our results gained from FLAMES/ARGUS 
     observations with the Subaru observations of Tajitsu \& Otsuka (\cite{TO.06}).
    
     Ps\,1 is a member of M\,15, hence its distance is well-known. We adopt here a
     distance of 10.2$\pm$0.4 kpc, the mean of the two most recent dynamical distance
     determinations (McNamara et al. \cite{mcnamara.04}; van den Bosch et al.
     \cite{vandenB.06}). The observed nebular dimension of 3\farcs1$\times$2\farcs7 
     (Alves, Bond \& Livio \cite{ABL.00})
     results in a mean shell radius of 0.07 pc, and combined with ${V_{\rm post} = 35}$ \kms, 
     a kinematic age of about 1900 years follows.     
 
    The real, i.e. post-AGB age, is difficult to estimate since Fig.~\ref{fig.kinages} 
    does not contain fast accelerating metal-poor models around low-mass central stars.\footnote
 {We note that metal-poor models reach high expansion rates already at rather low 
  stellar temperatures: Figure~10 in Sch\"onberner et al. (2010) shows a PN model with a 
  metallicity reduced by a factor of ten around a central star with  
  ${\Teff  \simeq41\,000}$ K and a post-shock velocity of already 40~\kms.}  
    It is clear that a rapid increase of shock speed following the rather long transition 
    phase during which the inner edge of the AGB wind envelope expands with about 10~\kms\ 
    leads to a large correction of the kinematical age.  
    Based on the 0.565 \Msun\ post-AGB sequence of Sch\"onberner (\cite{Sch.83}), 
    we can make the following estimate:  expansion of the AGB matter with 10 \kms\
    for 5000 years (${\Teff\simeq 30\,000}$ K), then formation of the shock and 
    linear increase of its velocity to 40 \kms\ for another 1500 years 
    (${\Teff\simeq 38\,000}$ K).  The kinematical age would then be 
    0.07\,pc/40\,\kms$\times$1.25 $\simeq$ 2100 years, i.e. smaller than the model's
    post-AGB age ($\simeq$\,6500 yr) by a factor of about 3!

    Because of the well-known distance to M\,15, the mass of the central star can be 
    determined rather precisely: Alves, Bond \& Livio (\cite{ABL.00})  
    found ${M= 0.60\pm0.02}$ \Msun, a value derived by comparing the observed
     central-star luminosity with existing post-AGB (core-)mass luminosity relations.  
     However, these authors used a distance to M\,15 of 12.3 kpc, and when reduced to 10.2 kpc, 
     a mass of only 0.57 \Msun\ follows.     
     Bianchi et al. (\cite{bietal.01}) derived directly a mass of 0.62~\Msun\ by combining
     stellar radius with photospheric gravity, but the error of ($\pm$0.10 \Msun) is too
     high as to make this result useful for further applications.
     Rauch, Heber \& Werner (\cite{rauchetal.02}) got ${M= 0.57^{+0.02}_{-0.01}}$~\Msun\ 
     by means of a NLTE determination of photospheric parameters 
     and a comparison with post-AGB evolutionary tracks in the $\Teff$-$\log g$ plane. 
     
     We conclude that the central star of Ps\,1 has a lower mass of only ${\simeq\!0.57}$
     \Msun, instead of the 0.60--0.62 \Msun\ as is usually assumed in the literature 
     where the Alves, Bond \& Livio and Bianchi et al. results were taken for granted. 
     With a correction factor of 3 (see above) we estimate the true, i.e. post-AGB age 
     of the nebula to ${\simeq 1900{\times}3 \approx 5700}$ years. 
     This value is consistent with the 6500 years predicted for a 0.565 \Msun\ 
     post-AGB model to reach about 40\,000~K 
     (Sch\"onberner \cite{Sch.83}) if one
     considers all the uncertainties concerning the underlying physics like, e.g., 
     post-AGB mass loss history or time of the cessation of the strong AGB wind.

\paragraph{DdDm 1}
     Otsuka et al. (\cite{otsetal.09}) observed also this object with Subaru/HDS and determined 
     an expansion velocity of 11$\pm$1 \kms\ from, e.g., the split \oiii\ line 
     (their Table~3).  This value corresponds, according to our 
     interpretation, to $V_{\rm rim}$.  A post-shock velocity cannot be derived from these
     profiles.   The 10\,\%-level velocity is between 25 and 30 \kms.
     Nevertheless, a rim velocity of 11 \kms\ fits well into the range observed for the other
     metal-poor objects.

\section{Discussion}                       
\label{dis.con}

   Based on high-resolution spectrograms of PNe from the Milky Way disk and halo   
   we performed a detailed study on how the expansion properties and internal kinematics 
   of PNe vary with evolution through the HR diagram, i.e. from the young, compact
   early phase horizontally across the HR diagram and finally down to a stage where the 
   central star is a very hot white dwarf with a comparatively low luminosity.   

   In particular,  we derived, for quite a large sample of objects (23 double-shell,
   5 evolved, and 4 metal-poor PNe),  accurate expansion velocities of the most conspicuous
   part of a PN, the rim. Except for the evolved objects,  we also determined, for many 
   objects for the first time, the gas velocity at the outer edge of the much fainter,
   attached shell, i.e. the post-shock velocity, by a method developed earlier by us.     
   The post-shock velocity is important because it is physically connected 
   to the outer shock's propagation which represents the \emph{true} expansion of a PN.
   The sample investigated here is significantly larger than the ones presented in Paper~II
   and Sch\"onberner et al. (\cite{SJSPCA.05}) and comprises objects with different
   metallicities and evolutionary stages, allowing a meaningful comparison with theory. 
    
    The kinematics of planetary nebulae opens the
    possibility to probe, although only indirectly, the mass-loss history during the 
    final AGB phase (upstream density and post-shock velocity) and the mass-loss 
    evolution during the whole post-AGB stage as well (rim strength and rim expansion). 
    We note explicitly
    that the very final, high mass-loss phase which removes nearly completely the AGB star's
    envelope is very brief and thus not directly observable. Also, any variations of
    the strength of the post-AGB mass loss with stellar temperature, as predicted
    by, e.g. Pauldrach \etal\ (\cite{pauldrachetal.04}), is difficult to measure
    because of the rather high uncertainties involved.  The predicted influence 
    of different AGB and post-AGB wind histories on the PN structure (surface brightness)
    can be seen in Steffen \& Sch\"onberner (\cite{St.Sch.06}, Figs. 5 and 6 therein).
  
   The existence of a nearly stalling rim during the early evolution of a PN sets also 
   a lower limit to the final mass-loss rate on the AGB which is necessary to build up
   sufficient thermal pressure
   as to decelerate the inner region of the newly formed ionised shell against the pressure of
   the wind-shocked hot bubble.  Based on our hydrodynamical simulations 
   we estimate a lower limit of the final AGB mass-loss rate of the order of 
   $1{\times}10^{-5}$ \Mdot. 

    Since our radiation-hydrodynamics simulation were performed without having any detailed
    comparison with observations in mind and without any fine-tuning, it is not 
    astonishing that none of our model sequences describes all aspects of the PN kinematics 
    quantitatively correct.  
    Taken all simulations together, however, we find a fair and satisfying 
    agreement of computed nebular quantities with their observed counterparts (cf. also
    Steffen \& Sch\"onberner \cite{St.Sch.06}).
   
   From all these evidences discussed here it is clear that the existence of two shell
   components with distinct properties, such as densities and
   flow velocities, as seen in nearly all PNe which are in the appropriate evolutionary stage, 
   is just a natural consequence of the prevailing physics and cannot be attributed to two
   mass loss events with different strengths and velocities, and which are separated in time.
   
   The high post-shock velocities found for the objects of our sample are in 
   big contrast to the canonical ``expansion'' velocity value of 20--25 \kms\ 
   usually used in the literature. 
   The mean post-shock velocity found by us is $\sim$\,40 \kms, which corresponds to a 
   true expansion rate of, on the average, 40$\times$1.25 = 50 \kms, i.e. about twice 
   of the generally adopted value!  
   The reason is that, erroneously, always the peak-to-peak separation of the strongest 
   line profile components were taken as the typical expansion velocity.  
   The mean rim velocity of
   our sample in Table~\ref{tab.gesamt} is 18~\kms, i.e. close to the canonical value.      
   
   In the study on Magellanic Cloud PNe, Dopita \etal\ (\cite{dopetal.85}) introduced 
   an ``expansion'' velocity based on the 10\,\% level of the line profile.  
   This method is much easier to perform,
   even with data of poorer quality, and yields results much closer to the post-shock velocity. 
   However, the velocity at the 10\,\% level is, of course, not necessarily equal to the 
   post-shock velocity because it has no physical meaning.  
   Rather this choice is a compromise in 
   the sense of being close to the maximum nebular flow velocity with still sufficient signal 
   above the background.  Additionally, the line profile, and the flux at the 10\,\% level,
   depends on evolutionary stage,  the ion used (cf. Fig.~\ref{decomposition}), and the 
   spectral resolution.   
   
   As example we can use the line profiles of \object{NGC 6826} shown in
   Fig.~\ref{line.analyse}.
   The 10\,\% velocities are 38 \kms\ for \nii\ and 32 \kms\ for \oiii, while  
   $V_{\rm post} = 34$ \kms\ (cf. Table~\ref{tab.gesamt}).
   Additional examples are provided by the work of Medina \etal\ (\cite{medetal.06})
   who applied the 10\,\%-level method as well.     
   For NGC 7009, Medina et al. got ${V_{10\%} = 33}$ \kms\ (\oiii) and 38 \kms\ (\nii),
   while $V_{\rm post} = 36$ \kms, and for NGC 6543 
   $V_{10\%} = 43$ \kms\ from both ions, but  ${V_{\rm post} = 47}$ \kms. 
   Anyway, the 10\,\% velocities are much closer to $V_{\rm post}$ than the
   results from the line-peak separation which are for the three objects 8, 16, and 18 \kms,
   respectively (see Table~\ref{tab.gesamt}).  

   In this context we warn the reader to put too much confidence into any
   kinematical age determination of a PN and to use this value for making 
   quantitative statements on, e.g., central-star masses. Even with a precise value of
   an ``expansion'' velocity and proper nebular radius, a correction must be applied that
   accounts for the difference between measured and true expansion velocity and its
   acceleration during the past.  Due to the complex hydrodynamical processes involved,
   this correction is rather uncertain and depends as well on the evolutionary state.
   Without such a correction, any determination of a post-AGB age from the nebular size
   and expansion is completely useless.

    We could not detect any significant deceleration of nebular matter during the evolution 
    around maximum stellar temperature and down to low stellar luminosities, 
    as recently reported by Pereyra et al. (\cite{pereyra.13}). 
    However, the number of objects investigated here is too small as to make a firm conclusion 
    on this matter. We only remark that nebulae in such phase are prone to 
    ionisation stratification, making any velocity measurements ion-dependent! 
    The numerous PNe which do not recombine because they have more slowly evolving nuclei 
    will certainly not decelerate their expansion if the nucleus fades.

    Finally, we repeat that all statements made in this study refer to the expansion 
    properties of PNe
    with central stars of normal surface composition only. Nebula around Wolf-Rayet
    central stars have a faster expanding rim, obviously because of the much stronger 
    stellar wind (cf. Medina \etal\ \cite{medetal.06}). It is expected that
    the expansion properties of PNe with a Wolf-Rayet central star, if coupled
    to hydrodynamical simulations, could establish very useful constraints for Wolf-Rayet
    winds.  However, because the formation
   mechanism and the evolution of hydrogen-deficient post-AGB stars are not known to date, 
   hydrodynamic simulations are impossible.

\section{Results and conclusion}    
\label{sect.res.concl}

   The results of our study are in line with
   the prediction of radiation-hydrodynamic simulations in that 
   the interaction of the fast central-star wind with the slow AGB wind via a
   hot bubble of shock-heated wind gas is not the main driving agent of a PN:   At the
   beginning of the post-AGB evolution and during the late evolution with a faint
   central star the wind power is just sufficient to avoid the collapse of the PN
   towards the star.  Only during the middle of the nebular evolution, if the 
   wind power increases to its maximum (cf. Fig.~\ref{mod.prop}, middle panel),
   the winds interaction gets important and forms the conspicuous nebular rims.
   But even then most of the nebular mass is still contained in the less prominent
   shells, and the expansion speed of their leading shocks is not influenced at all 
   by the stellar wind.

   Our main results/conclusions are as follows:
\begin{itemize}  
\item
   There exists a rather tight correlation of the internal kinematics 
   with age/stellar temperature which, however, is
   distinct between rim and shell. The shell's leading shock accelerates 
   from the AGB wind speed of ${\simeq\! 10}$ \kms\
   to about 50 \kms\ (${V_{\rm post}\simeq 40}$ \kms), first as  
   a D-type ionisation front and then in a ``champagne''-flow configuration,  
   because (i) the electron 
   temperature increases in line with the growing energy of the ionising photons, and (ii) 
   the density gradient of the halo may steepen with distance from the star.      
   The rim, on the other hand, is driven by the stellar wind power which increases as 
   the star  shrinks. For young objects the rim matter expands even \emph{slower} than the 
   former AGB wind, 
   but at the end of the horizontal evolution values of about 30 \kms\ are reached.
 
\item   
   Since both rim and shell obviously expand outwards independently of each other,
   the expansion of a PN is by no means ballistic. Rather, 
   \emph{the whole PN is a dynamical system throughout its entire evolution, 
   without reaching a self-similar stage of expansion at any time.}    
\item   
   The difference between rim expansion and post-shock velocity
   \emph{decreases} with progress of evolution from about 25 \kms\
   for the youngest objects till about 15 \kms\ for the most evolved objects of our sample.
\item
   The high expansion velocities of the outer shock indicate a steep gradient in the radial 
   upstream density profile, i.e. in the halo, which is only possible if the mass loss 
   during the final AGB evolution ``accelerates'' continuously.   
   This may be an indication that in the normal case the PN precurser left the tip of the AGB
   quite soon after the last thermal pulse on the AGB, i.e. during recovering from the pulse.
   The only exception in our sample is NGC 2022 whose unusual low post-shock velocity
   hints to a period of constant final AGB mass-loss rate.
\item 
   The line profiles of four targets with intrinsically faint but very hot central stars 
   could be interpreted in terms of a reionisation process after a 
   rather severe but short recombination phase caused by the rapid luminosity drop of the star.  
   We found in all cases quite high flow velocities of 40--50 \kms\
   at the outer edge of the newly created reionised shell which are comparable to
   those measured for objects which are still in the high-luminosity phase of their evolution.
\item  
  The expansion behaviour of the four metal-poor objects is as expected from theory:
  Compared to objects with normal (i.e. Galactic disk) metallicity,
  post-shock velocities are generally higher because of the higher electron
  temperature,  and rim velocities lower because of the weaker winds interaction.
\end{itemize}
   To summarise up, 
   our study demonstrates, in line with our previous ones, that the
   expansion of a PN is much faster than assumed so far.  The reason is an incorrect
   assignment of a measured velocity, usually that of the most conspicuous part
   of the nebulae (i.e. the rim), to be the ``expansion velocity'' of a PN.  
   Instead, it is the outer shell's shock whose propagation determines the true
   expansion of a PN.  The only velocity which is physically related to the shock 
   expansion is the gas velocity immediately behind the shock, the post-shock velocity.
   But this velocity is, in general, higher than the velocity measured for the
   rim by factors     
   up to about 7 for very  young objects.

   In closing we state that the
   \emph{expansion velocities of PNe quoted in the literature and frequently used in 
   statistical studies must be treated with utmost care.  It is absolutely necessary
   to check which kind of velocity is measured and how this velocity refers to
   the real expansion velocity of the object in question.}    

\acknowledgement
   We are grateful to the referee for his careful reading of the manuscript and the 
   very short reviewing time.  C.\,S. acknowledges support by DFG grant SCHO 394/26.


\vspace*{-1mm}

\appendix

\vspace*{-2mm}
\section{Compilation of line profiles and velocity data}
\label{comp.lines}

\begin{figure*}
\includegraphics[width=0.99\linewidth, clip]{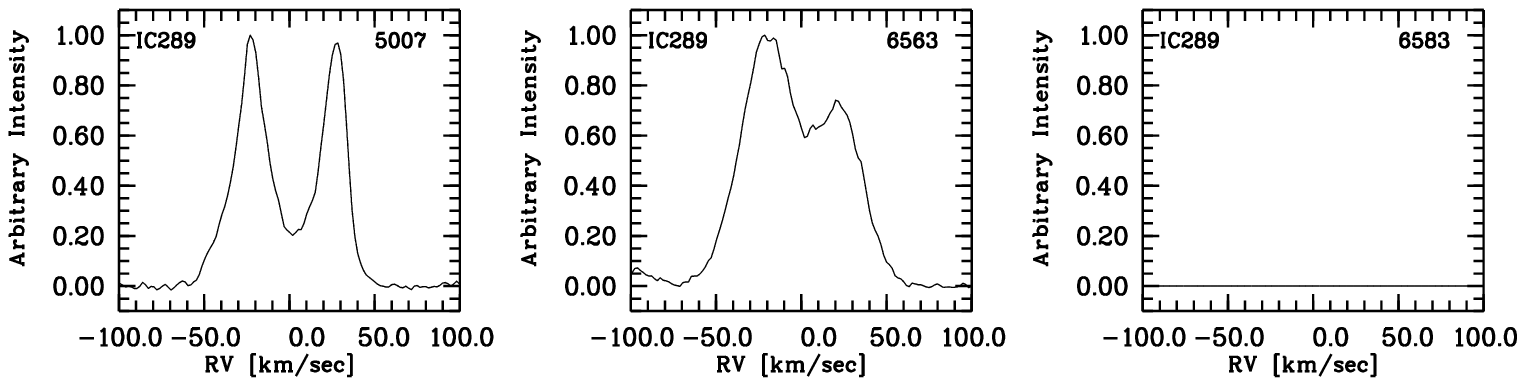}
\vskip3mm
\includegraphics[width=0.99\linewidth, clip]{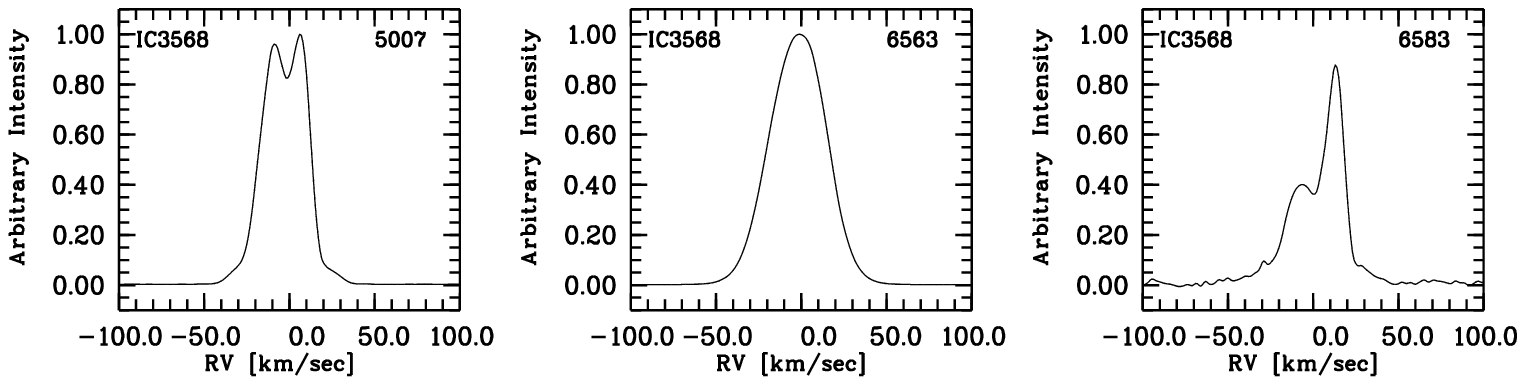}
\vskip3mm
\includegraphics[width=0.99\linewidth, clip]{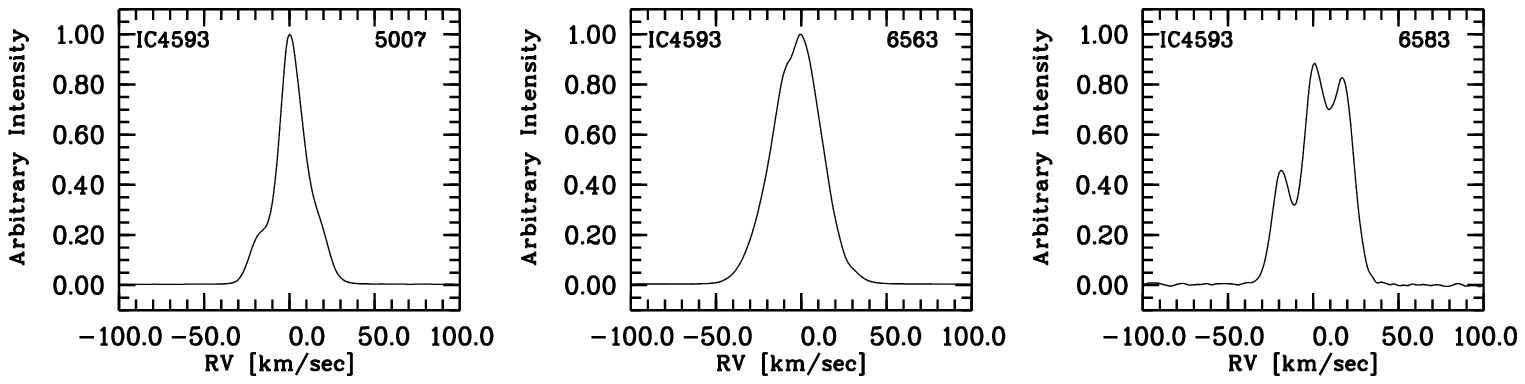}
\vskip3mm
\includegraphics[width=0.99\linewidth, clip]{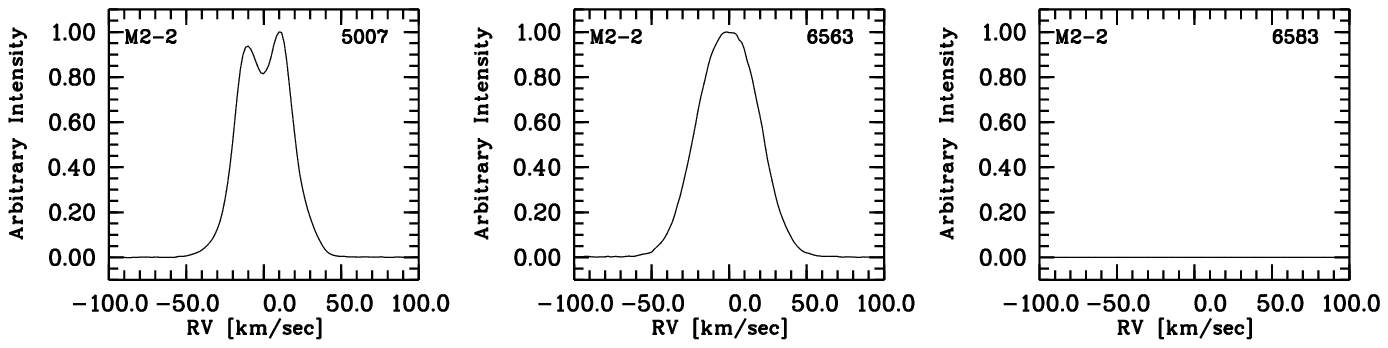}
\vskip3mm
\includegraphics[width=0.99\linewidth, clip]{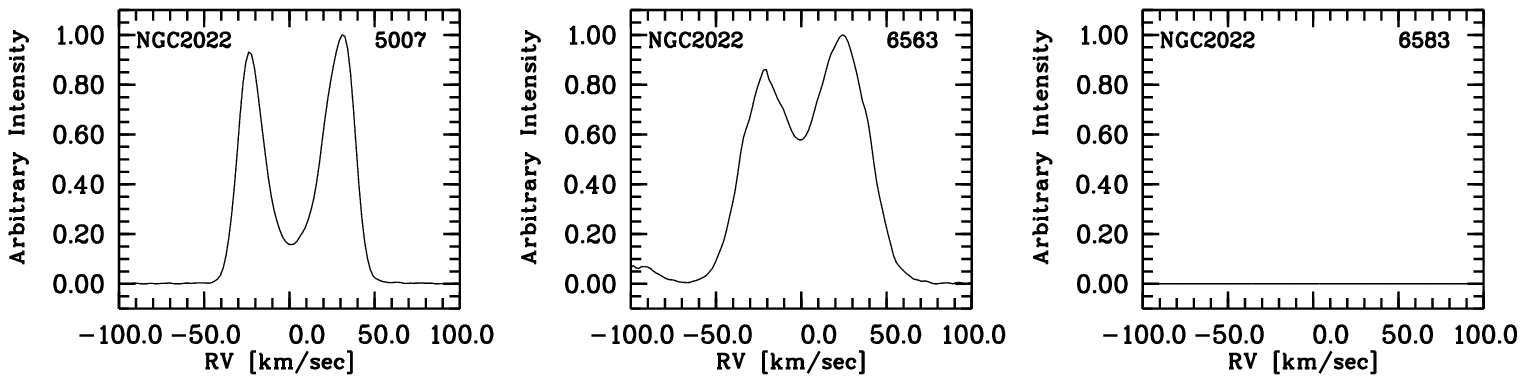}
\caption{\label{multi.1}
         Normalised, central line-of-sight emission line profiles of \oiii $\lambda$5007\,{\AA},
         H$\alpha$, and \nii\ $\lambda$6583\,{\AA}. 
	     Empty panels: line not observed.  \vspace{1cm}
        }
\end{figure*}

  The finally extracted line profiles of 17 objects are collected in Fig. \ref{multi.1}.  
  In cases of multiple observations, only one has been 
  selected. The individual velocities $V_{\rm rim}$ and $V_{\rm post}$ based on the 
  strong [\ion{O}{iii}] and [\ion{N}{ii}] lines
  of our target objects are compiled in Table \ref{data.tautenburg}.
  Data for H$\alpha$ are not given because shell and rim are rarely resolved.
  For comparison, the last two columns give values from the literature.

%
\addtocounter{figure}{-1}
\begin{figure*}
\includegraphics[width=0.99\linewidth, clip]{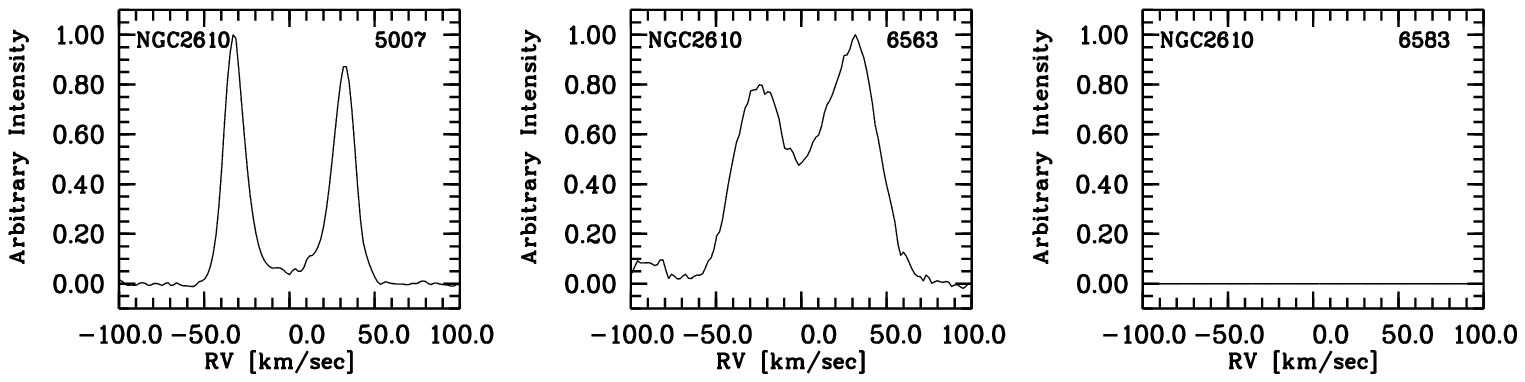}
\vskip3mm
\includegraphics[width=0.99\linewidth, clip]{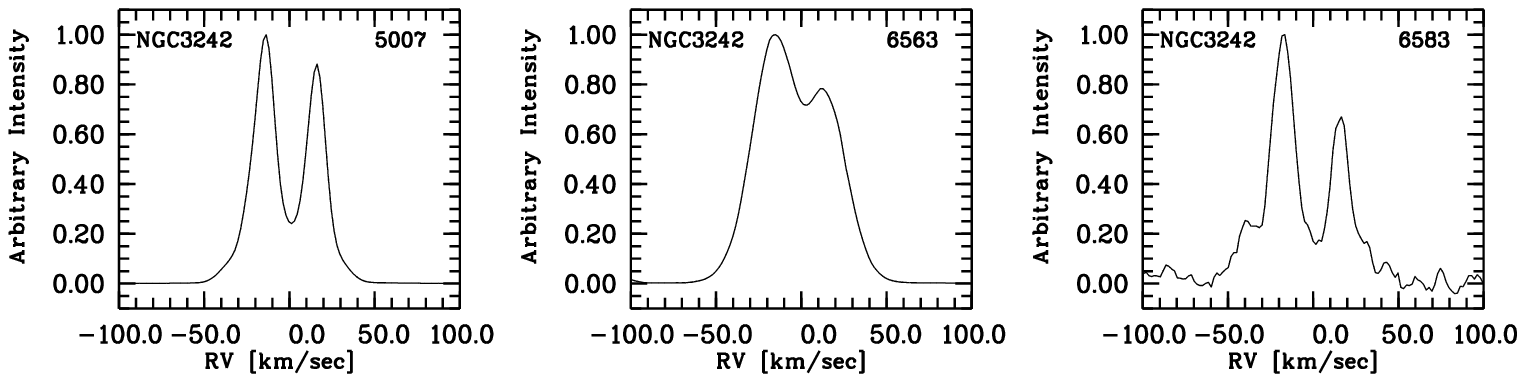}
\vskip3mm
\includegraphics[width=0.99\linewidth, clip]{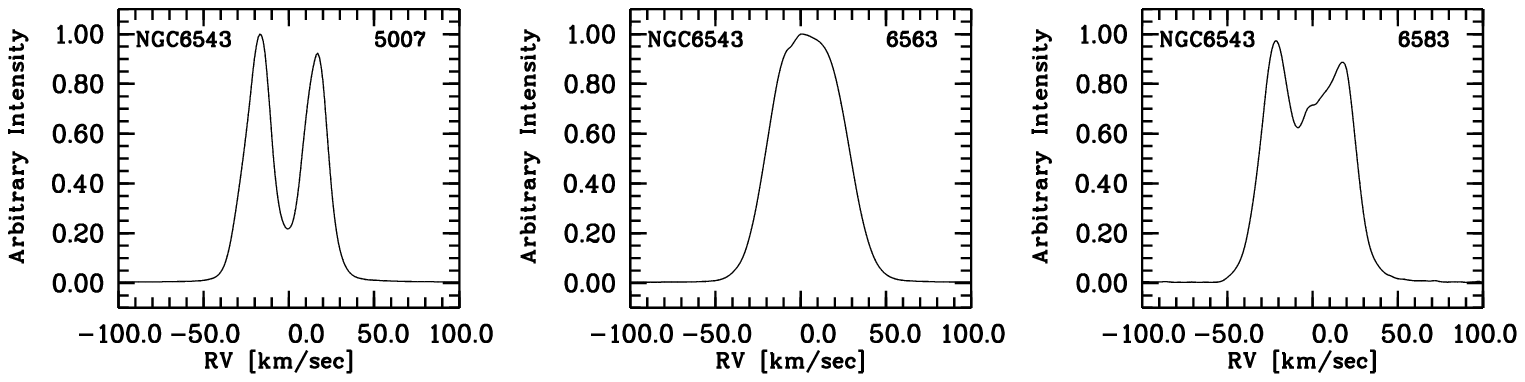}
\vskip3mm
\includegraphics[width=0.99\linewidth, clip]{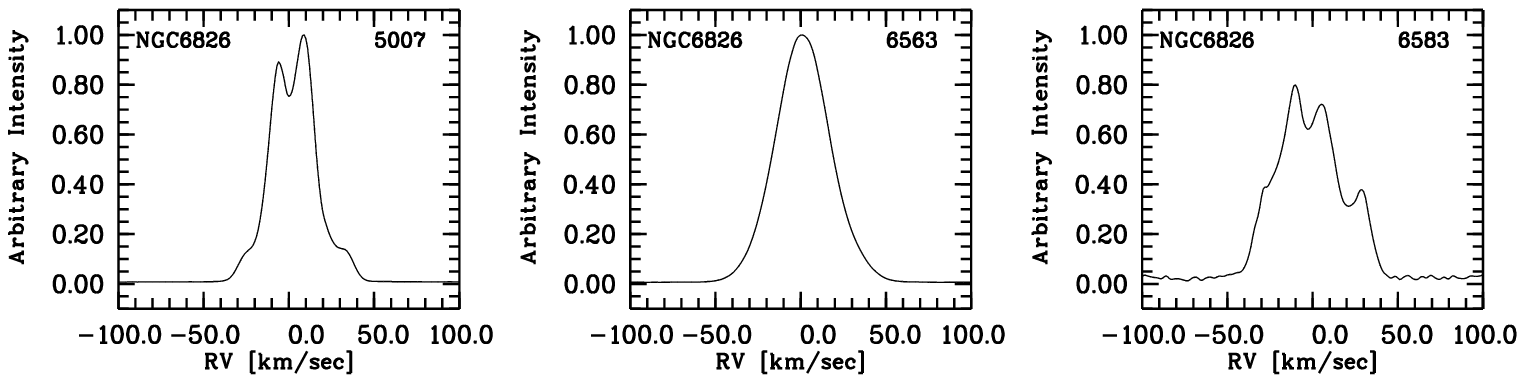}
\vskip3mm
\includegraphics[width=0.99\linewidth, clip]{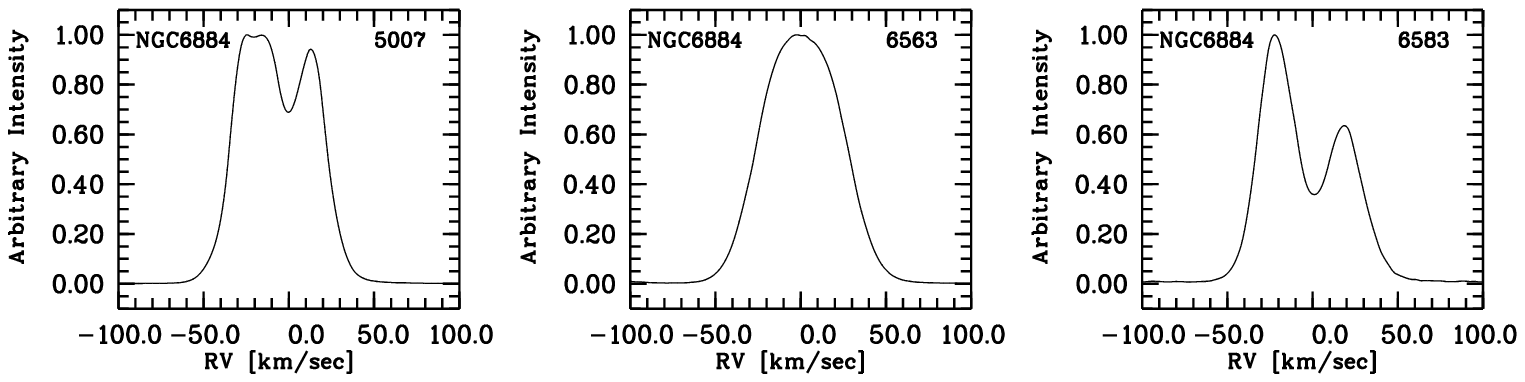}
\caption{\label{multi.2}
          Figure A1 continued.
          \vspace{1cm}
        }
\end{figure*}
%
\addtocounter{figure}{-1}
\begin{figure*}
\includegraphics[width=0.99\linewidth, clip]{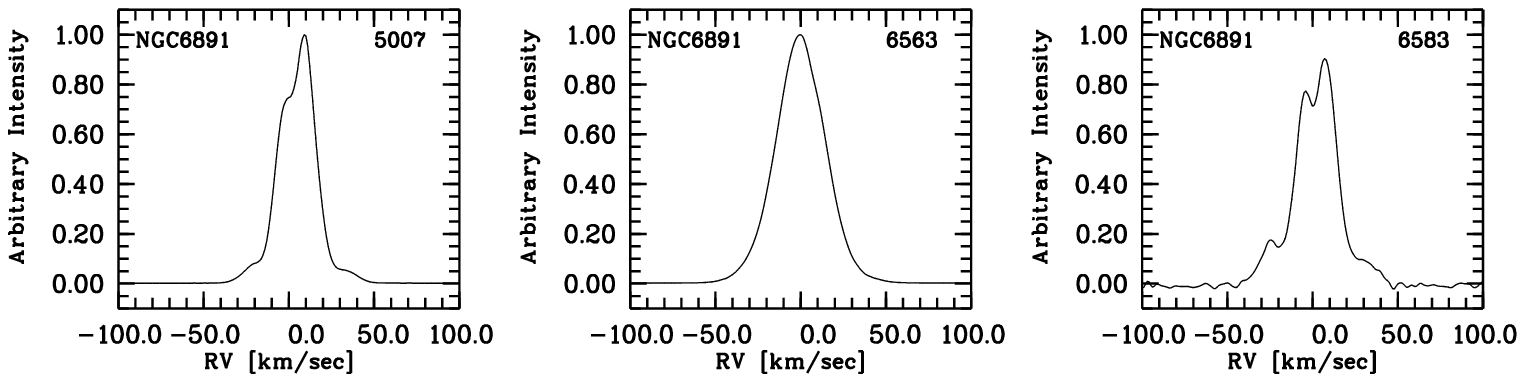}
\vskip3mm
\includegraphics[width=0.99\linewidth, clip]{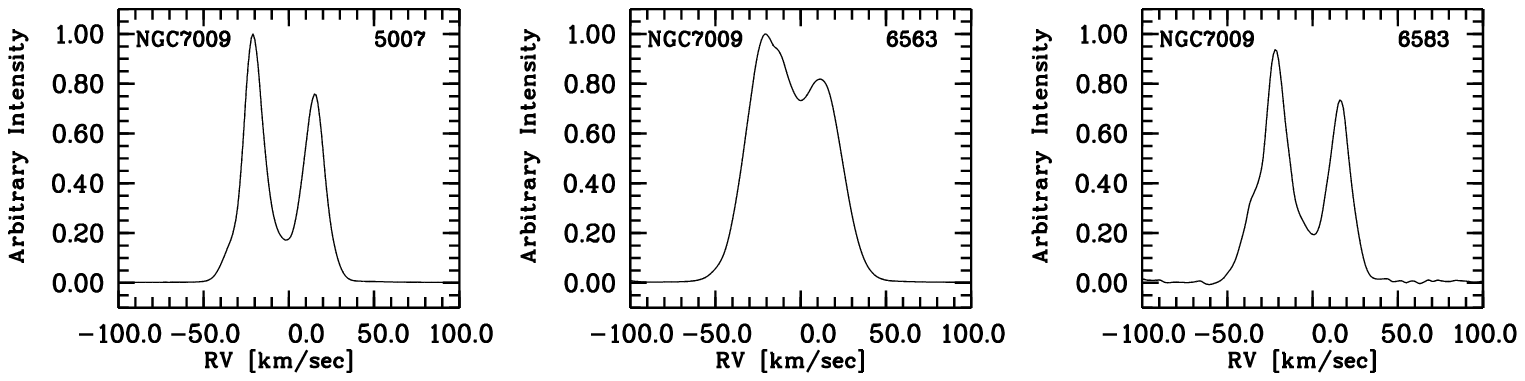}
\vskip3mm
\includegraphics[width=0.99\linewidth, clip]{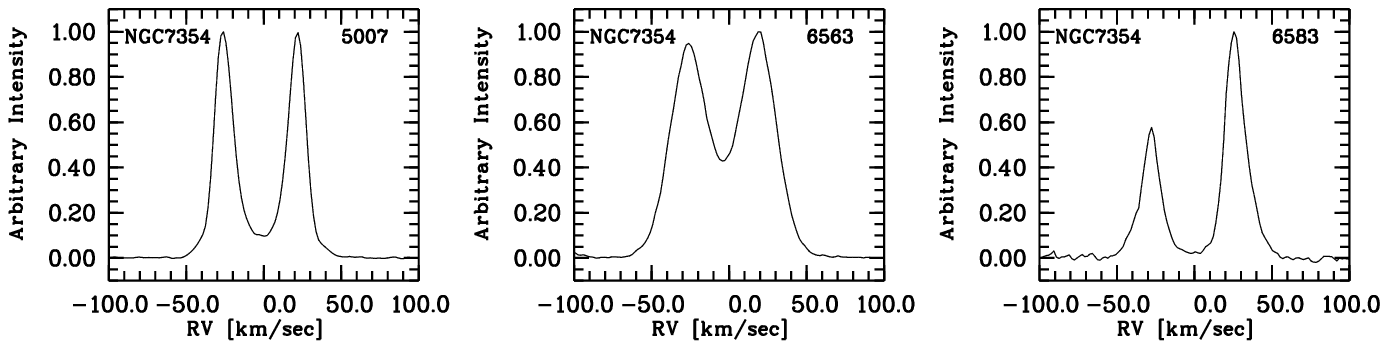}
\vskip3mm
\includegraphics[width=0.99\linewidth, clip]{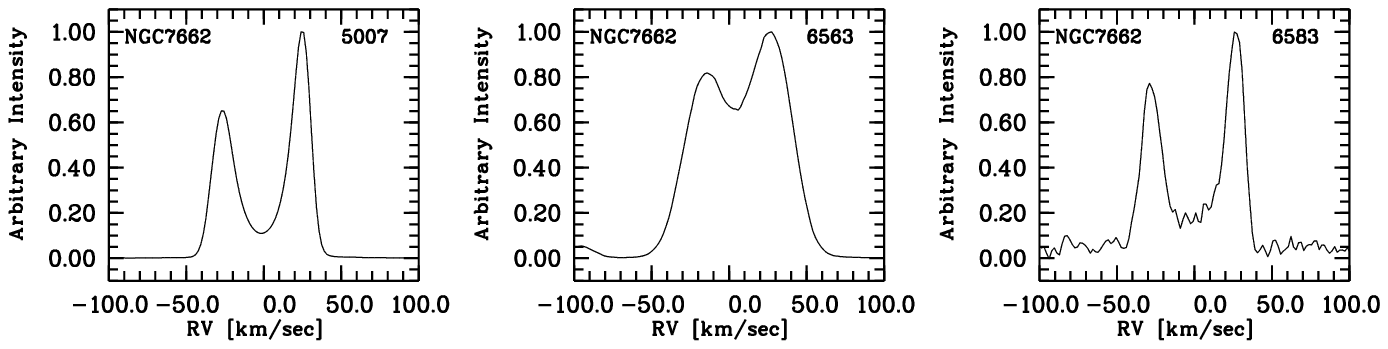}
\vskip3mm
\includegraphics[width=0.99\linewidth, clip]{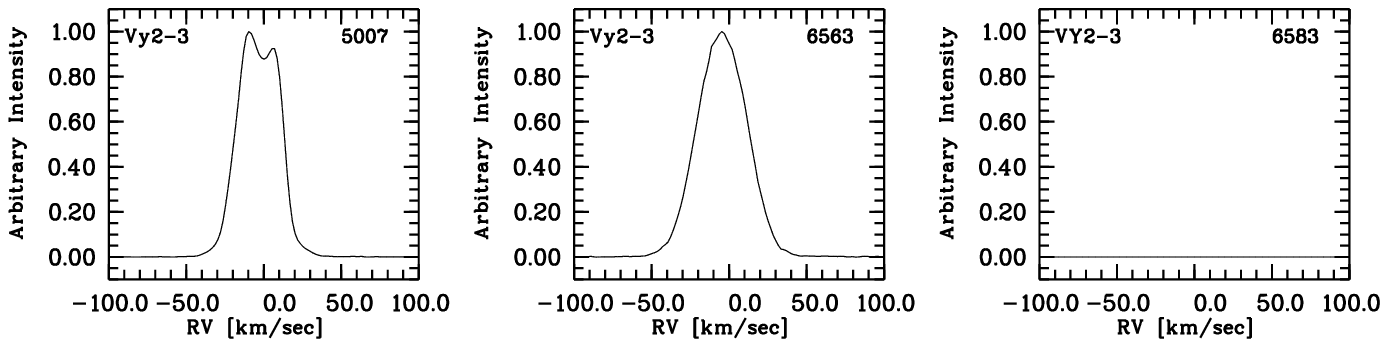}
\caption{\label{multi.3}
         Figure A1 continued.
          \vspace{1cm}
        }
\end{figure*}

\begin{table*}[ht]
\caption{Individual velocity measurements for rims and shells from central [\ion{O}{iii}]
         $\lambda\lambda$ 4959, 5007 \AA\ and [\ion{N}{ii}] $\lambda\lambda$ 6548, 6583\,\AA\
         emission lines based on Tautenburg \'echelle spectrogrammes of PNe in their high
         central-star luminosity evolutionary stage. Rim velocities
         ($V_{\rm rim}$) are from the peak separarations of the two innermost Gaussians
         fitted to the strong line profile components, while shell velocities ($V_{\rm post}$)
         refer to the separations of the extremes of the line profile derivatives 
         (or sigmas of shell Gaussians) at
         the line wings/shoulders as described in Sect.~\ref{obs.red}.
         }
\label{data.tautenburg}
\tabcolsep=5.2pt
\scalebox{0.999}{
\begin{tabular}{lcccccccccccc} 
\hline
\noalign{\smallskip}
Object    &  \multicolumn{4}{c}{$V_{\rm rim}$ (km\,s$^{-1}$)}
           &
           &  \multicolumn{4}{c}{$V_{\rm post}$ (km\,s$^{-1}$)}  &
           &  \multicolumn{2}{c}{Literature\,$^{\rm a}$}    \\[1.5pt]
\cline{2-5}
\cline{7-10} \cline{12-13}
\noalign{\smallskip}
           &  4959 \AA  &  5007 \AA  &  6548 \AA  &  6583 \AA  & &  4959 \AA  &  5007 \AA  &  6548 \AA  &  6583 \AA  &
           & \oiii  & \nii \\[1.5pt]
\hline
\noalign{\smallskip}
IC 289     &  25.7  &  25.8  &    --  &    --  &   &    --  &  43.2  &    --  &    --  &   & 22, 27    &   --  \\
           &    --  &  25.2  &    --  &    --  &   &    --  &    --  &    --  &    --  &   &   --      &   --  \\
           &  25.2  &  24.9  &    --  &    --  &   &    --  &  50.7  &    --  &    --  &   &   --      &   --  \\
           &  26.1  &  25.3  &    --  &    --  &   &    --  &  45.5  &    --  &    --  &   &   --      &   --  \\
\noalign{\smallskip}
IC 3568    &    --  &    --  &\enspace9.3  &\enspace9.6  &   &    --  &    --  &    --  &  30.6  &   & 8\,$^2$   &       \\
           &\enspace8.1  &\enspace8.3  &    --  &    --  &   &  33.5  &  33.2  &    --  &    --  &   &           &       \\
           &\enspace9.3  &\enspace9.3  &    --  &    --  &   &  33.2  &  32.8  &    --  &    --  &   &           &       \\
\noalign{\smallskip}
IC 4593   &    --  &    --  &\enspace4.5  &7\,\rlap{$^{\rm b}$}  &   &    --  &    --  &  23.3  &  24.2  &   & 8\,$^3$, 12\,$^3$ &       \\
           &    --  &    --  & \enspace3.5  &7\,\rlap{$^{\rm b}$}  &   &    --  &    --  &  24.6  &  23.8  &   &                   &       \\
           &\enspace3.1  &\enspace3.6  &    --  &    --                  &   &  22.0  &  22.8  &    --  &    --  &   &                   &       \\
\noalign{\smallskip}
M\,2-2    &  11.1  &  11.3  &    --  &    --  &   &  33.4  &  37.4  &    --  &    --  &   &           & 17\,$^6$  \\
           &  11.6  &  11.6  &    --  &    --  &   &  36.9  &  35.1  &    --  &    --  &   &           &           \\
           &  10.9  &  11.0  &    --  &    --  &   &  34.0  &  37.6  &    --  &    --  &   &           &           \\
           &  11.2  &  11.0  &    --  &  11.0  &   &  37.6  &  37.4  &    --  &    --  &   &           &           \\
\noalign{\smallskip}
NGC 2022  &  28.7  &  28.9  &    --  &    --  &   &    --  &    --  &    --  &    --  &   & 29\,$^4$, 27\,$^5$ &     \\
           &  29.2  &  29.1  &    --  &    --  &   &    --  &    --  &    --  &    --  &   &                    &     \\
\noalign{\smallskip}
NGC 2610  &  32.6  &  32.0  &    --  &    --  &   &    --  &    --  &    --  &    --  &   &           &       \\
           &  32.3  &  32.5  &    --  &    --  &   &    --  &    --  &    --  &    --  &   & 14, 34    &       \\
\noalign{\smallskip}
NGC 3242  &  15.7  &  15.9  &  17.2  &  18.0  &   &  38.3  &  39.1  &    --  &    --  &   & 19.5\,$^5$, 35.7\,$^5$ & 19.7\,$^5$ \\
           &  16.6  &  16.6  &    --  &  18.4  &   &  40.3  &  40.0  &    --  &    --  &   &                        &            \\
           &  15.2  &  15.3  &    --  &  17.5  &   &  37.2  &  37.5  &    --  &  40.3  &   &                        &            \\
           &  14.9  &  15.0  &    --  &  17.1  &   &  37.9  &  38.4  &    --  &  38.9  &   &                        &            \\
\noalign{\smallskip}
NGC 6543  &    --  &    --  &  20.5  &  20.8  &   &    --  &    --  &  44.0  &  45.6  &   & 19.5, 43\,$^7$ & 20, 43\,$^7$ \\
           &  16.3  &  16.5  &    --  &    --  &   &    --  &    --  &    --  &    --  &   &                &              \\
\noalign{\smallskip}
NGC 6826  &    --  &    --  &\enspace8.9  &\enspace8.6  &   &    --  &    --  &  33.9  &  33.7  &   &    & 10.6\,$^6$, 28.3\,$^6$ \\
           &    --  &    --  &\enspace8.6  &\enspace8.2  &   &    --  &    --  &  33.1  &  33.7  &   &    &                        \\
           &\enspace7.6  &\enspace7.6  &    --  &    --  &   &  32.8  &  33.1  &    --  &    --  &   &    &                        \\
           &\enspace7.8  &\enspace7.8  &    --  &    --  &   &  31.5  &  32.6  &    --  &    --  &   &    &                        \\
\noalign{\smallskip}
NGC 6884  &    --  &    --  &  18.9  &  19.4  &   &    --  &    --  &    --  &    --  &   & 23 & 24.0\,$^6$ \\
\noalign{\smallskip}
NGC 6891  &    --  &    --  & \enspace7.5  &\enspace6.7  &   &    --  &    --  &  32.3  &  34.4  &   &  7 & 9.7\,$^6$, 30\,$^6$ \\
           &    --  &    --  &\enspace7.3  &\enspace6.8  &   &    --  &    --  &  34.4  &  34.0  &   &    &                     \\
           &   6.6  &  6.6   &    --  &    --  &   &  31.0  &  31.7  &    --  &    --  &   &    &                     \\
\noalign{\smallskip}
NGC 7009  &    --  &    --  &  18.4  &  18.6  &   &    --  &    --  &   38.3  &   35.4 &   & 20, 33\,$^7$ & 20, 38\,$^7$ \\
           &  17.8  &  17.8  &    --  &    --  &   &  35.9  &  36.1  &    --  &    --  &   &              &              \\
\noalign{\smallskip}
NGC 7354  &  --    &   --   & 27.2 & 27.2 &  & --   &  --  & 45.7 & 45.6 & &  25 &  28\,$^8$, 35\,$^8$          \\
           &  23.8  &  23.8  &  --  & 27.6 &  &  --  & 44.2 &  --  &  --  & &     &              \\
           &  23.6  &  24.0  &  --  & 26.5 &  &  --  &  --  &  --  & 46.1 & &     &              \\
           &  24.4  &  24.2  & 27.7 & 26.3 &  & 43.8 & 46.9 &  --  &  --  & &     &              \\
           &  23.5  &  23.7  & 26.9 & 26.8 &  &  --  &  --  & 42.7 & 46.3 & &     &              \\
\noalign{\smallskip}
NGC 7662  &    --  &    --  &  26.9  &  26.7  &   &    --  &    --  &    --  &    --  &   & 26.1\,$^5$ & 25.4\,$^6$ \\
           &  25.5  &  25.7  &  28.1  &  28.1  &   &    --  &    --  &    --  &    --  &   &            &            \\
\noalign{\smallskip}
Vy\,2-3   &\enspace8.8  &\enspace8.7  &    --  &    --  &   &  34.9  &  34.6  &    --  &    --  &   & 16, 12     & 13.0\,$^6$ \\
           &\enspace8.7  &\enspace8.7  &    --  &    --  &   &  34.6  &  35.4  &    --  &    --  &   &            &            \\
\hline
\end{tabular}
}
\\[3pt]
\emph{Notes}: Uncertain measurements are rounded to full \kms.
$^{\rm a}$\,If not noted otherwise, from Weinberger (\cite{We.89}) which incorporates
           all the previous references. 
$^{\rm b}$\,From half width at half maximum of the central Gauss profile. \\
$^1$\,Flannery \& Herbig (\cite{FH.73}). 
$^2$\,The \oiii\ line profile presented in Gesicki \etal\ (\cite{GAZ.96})
   shows weak wings similar to our observation, indicating a comparatively high shell
   velocity (Tab.\,\ref{multi.1}). 
$^3$\,From Weinberger (\cite{We.89}), but for H$\alpha$; 
$^4$\,Sabbadin, Bianchini \& Hamzaoglu (\cite{SBH.84}); 
$^5$\,Sch\"onberner \etal\ (\cite{SJSPCA.05}); 
$^6$\,Guerrero \etal\ (\cite{GVM.98});
$^7$\,$V_{10\%}$ velocity from Medina \etal\ (\cite{medetal.06});
$^8$\,Contreras et al. (\cite{contreras.10}).
\vspace{1cm}
\end{table*}

\end{document}